\documentclass[
	english,
	a4paper,
	pagesize,
	pdftex,
	12pt,
	twoside, % + BCOR darunter: für doppelseitigen Druck aktivieren, sonst beide deaktivieren
	BCOR=5mm, % Dicke der Bindung berücksichtigen (Copyshop fragen, wie viel das ist)
	ngerman,
	fleqn,
	final,
	]{scrartcl}
\usepackage{ucs}
\usepackage[utf8x]{inputenc} % Eingabekodierung: UTF-8
\usepackage{fixltx2e} % Schickere Ausgabe
\usepackage[T1]{fontenc} % ordentliche Trennung
\usepackage[ngerman]{babel}
\usepackage{lmodern} % ordentliche Schriften
\usepackage{setspace,graphicx,tikz,tabularx} % für Elemente der Titelseite
\usepackage[draft=false,babel,tracking=true,kerning=true,spacing=true]{microtype} % optischer Randausgleich etc.

\defcaptionname{ngerman}{\refname}{References} % rename "Literatur" to "References"
\defcaptionname{ngerman}{\figurename}{Figure} % rename "Abbildung" to "Figure"
\defcaptionname{ngerman}{\contentsname}{Table of Contents} % rename "Inhaltsverzeichnis" to "ToC"
\defcaptionname{ngerman}{\listfigurename}{List of Figures}
\defcaptionname{ngerman}{\tablename}{Table}
\defcaptionname{ngerman}{\listtablename}{List of Tables}
\addto\captionsngerman{} %alg list title english

\graphicspath{ {./plots/} }

\usepackage[nottoc,notlof,notlot,numbib]{tocbibind} % add "References" to table of contents
\usepackage{float} % to place figures at a fixed point
\usepackage{amsmath}
\usepackage{amssymb}
\usepackage{xcolor}
\usepackage[ruled, english]{algorithm2e}
\usepackage{setspace}

\usepackage{hyperref}
\hypersetup{
unicode=true,
	pdftitle={Efficient Approximation of Centrality Measures in Uncertain Graphs},
	pdfsubject={Algorithms for Centrality Measures in Uncertain Graphs},
	pdfauthor={Daniel Ketels},
	pdfkeywords={graph theory, network analysis, Uncertain Graphs, Approximation Algorithms,
	Heuristic Algorithms, Efficient Algorithms, Computational Complexity, 
	centrality, Betweenness centrality, closeness centrality, harmonic closeness centrality, centrality measures, possible shortest paths, probabilistic graphs,
	Generative Models, Dynamic Networks, Parallel Graph Algorithms, Big Data Analytics, 
		Graph Approximation, Algorithmic Graph Theory, Graph Algorithms, Randomized Algorithms, Optimization in Graphs, Algorithms, Uncertain Graph Distance,
		Parallelization, C++, OpenMP, algorithms, combinatorics, optimization algorithms}
}

\usepackage{mathtools}
\usepackage{subcaption}

\usepackage{setspace}

\DeclareRobustCommand{\bbone}{\text{\usefont{U}{bbold}{m}{n}1}}

\newtheorem{definition}{Definition}

\SetKwBlock{Repeat}{repeat}{}
\newcommand{\hrulealg}[0]{\vspace{1mm} \hrule \vspace{1mm}}

\usepackage{pdfpages}  % neues deckblatt

\usepackage[
    	type={CC},
    	modifier={by-sa},
    	version={3.0},
]{doclicense}

\begin{document}

%% Beispielhafte Nutzung der Vorlage für die Titelseite (bitte anpassen):
%\input{Institutsvorlage}
%\titel{Efficient Approximation of Centrality Measures in Uncertain Graphs} % Titel der Arbeit
%\typ{Bachelorarbeit} % Typ der Arbeit:  Diplomarbeit, Masterarbeit, Bachelorarbeit
%\grad{Bachelor of Science (B. Sc.)} % erreichter Akademischer Grad
%% z.B.: Master of Science (M. Sc.), Master of Education (M. Ed.), Bachelor of Science (B. Sc.), Bachelor of Arts (B. A.), Diplominformatikerin
%\autor{Daniel Ketels} % Autor der Arbeit, mit Vor- und Nachname
%\gebdatum{03.09.1998} % Geburtsdatum des Autors
%\gebort{Tondern} % Geburtsort des Autors
%\gutachter{Prof. Dr. Henning Meyerhenke}{Prof. Dr. Johannes Koebler} % Erst- und Zweitgutachter der Arbeit
%\mitverteidigung % entfernen, falls keine Verteidigung erfolgt

\includepdf[pages=-]{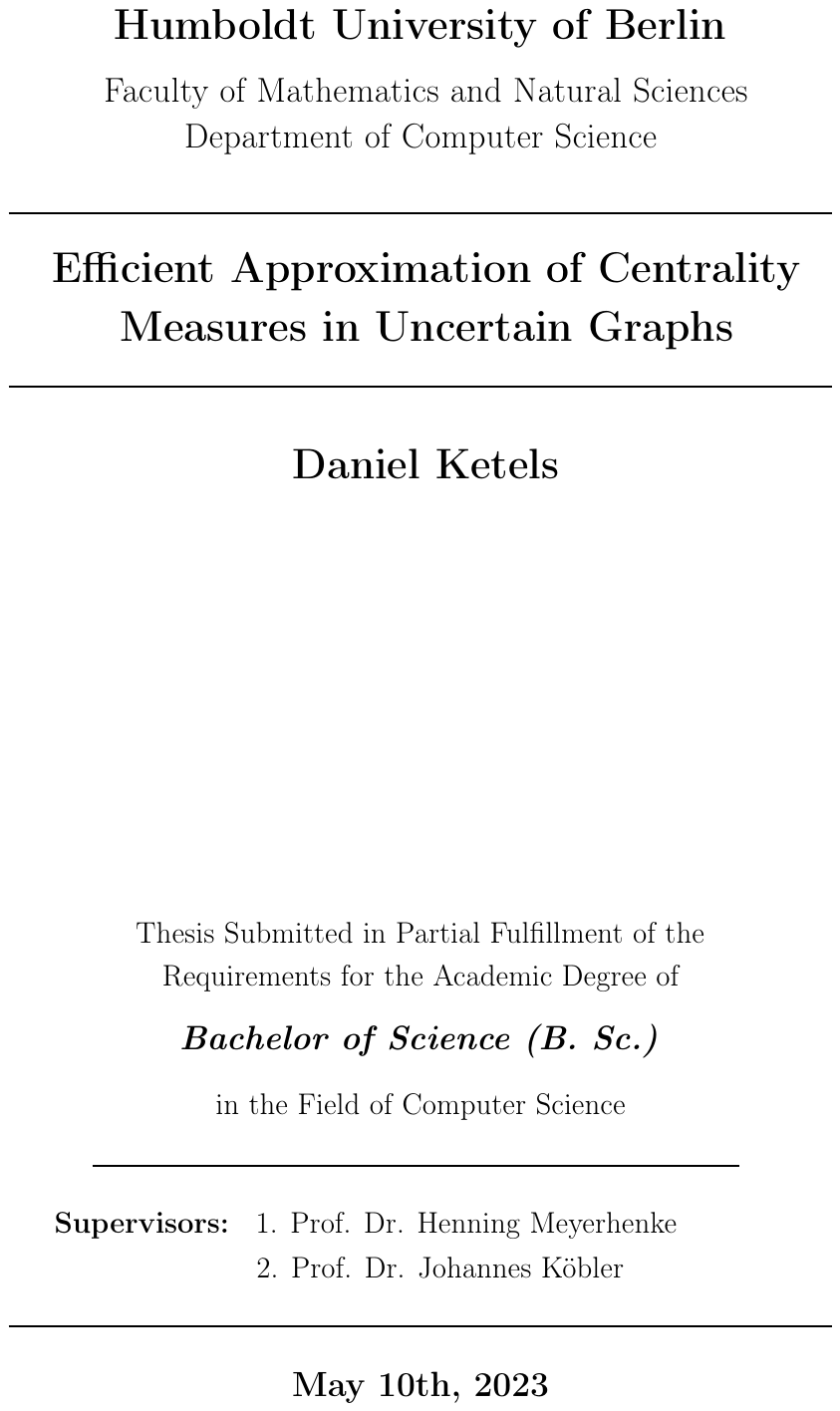}

% Hier folgt die eigentliche Arbeit (bei doppelseitigem Druck auf einem neuen Blatt):
\newpage

\pagenumbering{gobble}

\null \quad 
\vfill

 \begin{center}
		%\doclicenseImage  \\ \vspace{0.3cm} 
 	  '\textit{Efficient Approximation of Centrality Measures in Uncertain Graphs}' © 2023 by Daniel Ketels is registered for public access under the CC BY-SA 4.0. license. \\ 
	  More info can be found here: \url{https://creativecommons.org/}
	  %This grants everyone open access as well as permission for any use or alteration. The only conditions are that 1) the author is acknowledged and 2) if you adapt, build upon or otherwise use this material in any substantial way, it is mandatory that the resulting work is also licensed under the same Creative Commons conditions (CC BY-SA). For more information visit \\
	  %\vspace{0.1cm}
	  
 \end{center}

\newpage

\section{Abstract}
In this thesis I propose an algorithm to heuristically calculate different distance measures on uncertain graphs (i.e. graphs where edges only exist with a certain probability) and apply this to the heuristic calculation of harmonic closeness centrality. This approach is mainly based on previous work on the calculation of distance measures by Potamias et al. \cite{Potamias} and on a heuristic algorithm for betweenness centrality by Chenxu Wang and Ziyuan Lin \cite{PSP}. I extend on their research by using the concept of possible shortest paths proposed in \cite{PSP} to apply them to the distances proposed in \cite{Potamias}. To the best of my knowledge, this algorithmic approach has never been studied before. I will compare my  heuristic results for harmonic closeness against the Monte Carlo method both in runtime and accuracy. Similarly, I will conduct new experiments on the betweenness centrality heuristic proposed in \cite{PSP} to test its efficacy on a bigger variety of instances. Finally, I will test both of these algorithms on large scale graphs to evaluate the scalability of their runtime.

\newpage

\tableofcontents

\newpage

\section{Introduction}
\subsection{Uncertain Graphs and General Computational Problem}
Uncertain graphs are a generalization of the well studied (deterministic) graph model, where we assign probabilities of existence to all edges. We denote a graph by $G=(V, E)$, with $V$ being a finite set and $E\subseteq {V\choose2}$. Here, for any set $V$, ${V\choose2}$ denotes the set of all subsets of $V$ with cardinality two. We will sometimes use the terms \textit{graph}, \textit{deterministic graph} and \textit{uncertain graph} interchangeably, if no ambiguity arises. The algorithm presented in this thesis could be used if edges are directed and/or weighted as well, whereas loops are just a redundancy in the application of our distance and centrality measures (but would also not hinder the algorithm). Further experiments would though be needed to study its efficiency in these cases.

\begin{definition}{\textbf{Uncertain Graphs}}
\\ Uncertain graphs are defined as a tuple $\mathcal{G}=(V,E,P)$, where $(V,E)$ is a graph and $P:E\rightarrow [0,1]$ assigns a probability of existence to every edge. We assume independence of all edges, i.e. the event that some edge $e\in E$ is present has probability $P(e)$, and it is independent of the event that any different edge $e'\in E\setminus \{ e \}$ is present.
\end{definition}

The fact that this is a generalization of deterministic graphs stems from the identification of $G=(V,E)$ with $\mathcal{G}=(V,E,P)$ where $P(v)=1$ for all $v\in V$. Possible applications of this model are numerous. For instance, consider a road network where certain roads may be blocked, or a computer network where links could potentially fail.
One way that has been commonly used in research is to view uncertain graphs as a generative model for deterministic graphs. We refer to these deterministic graphs as possible worlds (or instances) of our uncertain graph.

\begin{definition}{\textbf{Possible Worlds}}
\\ Let $\mathcal{G}=(V,E,P)$ be an uncertain graph and let $E_{1} = \{ e \in E \mid P(e)=1\}$. We call a graph $G=(V,E_G)$ possible world (or instance) of $\mathcal{G}$ if $E_{1}\subseteq E_G\subseteq E$. This is denoted as $G\sqsubset \mathcal{G}$.
\end{definition}

As one can easily observe, there are $2^{|E|-|E_{1}|}\in\mathcal{O}(2^{|E|})$ possible worlds of $\mathcal{G}$.

\begin{definition}{\textbf{Probability of Possible Worlds}}
\\ For any possible world $G=(V,E_G)$ of an uncertain graph $\mathcal{G}=(V,E,P)$, we define the probability of $G$, denoted $Pr(G)$, as the probability of $G$ being the resulting graph when sampling each individual edge $e\in E$ with probability $P(e)$.
\end{definition}

Where, in the context of this thesis, sampling always refers to uniform random sampling. Hence, $\mathcal{G}$ can be identified with a random variable that can assume any of the values $ G_1, \cdots , G_{2^{|E|-|E_1|}}$ with respective occurrence probabilities $Pr(G_1),\cdots,Pr(G_{2^{|E|-|E_1|}})$. As we have independent existence probabilities for all edges, it is easy to see that the following equation holds: 
\[ Pr(G) = \left( \prod_{e\in E_G} P(e) \right) \left( \prod_{e\in E\setminus E_G} 1-P(e) \right) \]

Using this, we can define the expected value of some real valued measure $\phi$, defined for deterministic graphs (e.g. the chromatic number), on an uncertain graph $\mathcal{G}$ as
\[ \mathbb{E}[\phi(\mathcal{G})]=\sum_{G\sqsubset \mathcal{G}} Pr(G)\cdot \phi(G) \]

Note though, this cannot be applied in a meaningful way if $\phi$ could be infinite or undefined for certain instances of $\mathcal{G}$ (e.g. consider the distance between two nodes $s,t\in V$ that may be disconnected in at least one instance).
Moreover, calculating the exact expected value of $\phi(\mathcal{G})$ is of course not practically feasible if there is an even moderately large amount of edges $e$ with $P(e)<1$, even if $\phi$ can be efficiently calculated. \\

One way to tackle this problem is to use the Monte Carlo method, i.e. sample $r$ random instances $G_1,\cdots,G_r\sqsubset \mathcal{G}$ for some $r\in\mathbb{N}$ and set $\mathbb{E}[\phi(\mathcal{G})]\approx \frac{1}{r} \sum_{i=1}^{r}\phi(G_r)$. As the occurence of all possible worlds is independent and identically distributed, we have $\lim_{r\to\infty}\frac{1}{r} \sum_{i=1}^{r}\phi(G_r) = \mathbb{E}[\phi(\mathcal{G})]$ by the law of large numbers.\\

However, Monte Carlo Sampling is still computationally expensive. Creating a single sample takes $\mathcal{O}(|E|)$ time. Hence, if we can compute $\phi(G)$ in $\mathcal{O}(\varphi)$ for every possible world $G\sqsubset\mathcal{G}$, we need $\mathcal{O}(r\varphi|E|)$ time to compute $\frac{1}{r}\sum_{i=1}^{r}\phi(G_i)$. To avoid this computationally expensive process, we can apply different heuristic or approximative algorithms to estimate a solution. Both of these methods are not guaranteed to give an exact result. Approximation algorithms can guarantee a bound of proximity to the exact result though, while heuristic algorithms may produce arbitrarily bad results (although they might usually work well in practice). This is why, in the next section, we will investigate a heuristic algorithm for betweenness centrality in uncertain graphs without the need to extensively sample random instances. The original publication by Chenxu Wang and Ziyuan Lin claimed to give an approximation \cite{PSP}. Though, I will argue why I think their given proof is incomplete. I will then present a novel heuristic algorithm to calculate harmonic closeness centrality and different notions of distance in uncertain graphs. Note, both of these algorithms could potentially give approximations and I can neither present a proof nor a disproof of this statement at this point. Finally, I will present experimental results on the efficacy and runtime of both algorithms, i.e. I will test my new algorithm and also conduct further experiments on the algorithm presented in \cite{PSP} to examine its efficacy and runtime on a different set of graphs.

\subsection{Centrality Measures}
Centrality measures are different ways to assign numbers to nodes (or edges, but we only consider node centrality) within a graph $G=(V,E)$. Namely, some function $\phi:V\to \mathbb{R}$ such that, for $v_1, v_2\in V$, a large value of $\phi(v_1)$ indicates that $v_1$ is in some way central in the graph structure, and $\phi(v_1)>\phi(v_2)$ indicates that $v_1$ is more central than $v_2$. This is of course just a very loose way to define the semantics of centrality. It is not unambiguous to rigorously define centrality though and multiple different measures have been studied. In this thesis, we will only consider two different notions of centrality: betweenness centrality, first proposed by Linton Freeman in 1977 \cite{Betweenness}, and harmonic closeness centrality, first proposed by Massimo Marchiori and Vito Latora in 2000 \cite{Harmonic}.

\begin{definition}{\textbf{Betweenness Centrality}}
\\ Let $G=(V,E)$ be a graph with $|V|\geq 3$. For pairwise distinct nodes $s,t,v\in V$ let $\sigma(s,t)$ equal the amount of shortest paths between $s$ and $t$, and let $\sigma(s,t | v)$ equal the amount of shortest paths between $s,t$ on which $v$ is an intermediary node. Using this, the (normalized) betweenness centrality of $v$ is defined as
\[ B(v) = \frac{2}{(|V|-1)(|V|-2)} \sum_{s\neq v \neq t} \frac{\sigma(s,t | v)}{\sigma(s,t)} \]
\end{definition}

For the sake of simple notation, we agree upon the convention that for $\sigma(s,t)=0$ we have $\frac{\sigma(s,t | v)}{\sigma(s,t)} = 0$ and that, in the summation, $s\neq v \neq t$ is equivalent to $\{s,t\}\in {V\choose2}$ and $v\notin \{s,t\} $. The normalization factor of $\frac{2}{(|V|-1)(|V|-2)}$ is needed to guarantee $B(v)\in [0,1]$ independently of the size of the graph, where the value 0 corresponds to the lowest possible centrality and the value 1 to the highest possible centrality respectively. This stems from the fact that we have $(|V|-1)(|V|-2)$ possibilities to choose nodes $s\neq v\neq t$. In the undirected case though, we are not separately considering the pairs $(s,t)$ and $(t,s)$, hence the factor $2$. \\

So, in the notion of betweenness centrality, a node $v$ is central if the proportion of shortest paths on which $v$ is an intermediary node is high when considering all possible choices for distinct starting and ending nodes of shortest paths different from $v$. 

\begin{figure}[H]
\centering
	\begin{tikzpicture}%[node distance=3cm, auto]
	\tikzstyle{vertex} = [circle, draw=black]

	\node[vertex] (v1) at (1.5,1.5) {$v_1$};
	\node[vertex] (v2) at (3,3) {$v_2$};
	\node[vertex] (v3) at (0,3) {$v_3$};
	\node[vertex] (v4) at (3,0) {$v_4$};
	\node[vertex] (v5) at (0,0) {$v_5$};
	
	\draw[] (v2) -- (v1);
	\draw[] (v3) -- (v1);
	\draw[] (v4) -- (v1);
	\draw[] (v5) -- (v1);
\end{tikzpicture}
\caption{Star Graph with 4 Outer Nodes}
\label{star_graph}
\end{figure}
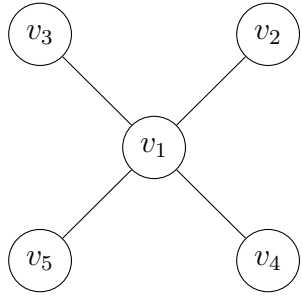

A simple example would be a star graph with $n\geq 3$ nodes $v_1, \cdots, v_n$ and central node $v_1$ (e.g., for $n=5$, the graph shown in figure \ref{star_graph}). As $v_1$ is the only intermediary node on every shortest path between two outer nodes we have $B(v_1)=1$. However, as any outer node $v_k$ is never an intermediary node on the shortest path between two different nodes, every choice of $2\leq k \leq n$ gives $B(v_k)=0$.

\begin{definition}{\textbf{Harmonic Closeness Centrality}}
\\ Let $G=(V,E)$ be a graph with $|V|\geq 2$. The (normalized) harmonic closeness centrality of a node $v\in V$ is defined as
\[ H(v)= \frac{1}{|V|-1} \sum_{s\neq v} \frac{1}{d(s,v)} \]
\end{definition}

Similarly to the definition of betweenness, we agree upon the convention that if $d(s,t)=\infty$, i.e. $s$ and $t$ are not connected, we have $\frac{1}{d(s,v)}=0$ and, in the summation, $s\neq v$ is short for $s\in V\setminus \{v\}$. Hence, a node $v$ has a high harmonic closeness if the average distance $d(v,s)$ to all other nodes $s$ is relatively short. Again, the factor $\frac{1}{|V|-1}$ guarantees $H(v)\in[0,1]$, as the sum is maximized iff $d(v,s)=1$ for all $s\in V\setminus \{ v\}$.

Let us once more consider the example of a star graph with $n\geq 2$ nodes $v_1, \cdots , v_n$ and center $v_1$. For the center node we get the same (maximal) value of 
\[ H(v_1) = \frac{1}{n-1} \sum_{i=2}^{n} \frac{1}{d(v_1, v_i)} = \frac{1}{n-1} \sum_{i=2}^{n} 1 = 1 \]
and for the outer nodes $v_j$, $2\leq j \leq n$, we have
\begin{align*}
H(v_j) 	&= \frac{1}{n-1}\left( \frac{1}{d(v_1,v_j)} + \sum_{\substack{i=2 \\ i \neq j}}^{n} \frac{1}{d(v_j, v_i)}\right) = \frac{1}{n-1} \left( 1 + \frac{(n-2)}{2} \right)\\
 		&= \frac{1}{n-1} \left( \frac{2+n-2}{2} \right) = \frac{n}{2(n-1)}
\end{align*}
As expected, for $n=2$ the nodes $v_1, v_2$ behave symmetrical, i.e. $H(v_2)=\frac{2}{2(2-1)}=1$. If we however increase the number of outer nodes, i.e. we increase $n$, the harmonic closeness of the outer nodes is strictly decreasing and bounded from below by $\frac{1}{2}$ with $\lim_{n\to\infty} \frac{n}{2(n-1)}=\frac{1}{2}$. This intuitively makes sense, as the harmonic closeness of $v$ is nothing but the arithmetic mean of the reciprocal distance of $v$ to all other nodes. Once the amount of outer nodes increases, the average reciprocal distances tends towards $\frac{1}{2}$, which is the pairwise distance of all nodes when leaving out the center node, which has less and less impact for greater $n$.

So, in these two centrality measures, we get vastly different values for the outer nodes in our star graph example, but it always holds that $v_1$ has the highest centrality (excluding the case of $n=2$), and the outer nodes behave symmetrical. If we are interested in ranking the centrality of nodes, they both yield the same ranking. \\

The choice of a centrality measure depends on the structural characteristics in a graph one wants to investigate. For example, if we consider a train network, represented as a graph, a high betweenness centrality value of a node $v$ (station) would indicate that many shortest train connections between different stations lead over the station represented by $v$. The distance between those stations is however not important (just the fact that we pass $v$). If we instead use harmonic closeness, a high value for the station represented by $v$ would indicate that, on average, all other stations are reachable from $v$ by traveling a relatively short distance. This may be the case if $v$ is some kind of hub (central station) with short distances to other stations, which could indicate a high betweenness centrality as well, but it could also mean that we can just reach such a hub quickly. Trains traveling between two other stations may nonetheless never pass the station $v$. Hence, $v$ could even have a betweenness value of zero.\\

Though, both of these centrality measures are not directly applicable to uncertain graphs. For betweenness centrality, we are only considering the amount of shortest paths between two nodes $s,t$  and the proportion of which a node $v$ is passed on them. Yet, in uncertain graphs, each potential shortest path only exists with a certain probability. Furthermore, even if a given path exists in some instance, it might not always be a shortest path in that instance. In the next section, to deal with this problem, we will see an alternative definition of betweenness in uncertain graphs.

Similarly, for harmonic closeness, we would need to define what the distance between two nodes $s$ and $t$ even means, once the existence of edges is uncertain. The canonical definition (the length of a shortest path between $s$ and $t$) cannot be directly applied for the same reason as in betweenness centrality. In section 4, I will propose a novel way to estimate distance in uncertain graphs, which we can then use to get an alternative notion of harmonic closeness.

\section{Betweenness Heuristic}
In 2019, Chenxu Wang and Ziyuan Lin proposed a method to heuristically estimate betweenness centrality in uncertain graphs \cite{PSP}. In experiments, they achieved results similar in quality compared to the Monte Carlo method for estimating the expected betweenness centrality. For this, they first introduced the concept of possible shortest paths.

\subsection{Possible Shortest Paths}

\begin{definition}{\textbf{Possible Shortest Paths}}
\\Let $\mathcal{G}=(V,E,P)$ be an uncertain graph. For two distinct nodes $s,t\in V$, we define the set $\mathcal{P}(s,t)=\{ \pi \in E^n \mid n \in \mathbb{N}, \exists G \sqsubset \mathcal{G}$ such that $\pi$ is a shortest $s-t$ path in $G\}$ as the set of all possible shortest paths (PSP) between the nodes $s$ and $t$.
\end{definition}

Given a path $\pi\in\mathcal{P}(s,t)$, we can calculate the \textit{absolute existence probability of $\pi$}, i.e. the probability that an instance $G\sqsubset\mathcal{G}$ is sampled in which all edges appearing on $\pi$ exist, as

\[ Pr(\pi) = \prod_{e\in \pi} P(e) \]

Now, as a possible shortest path $\pi\in\mathcal{P}(s,t)$ that exists in some instance $G\sqsubset\mathcal{G}$ might still not be a shortest path between $s$ and $t$ in $G$, we are also interested in the \textit{relative probability} that we sample an instance where $\pi$ does exist and is also a shortest $s-t$ path. Calculating this exactly is however known to be $\#P$-hard \cite{MPSP}. So, under the assumption that $P\neq NP$, we cannot calculate this efficiently. To overcome this limitation, Chenxu Wang and Ziyuan Lin defined an estimate for the relative probability of possible shortest paths.

\begin{definition}{\textbf{Estimated Relative Probability of a Shortest Path}}
\\ Let $\mathcal{G}=(V,E,P)$ be an uncertain graph with distinct $s,t\in V$. To substitute for the exact relative probability of $\pi\in\mathcal{P}(s,t)$ being a shortest path between $s$ and $t$, we use the following estimate:
\[ \overline{Pr}(\pi) = Pr(\pi) \prod_{\substack{\gamma\in\mathcal{P}(s,t) \\ \ |\gamma|<|\pi| }} \left( 1-Pr(\gamma) \right)  \]
\end{definition}

So, we multiply the absolute existence probability of $\pi\in\mathcal{P}(s,t)$ with the negated absolute existence probabilities of all $\gamma\in\mathcal{P}(s,t)$ that are shorter than $\pi$. Note, this is not exact, as any of these paths can share an arbitrary amount of edges. However, $1-Pr(\gamma)$ is just the probability that not all of the edges on $\gamma$ exist. Then, using this estimate of the relative probability of shortest paths, we can also get an estimated probability that two given nodes $s$ and $t$ are connected.

\begin{definition}{\textbf{Estimated Probability of Connection}}
\\ Let $\mathcal{G}=(V,E,P)$ be an uncertain graph with distinct $s,t\in V$. We use the following estimate for the probability of $s$ and $t$ being connected:
\[ \varphi_{st} = 1 - \prod_{\pi \in \mathcal{P}(s,t)} (1-Pr(\pi)) \]
\end{definition}

Again, this is not exact for the same reason (the paths may share edges). Now, using the aforementioned definitions, Chenxu Wang and Ziyuan Lin proposed the following alternative notion of betweenness centrality in uncertain graphs.

\begin{definition}{\textbf{Possible Shortest Path Betweenness}}
\\ In an uncertain graph $\mathcal{G}=(V,E,P)$ with $|V|\geq3$, we define the PSP-betweenness of $v\in V$ as
\[ \overline{B}(v) = \frac{2}{(|V|-1)(|V|-2)} \sum_{s\neq v \neq t} \frac{\overline{\sigma}(s,t | v)}{\overline{\sigma}(s,t)}\varphi_{st}\]
where, for $\pi\in\mathcal{P}(s,t)$, we define 
\[ \bbone_{\pi}(v) = \begin{cases} 1, \text{if $v$ lies on $\pi$} \\ 0, \text{otherwise} \end{cases} \]
as the indicator function whether $v$ lies on a given possible shortest path,
\[\overline{\sigma}(s,t | v) 	= \sum_{\pi \in \mathcal{P}(s,t)} \bbone_{\pi}(v) \overline{Pr}(\pi) \]
as the sum of all (estimated) relative probabilities of shortest paths on which $v$ lies,
\[ \overline{\sigma}(s,t)	 = \sum_{\pi \in \mathcal{P}(s,t)} \overline{Pr}(\pi) \]
as the sum of all (estimated) relative probabilities of possible shortest $s-t$ paths and, as stated earlier, $\varphi_{st} = 1 - \prod_{\pi \in \mathcal{P}(s,t)}(1-Pr(\pi))$ is our estimated probability of $s$ and $t$ being connected.
\end{definition}

Now, as the amount of possible shortest paths in $\mathcal{G}$ can grow exponentially in the size of the graph, we generally cannot enumerate all possible shortest paths between two given nodes. To overcome this, Chenxu Wang and Ziyuan Lin proposed the following algorithm to heuristically explore possible shortest paths in uncertain graphs. \\

\subsection{Exploration Algorithm}
The basic idea of this algorithm is to treat an uncertain graph $\mathcal{G}=(V,E,P)$ as a deterministic graph $G=(V,E)$. In $G$, we calculate all shortest paths between two given nodes $s$ and $t$ with some augmented shortest path algorithm (e.g. breadth-first search). Then, we remove the (or one of the) edge(s) with minimal probability on each path. This is repeated until either $s$ and $t$ are disconnected or the estimated connection probability $\varphi_{st}$ reaches a certain threshold $\phi$, which is given as a hyperparameter.

\begin{algorithm}[H]
	\SetAlgorithmName{Algorithm}{algorithm}{List of Algorithms}
	\LinesNumbered
	\caption{$ExplorePSP(\mathcal{G},s,t,\phi$)}
	\KwData{Uncertain Graph $\mathcal{G}=(V,E,P)$, threshold $\phi$, nodes $s,t\in V$}
	\KwResult{Set $P_{st}\subseteq\mathcal{P}(s,t)$ of possible shortest paths between $s$ and $t$}
	Set $P_{st}=\emptyset$ and $\varphi_{st}=0$ \\
	Set $G=(V,E)$ to be the graph that results from ingoring probalilities in $\mathcal{G}$ \\
	Set $SP$ to be all the shortest paths betweens $s$ and $t$ in $G$ \\
	\While{$\varphi_{st}<\phi$ and $SP\neq\emptyset$}{
		\ForEach{$S\in SP$}{
			Add $S$ to $P_{st}$ \\
			Set $e_{min}$ to be an edge with minimal probability in $S$ \\
			Remove $e_{min}$ from $G$	
		}
		Set $SP$ to be all the shortest paths betweens $s$ and $t$ in $G$ \\
		Set $\varphi_{st}=1-\prod_{P\in P_{st}}(1-Pr(P))$
	}
	\Return{$P_{st}$}
\end{algorithm}

The selection of the $e_{min}$ edges may not be unambiguous if, on a given path, at least two edges have the same (minimal) probability. In the original publication, it was not further specified what kind of tie-breaking mechanism should be applied in such a situation. I decided to just delete the edge that was found least recently (i.e. the one that is closest to the node $t$), hoping to reach the condition of $s$ and $t$ being disconnected relatively fast. However, multiple other approaches would be possible.

Concerning the running time of this algorithm, Chenxu Wang and Ziyuan Lin claimed that if we calculate all shortest $s-t$ paths $k$ times for some $k\in\mathbb{N}$ and use breadth-first search, the algorithm terminates in $\mathcal{O}(k(|V|+|E|))$. I object their claim though. This would only be true if we use normal breadth-first search in lines 3 and 10 respectively. However, as we explicitly calculate \textit{all} shortest paths, we cannot guarantee a running time of $\mathcal{O}(|V|+|E|)$ anymore, as there are canonical and easy to construct graphs with the amount of shortest paths between two given nodes being exponential in the size of the graph. It is easy to augment breadth-first search to calculate the amount of shortest paths between two nodes without affecting its running time of $\mathcal{O}(|V|+|E|)$. This is not enough here though, as we explicitly need the (estimated) relative probability $\overline{Pr}(S)$ and the traversed inner nodes for every retrieved shortest path $S$.

\subsection{Approximation Guarantee}
In the original publication, Chenxu Wang and Ziyuan Lin claimed to have an approximation guarantee using their exploration algorithm for possible shortest paths. To be precise, they claimed that when comparing the value of the PSP-betweenness that is computed using their exploration algorithm for any given node $v$, $B^{PSP}_{explore}(v)$, against the theoretical exact value, $B^{PSP}_{exact}(v)$, the mean absolute error is up bound by $\frac{1}{12}$. So, for any uncertain graph $G=(V,E,P)$, we would have 

\[ \frac{1}{|V|} \sum_{v\in V} |B^{PSP}_{explore}(v) - B^{PSP}_{exact}(v)| < \frac{1}{12} \approx 0.08333 \]

The presented proof of this claim however was based on the assumption that "(...)[T]he values of betweenness centrality are inversely proportional to and dominated by the square of the number of nodes. Thus, the errors are also inversely proportional to and dominated by the square of the number of nodes. That is, networks with fewer nodes have more substantial calculation errors."\cite{PSP}. So, as far as I understand their claim, they factored out the common normalization term of $\frac{2}{(|V|-1)(|V|-2)}$ and then argued that the absolute value $|B^{PSP}_{explore}(v) - B^{PSP}_{exact}(v)|$ is dominated by the term $|V|^2$ in the denominator. Their whole proof was based upon this assumption, and thus they only checked graphs with up to 4 nodes. However, when we factor out the normalization term from the absolute value, the difference could also be arbitrarily large, as the non-normalized betweenness centrality is not bound from above by 1 and can tend towards infinity as $|V|$ tends towards infinity.  Though, I cannot present a proof that the PSP-betweenness calculated with the exploration algorithm does not yield an approximation (or the other way around).

\section{Harmonic Closeness Heuristic}
Because of the promising experimental results in their publication, I decided to use the idea of PSP-betweenness and adapt it to harmonic closeness. To the best of my knowledge, no algorithm with a similar approach for harmonic closeness in uncertain graphs has ever been studied. Recall, in graphs $G=(V,E)$, we defined the harmonic closeness of $v\in V$ as 

\[H(v) = \frac{1}{|V|-1} \sum_{s\neq v} \frac{1}{d(s,v)}\]

So, to redefine this for uncertain graphs, we first need to define what distance even means in the case of edges being uncertain. 

\subsection{Distance in Uncertain Graphs}
The notion of distance I used is based upon a publication on $k$-nearest neighbors in uncertain graphs by Potamias et al. \cite{Potamias}. To define different distance metrics in an uncertain graph, $\mathcal{G}=(V,E,P)$, they first observed that, for all $\{s,t\}\in{V\choose2}$, in any instance $G\sqsubset\mathcal{G}$ we have $d_G(s,t)\in \{1, \cdots , |V|-1, \infty\}$. Using this, they defined the following probability distribution.

\newpage
\begin{definition}{\textbf{Distance Probability Distribution in Uncertain Graphs}}
\\ Let $\mathcal{G}=(V,E,P)$ be an uncertain graph, and let $\{s,t\}\in{V\choose2}$. We define the probability of the distance between $s$ and $t$ being equal to $k\in\{1, \cdots , |V|-1, \infty \}$ as
\[ p_{s,t}(k) = \sum_{\substack{G\sqsubset \mathcal{G} \\ d_G(s,t)=k}} Pr(G) \]
\end{definition}

So, the probability of the distance in $\mathcal{G}$ being equal to $k$ is exactly the probability that an instance $G\sqsubset\mathcal{G}$ with $d_G(s,t)=k$ is sampled. Now, using this probability distribution, Potamias et al. defined the following three distance metrics for all $\{s,t\}\in {V\choose2}$.

\begin{definition}{\textbf{Median-Distance}}
\[ d_{Med}(s,t) = \arg\max_{D}\left\{ \sum_{k=1}^{D} p_{s,t}(k)\leq \frac{1}{2} \right\}, \ D\in\{1, \cdots , |V|-1\} \]
\end{definition}

\begin{definition}{\textbf{Majority-Distance}}
\[ d_{Maj}(s,t)= \arg\max_{D} p_{s,t}(D), \ D\in\{1, \cdots , |V|-1, \infty\} \]
\end{definition}

\begin{definition}{\textbf{Expected-Reliable-Distance}}
\[ d_{ER}(s,t) = \frac{1}{1-p_{s,t}(\infty)} \sum_{k=1}^{|V|-1}p_{s,t}(k)k\]
\end{definition}

For $d_{Maj}$ we could of course get ambiguous values, i.e. some kind of tie breaking mechanism would be needed. However, this was not further specified. Furthermore, for $d_{ER}$ we could run into the case of $p_{s,t}(\infty)=1$ if $s$ and $t$ are disconnected in every possible world of $\mathcal{G}$. This was also not rigorously taken care of in the original publication. I just decided to use the semantically natural extension of the definition with $d_{ER}(s,t)=\infty$ if $p_{s,t}(\infty)=1$. \\

Potamias et al. approximated their distance probability distribution $p_{s,t}$ using the Monte Carlo method. They tested the accuracy of these three distance metrics compared to other common notions of distance in uncertain graphs. This was done by using uncertain graphs with known ground-truth neighbors, selecting random ground-truth neighbor pairs $(A,B_0)$ and corresponding pairs $(A,B_1)$ that are not ground-truth neighbors. A random permutation of $B_0$ and $B_1$ created a triplet of either type $(A,B_0,B_1)$ or $(A,B_1,B_0)$. Then, they used all tested distance metrics to determine whether the given triplets are of the first or latter type by calculating the distance between $A$ and the other two respective nodes. All three novel distance metrics based upon the probability distribution $p_{s,t}$ dominated the other tested distance metrics in this classification task. $d_{ER}$ however always had the most accurate classification scores in the experiments, and it was the fastest to converge in the Monte Carlo method. Furthermore, I think to define a distance metric that can produce finite, non-integer values is the most natural way to measure distance in uncertain graphs. Hence, I decided to stick to $d_{ER}$ when testing my algorithm. The same algorithmic approach could easily be used to estimate the other two distance metrics though, as it is based on estimating $p_{s,t}$.

\subsection{Estimating Distances with Possible Shortest Paths}
Now, I wanted to avoid the Monte Carlo method when calculating the above mentioned probability distribution. So, here the concept of possible shortest paths and the heuristic exploration algorithm by Chenxu Wang and Ziyuan Lin come into play. \\
Recall, in the exploration algorithm we calculate a subset $P_{st}$ of all possible shortest paths between two distinct nodes $s,t\in V$ in an uncertain graph $\mathcal{G}=(V,E,P)$. For each $\pi\in P_{st}$, we estimate the relative probability of $\pi$ being a shortest path as 
\[ \overline{Pr}(\pi)=Pr(\pi)\prod_{\substack{\gamma\in P_{st} \\ |\gamma|<|\pi|}}(1-Pr(\gamma)) \]
with $Pr(\gamma)=\prod_{e\in \gamma}P(e)$ being the absolute existence probability of any $\gamma\in P_{st}$. Using this, I decided to estimate the probability distribution $p_{s,t}(k)$ for $1\leq k \leq |V|-1$ as

\[ \overline{p}_{s,t}(k) = \sum_{\substack{\pi\in P_{st} \\ |\pi|=k}}\overline{Pr}(\pi) \]

Once $s$ and $t$ are disconnected or we reach the threshold $\phi$ for the estimated connection probability $\varphi_{st}$ (i.e the exploration terminates), the estimate of $p_{s,t}(\infty)$ is given by

\[ \overline{p}_{s,t}(\infty) = 1 - \sum_{k=1}^{|V|-1}\overline{p}_{s,t}(k) \]

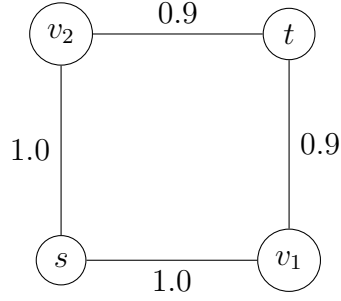
\begin{figure}[H] 
	\begin{center}
	\begin{tikzpicture}[node distance=3cm, auto]
		\tikzstyle{vertex} = [circle, draw=black]

		\node[vertex] (s) at (0,0) {$s$};
		\node[vertex] (v1) at (3,0) {$v_1$};
		\node[vertex] (v2) at (0,3) {$v_2$};
		\node[vertex] (t) at (3,3) {$t$};
	
		\draw[] (s) -- (v1) node [midway, below] {1.0};
		\draw[] (s) -- (v2) node [midway] {1.0};
		\draw[] (v1) -- (t) node [midway, right] {0.9};
		\draw[] (v2) -- (t) node [midway] {0.9};
	\end{tikzpicture}
	\end{center}
	\caption{Uncertain Graph with an Exceptional Case for PSP Distance (1)}
	\label{first exception}
\end{figure}

Upon further inspection though, I noticed that there are special cases where this would yield $\sum_{D=1}^{|V|-1} \overline{p}_{s,t} (D) > 1$. In such cases, I decided to stop the exploration and set

\[ \overline{p}_{s,t}(k) = 1 - \sum_{j=1}^{k-1}\overline{p}_{s,t}(j) \quad \text{ and } \quad \overline{p}_{s,t}(d)=0 \text{ for } d\in\{1,\cdots, |V|-1, \infty\}\setminus\{1,\cdots,k\} \]

with $k$ being the length of the longest retrieved path(s) once this situation occurred. A simple example of such an exception is the uncertain graph in figure \ref{first exception}.

Here, for $s$ and $t$, the exploration algorithm first finds the paths $S_1=(\{s,v_1\},\{v_1,t\})$ and $S_2=(\{s, v_2\}, \{v_2,t\})$. We get $\overline{Pr}(S_1)=0.9$ and $\overline{Pr}(S_2)=0.9$. So, instead of setting $\overline{p}_{s,t}(2)=1.8$, we set $\overline{p}_{s,t}(2)=1$. Then, all other values of $\overline{p}_{s,t}$ are set to 0 and we terminate the exploration. The theoretical exact values are however $p_{s,t}(1)=p_{s,t}(3)=0$,
\[ p_{s,t}(2) = (0.9)^2 + 2(0.9)(1-0.9) = 0.99  \]
and
\[ p_{s,t}(\infty) = (1-0.9)^2 = 0.01 \]
The Expected-Reliable-Distance with our estimated probability distribution $\overline{p}_{s,t}$ gives

\[ \frac{1}{1-\overline{p}_{s,t}(\infty)} \sum_{k=1}^{3} \overline{p}_{s,t}(k)k = \overline{p}_{s,t}(2)2 = 2\]

Maybe suprisingly though, the theoretical value also equates to 2:

\[ \frac{1}{1-p_{s,t}(\infty)} \sum_{k=1}^{3}p_{s,t}(k)k = \frac{1}{1-0.01} 2 \cdot 0.99 = \frac{0.99}{0.99}2 = 2 \]

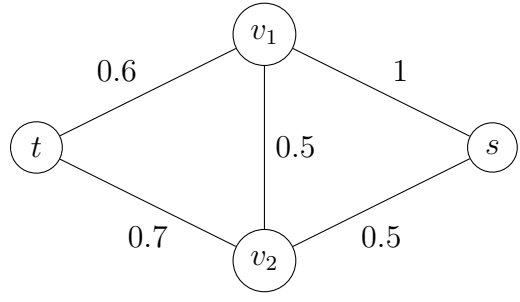
\begin{figure}[H] 
	\begin{center}
	\begin{tikzpicture}[node distance=3cm, auto]
		\tikzstyle{vertex} = [circle, draw=black]

		\node[vertex] (s) at (6, 1.5) {$s$};
		\node[vertex] (v2) at (3,0) {$v_2$};
		\node[vertex] (v1) at (3,3) {$v_1$};
		\node[vertex] (t) at (0,1.5) {$t$};
	
		\draw[] (v2) -- (s) node [midway, below=5] {0.5};
		\draw[] (v1) -- (s) node [midway] {1};
		\draw[] (v1) -- (v2) node [midway, right] {0.5};
		\draw[] (t) -- (v2) node [midway, below=5] {0.7};
		\draw[] (t) -- (v1) node [midway] {0.6};
		
	\end{tikzpicture}
	\end{center}
	\caption{Uncertain Graph with an Exceptional Case for PSP Distance (2)}
	\label{second exception}
\end{figure}

This must not be the case though. E.g. consider the uncertain graph in figure \ref{second exception}.

Here, for $s$ and $t$, the exploration algorithm first finds the two paths $S_1=(\{s,v_1\},\{v_1,t \})$ and $S_2=(\{s,v_2\},\{v_2,t \})$. We get $\overline{Pr}(S_1)=0.6$ and $\overline{Pr}(S_1)=0.35$. Hence, we set 
\[ \overline{p}_{s,t}(2)=\overline{Pr}(S_1)+\overline{Pr}(S_2)=0.95 \]
and then delete the two respective minimal edges $\{v_1,t\}$ and $\{s,v_2\}$. Next, we find the path $S_3=(\{s,v_1\}, \{v_1,v_2\}, \{v_2,t\})$ with $Pr(S_3)=0.35$. This gives us

\[ \overline{Pr}(S_3) = Pr(S_3)(1-Pr(S_1))(1-Pr(S_2) = 0.091 \]

But, as $0.95+0.091=1.041>1$, we once again stop the exploration and set 

\[ \overline{p}_{s,t}(3)=1-0.95=0.05 \]
and $p_{s,t}(1)=p_{s,t}(\infty)=0$. Now, for the estimated Expected-Reliable-Distance we get a value of

\[ \frac{1}{1-\overline{p}_{s,t}(\infty)} \sum_{k=1}^{3} \overline{p}_{s,t}(k)k = 0.95\cdot 2 + 0.05 \cdot 3 = 2.05 \]

However, considerung all 16 possible worlds, we get exact values of

\[ p_{s,t}(1)=0, \quad p_{s,t}(2)=0.74, \quad p_{s,t}(3)=0.07, \quad p_{s,t}(\infty) = 0.19 \]
which in turn yields an Expected-Reliable-Distance of

\[ \frac{1}{1-p_{s,t}(\infty)} \sum_{k=1}^{3}p_{s,t}(k)k = \frac{1}{0.81}(2\cdot0.74+3\cdot0.07)\approx 2.0865 > 2.05\]

Unfortunately, I cannot provide any approximation guarantee for this as of yet.

\subsection{Heuristic Algorithm for Harmonic Closeness}
Now, using this method of calculating the estimated probability distribution $\overline{p}_{s,t}$, we can finally define PSP-harmonic closeness and give an algorithm for it.

\begin{definition}{\textbf{PSP Harmonic Closeness}}
\\ Let $\mathcal{G}=(V,E,P)$ be an uncertain graph with $|V|\geq2$. Using the estimated probability distribution $\overline{p}_{s,t}$ for all $\{s,t\}\in{V\choose2}$, we define the PSP-harmonic closeness of $v\in V$ as
\[ \overline{H}(v) = \frac{1}{|V|-1} \sum_{s\neq v} \frac{1}{d_{\overline{ER}}(s,v)} \text{ \quad where \quad } d_{\overline{ER}}(s,v)=\begin{cases} \infty, \text{\qquad\qquad \ \ \ if }\overline{p}_{s,v}(\infty)=1 \\
\frac{1}{1-\overline{p}_{s,v}(\infty)} \sum\limits_{k=1}^{|V|-1} \overline{p}_{s,v}(k)k, \text{ \quad else} \end{cases}\]
\end{definition}

This can be calculated for a single node $v\in V$ using the following algorithm. \medskip %We iterate over all $s\in V\setminus\{v\}$ and calculate $d_{\overline{ER}}(s,v)$ to successively get $\sum_{s\neq v}\frac{1}{d_{\overline{ER}}(s,v)}$.
%Given distinct $s,t\in V$, a fixed explored PSP set $P_{st}$, and any potential finite $s,t$ distance $k=1,\cdots ,|V|-1$, we use the following notation
%\[ \sigma(k) \ = \sum_{\pi \in P_{sv}, \ |\pi|=k}  \overline{Pr}(\pi) \]

\begin{algorithm}[H]
	\setstretch{1}
	\SetAlgorithmName{Algorithm}{algorithm}{List of Algorithms}
	\LinesNumbered
	\DontPrintSemicolon
	\caption{$PSP$-$Harmonic(\mathcal{G},v,\phi$)}
	\KwData{Uncertain Graph $\mathcal{G}=(V,E,P)$, threshold $\phi$, node $v\in V$}
	\KwResult{PSP-harmonic closeness centrality $\overline{H}(v)$}
	Set $\overline{H}(v)$ = 0 \\
	\ForEach{$s\in V\setminus\{v\}$}{
		Set $\Delta=0$\\ %\quad \tcp*[r]{$\Delta=\sum_{d=1}^{k}\overline{p}_{s,v}(d)d$ after iteration $k$}
		Set $\Gamma=0$ \\ %\quad \tcp*[r]{$\ \Gamma=\sum_{d=1}^{k}\overline{p}_{s,v}(d) \ $ after iteration $k$}			
		Set $P_{sv}$ to $Explore$-$PSP(\mathcal{G},s,v,\varphi)$ \\
		\For{$k = 1$ \KwTo $k=|V|-1$}{
			Set $\delta$ to the sum over $\overline{Pr}(\pi)$ for all $\pi\in P_{sv}$ with $|\pi|=k$ \\
			\If(\tcp*[f]{expectional case}){$\Gamma+ \delta \geq 1$} {
				Set $\Delta=\Delta + k \ (1-\Gamma)$ \\
				Set $\Gamma=1$\\
				\textbf{break}
			}
			Set $\Delta = \Delta + k \ \delta$ \tcp*{$\Delta=\sum_{d=1}^{|V|-1}\overline{p}_{s,v}(d)d$ in line 16}
			Set $\Gamma = \Gamma + \delta$ \qquad  \tcp*{$ \Gamma=\sum_{d=1}^{|V|-1}\overline{p}_{s,v}(d) \ $ in line 16}
		} 
		\lIf{$\Gamma \neq 0$}{ Set $\overline{H}(v) = \overline{H}(v) + \text{\large{$\frac{\Gamma}{\Delta}$}} $ \tcp*[f]{$\frac{1}{d_{\overline{ER}}(s,v)} = \left( \text{\large{$\frac{\Delta}{\Gamma}$}}  \right)^{-1}$}}
	}
	\Return  \text{\Large{$\frac{1}{|V|-1}$}} $\overline{H}(v)$
	\smallskip
		\hrulealg
		\tcc{$\overline{p}_{s,v}(\infty)= 1 -  \sum\limits_{d=1}^{|V|-1}\overline{p}_{s,v}(d) = 1-\Gamma \ \quad 					\Longrightarrow \ \quad \frac{1}{d_{\overline{ER}}(s,v)} = 0$ if $\Gamma=0$}
		\bigskip
		\tcc{$d_{\overline{ER}}(s,v)=\frac{1}{1- \overline{p}_{s,v}(\infty)}\sum\limits_{d=1}^{|V|-1}\overline{p}_{s,v}(d)d = 			\frac{1}{1-(1-\Gamma)}\Delta = \text{\large{$\frac{\Delta}{\Gamma}$}}$}
\end{algorithm}

\newpage
\section{Implementation}
All algorithms where implemented in C++ and parallelized using the OpenMP Library, a library for shared memory parallelization on the basis of compiler directives \cite{OMP}. The data structure for uncertain graphs was implemented using adjacency lists to save memory in large graphs (realized as vectors of node indices), with the probability of each edge $\{u,v\}$ being stored in a hash map, which maps the pair $(\min(u,v), \max(u,v))$ to the probability $P(\{u,v\})$, where for the remainder of this section we assume that $V=\{0, \cdots , |V|-1\}$. The hashing of integer pairs was always realized using the corresponding hash function from the boost C++ library \cite{boost}. Whenever possible, book-keeping techniques for already calculated values where employed, move-semantics where prioritized over copying and memory was reserved beforehand, if it was known prior how much memory was needed. During the parallelization, each thread operated on immutable, shared data (e.g. the uncertain graph data structure) or on data that was private for the given thread (e.g. partial results or a hash set of sampled/deleted edges), which was then joined (if needed) after all threads finished. So, no deadlocks or data-races could occur. I implemented both my novel algorithm for PSP-harmonic as well as the algorithm for PSP-betweenness by Chenxu Wang and Ziyuan Lin.

\subsection{All Shortest Paths Algorithm for Harmonic}

Algorithm \ref{alg_asph} is for the exploration of all shortest $s,t$ paths in the case of PSP-harmonic, and algorithm \ref{alg_e_min} is for the retrieval of the edges with minimal probability, which are to be deleted from the shortest paths (just separated for spacing). Each such set of shortest paths, together with the product over $(1-Pr(S))$ for all previously found shortest $s,t$ paths $S$, successively yields the values of $\overline{p}_{s,t}$ for strictly increasing distances. This was realized using an augmented version of breadth-first search. In the case of harmonic, we do not need to know which nodes where traversed on the shortest paths. So, we do not even actually calculate sets of paths, instead only the respective existence probabilities are stored and returned. The already deleted edges are kept in a hash set of integer pairs. During the breadth-first search, for every traversed node $v$, we store a vector $path$-$probs[v]=[p_1,\cdots ,p_k]$ of the existence probabilities of all already found shortest $s,v$ paths. Additionaly, the least recently found edge $e$ with minimal probability on any of those paths, its probability ($p_e$) and it's distance from $s$ ($d_e$) is stored in a vector, i.e. $min$-$edges[v]=(e,p_e,d_e)$ for every node $v$. This tuple is updated once an edge with smaller probability on an alternative shortest $s,v$ path is found, or once an alternative shortest $s,v$ path is detected, for which an edge with the same probability but greater distance from $s$ is known (to delete edges closest to $t$ as the tie breaking mechanism). The vector $E_{min}$ of edges to remove is then retrieved by doing another breadth-first traversal, backwards from $t$ and only utilizing edges on found shortest $s,t$ paths, where we add the edges from $min$-$edges$[ ] to $E_{min}$, but only once they are traversed again (and stop the traversal at those points). This assures the deletion of at least one edge on every shortest paths (and not more than necessary), without the need to keep track of the minimal edges on each path.

\scalebox{1}{\begin{minipage}{1.0\textwidth}
\begin{algorithm}[H]
	\DontPrintSemicolon
	\SetAlgorithmName{Algorithm}{algorithm}{List of Algorithms}
	\LinesNumbered
	\caption{$AllShortestPathsHarmonic(\mathcal{G},s,t,D)$}\label{alg_asph}
	\KwData{Uncertain graph $\mathcal{G}=(V,E,P)$, distinct nodes $s,t$, deleted edges $D\subseteq E$}
	\KwResult{$(P_{st}, l, E_{min})$: List $P_{st}$ of $Pr(S)$ for all shortest $s,t$ paths $S$ in $(V,E\setminus D)$, their length $l$, list $E_{min}$ of edges to remove from paths in $P_{st}$}
	$dist$[ ] $\leftarrow$ length $|V|$, init. all $\infty$, except $dist[s]=0$ \\
	$path$-$probs$[ ] $\leftarrow$  size $|V|$, init. all $path$-$probs[v] = $[ ], except $path$-$probs[s] = [1]$ \\
	$min$-$edges$[ ] $\leftarrow$ list of $|V|$ tuples $(edge, prob, depth)$, init. all ($\emptyset$, $\infty$, 0)\\ 
	Queue $Q$ = $\{s\}$ \\
	\While{$Q$ is not empty}{
		$curr$ = $Q$.pop() \\
		\lIf{curr = target}{\textbf{break}}
		\ForEach{child of curr}{
			\lIf{$\{curr, child\}\in D$}{\textbf{continue}}
			\If{dist[child]=$\infty$}{
				$dist[child]=dist[curr] + 1$\\ 
				$Q$.enqueue($child$) \\
				$path$-$probs[child] = [ \ P(curr,child)\cdot p \mid p \in  path-probs[curr] \ ]$ \\
				\If{$min$-$edges[curr].prob \geq P(curr,child)$}{
					$min$-$edges[child]= (\{curr, child\}, P(curr,child), dist[child])$
				}
				\Else{
					$min$-$edges[child] = min$-$edges[curr]$
				}
			}
			\ElseIf{dist[child] = dist[curr] + 1}{
				$path$-$probs[child]$.add( $P(curr,child) \cdot p \mid p \in path$-$porbs[curr] $ )\\
				$min$-$prob = \min(min$-$edges[child].prob, \ min$-$edges[curr].prob)$ \\
				\If{$min$-$prob \geq P(curr,child)$}{
					$min$-$edges[child]=(\{curr, child\}, P(curr,child), dist[child])$
				}
				\ElseIf{$min$-$edges[child].prob > min$-$edges[curr].prob$}{
					$min$-$edges[child] = min$-$edges[curr]$
				}
				\ElseIf{$min$-$edges[curr].prob = min$-$edges[child].prob$}{
					\If{$min$-$edges[curr].depth > min$-$edges[child].depth$}{
						$min$-$edges[child] = min$-$edges[curr]$
					}				
				}
			}
		}
	}
	\lIf{dist$[t]=\infty$}{\Return{$(\emptyset,\infty,\emptyset$)}}
	$E_{min}$ = $RetrieveMinEdges(\mathcal{G}$, $t$ ,dist[ ],min-edges[ ], $D$) \\
	\Return{$(path$-$probs[t], dist[t], E_{min} )$}
\end{algorithm}
\end{minipage}} %scalebox,minipage

\scalebox{1}{\begin{minipage}{1\textwidth}
\begin{algorithm}[H]
	\SetAlgorithmName{Algorithm}{algorithm}{List of Algorithms}
	\LinesNumbered
	\DontPrintSemicolon
	\caption{$RetrieveMinEdges(\mathcal{G},t,dist$[ ]$,min$-$edges$[ ])}\label{alg_e_min}
	\KwData{ Uncertain graph $\mathcal{G}=(V,E,P)$, node $t$, list of distances $dist$[ ] where $dist[v]$ equals the $s,v$ distance in $(V,E\setminus D)$ for all $v\in V$, \\ \qquad\quad \ list $min$-$edges$[ ] of kind $(edge, prob, depth)$, deleted edges $D\subseteq E$ }
	\KwResult{ List $E_{min}$ of edges to remove}
	$visited$[ ] $\leftarrow$ size $|V|$, init. all $False$ except $visited[t]=True$ \\
	$E_{min}$ = [ ] \\
	Queue $Q = \emptyset$ \\
	\ForEach{child of t}{
		\lIf{$\{t, child\}\in D$}{\textbf{continue}}
		\If{$dist[child] = dist[t]-1$}{
				$visited[child] = True$\\
				\If{$P(t,child)\leq min$-$edges[child].prob$}{
					$E_{min}$.add$(\{ t, child \})$ \\
				}
				\Else{
					$Q$.enqueue($child$)
				}
		}
	}
	\While{$Q$ is not empty}{
		$curr = Q$.pop() \\
		\ForEach{child of curr}{
			\lIf{$\{curr, child\}\in D$}{\textbf{continue}}
			\If{$dist[child] = dist[curr] -1 $}{
				\If{$min$-$edges[curr].edge = \{ curr, child \}$}{
					$E_{min}$.add($\{child, curr\}$)
				}
				\ElseIf{$visited[child] = False$}{
					$visited[child] = True$ \\
					$Q$.enqueue($child$)
				}
			}
		}
	}
	\Return{$E_{min}$}
\end{algorithm}
\end{minipage}} %scalebox,minipage

\subsection{PSP Distance Distribution}

The estimated probability distribution $\overline{p}_{s,t}$ is then calculated by succesive calls of algorithm \ref{alg_asph}, utilizing the recursive nature of our estimated relative probability $\overline{Pr}(S)$, until either $s$ and $t$ are disconnected or the estimated connection probability $\varphi_{st}$ reaches the threshold $\phi$. The implementation of this is shown in algorithm \ref{alg_dist_distrib}.

\scalebox{1}{\begin{minipage}{1\textwidth}
\begin{algorithm}[H]
	\LinesNumbered
	\DontPrintSemicolon
	\SetAlgorithmName{Algorithm}{algorithm}{List of Algorithms}
	\caption{$PSP$-$DistanceDistribution(\mathcal{G},s,t,\phi)$}\label{alg_dist_distrib}
	\KwData{ Uncertain graph $\mathcal{G}=(V,E,P)$, nodes $s$ and $t$, threshold $\phi$}
	\KwResult{ PSP-distance distribution $\overline{p}_{s,t}$}
		$D = \emptyset$ \\
		$\overline{p}_{s,t}(k)=0$ for $k=0,\cdots ,|V|-1, \infty$ \\
		$negatedProbabilityProd = 1$ \tcp*{current product over all $(1-Pr(S))$}
		$probabilitySum = 0$ \tcp*{current sum over all $\overline{Pr}(S)$}
		$\varphi_{st} = 0$ \\
		\While{$\varphi_{st}<\phi$}{
			$(pathProbs, dist, E_{min}) = AllShortestPathsHarmonic(\mathcal{G},s,t,D)$ \\
			\lIf{dist = $\infty$}{\textbf{break}}
			$newProb = negatedProbabilityProd \cdot \sum_{p \in pathProbs} p$ \\
			\If {$probabilitySum + newProb \geq 1$}{
				$\overline{p}_{s,t}(dist) = 1 - probabilitySum$ \\
				\Return{$\overline{p}_{s,t}$}
			}
			$\overline{p}_{s,t}(dist)= newProb$ \\
			$probabilitySum = probabilitySum + newProb$ \\
			$negatedProbabilityProd =  negatedProbabilityProd \cdot \prod_{p\in pathProbs}(1-p)$ \\
			$\varphi_{st} = 1- negatedProbabilityProd$ \\
			$D = D \cup E_{min}$
		}
		Set $\overline{p}_{s,t}(\infty)=1- probabilitySum$ \\
		\Return{$\overline{p}_{s,t}$}
	\end{algorithm}
\end{minipage}} %scalebox,minipage

\subsection{PSP-Harmonic for All Nodes}

Now, once $\overline{p}_{s,t}$ is returned for some $s$ and $t$, it can be used to calculate $d_{\overline{ER}}(s,t)$. This is done in parallel for all choices of $\{s,t\}\in {V\choose2}$. The estimated distances are stored in a triangular matrix, i.e. a vector of vectors $D$ of size $|V|-1\times \cdots \times 1$, where we have $D$[$s$][$t-s-1]=d_{\overline{ER}}(s,t)$ for $s<t$. Finally, $D$ is then used to calculate the PSP-harmonic closeness for all nodes. The implementation is shown in algorithm \ref{alg_psp_har}. Note that no nested parallelism was used. I tested both variants, and the one without nesting was faster.  The same did hold for all following parallelized algorithms.

Here, it would be more efficient if algorithm \ref{alg_dist_distrib} directly returned $d_{\overline{ER}}(s,t)$ instead of $\overline{p}_{s,t}$. The successive calculation of $\overline{p}_{s,t}(k)$ for increasing $k$ allows to easily sum up $\overline{p}_{s,t}(k)k$ as well. Though, I choose to calculate $\overline{p}_{s,t}$ during the implementation phase, as I did not initially settle exclusively on $d_{\overline{ER}}$. After I decided to only use $d_{\overline{ER}}$, this optimization came into my mind too late to be included. Still, redundantly calculating $\sum_{k=1}^{|V|-1}\overline{p}_{s,t}(k)k$ in algorithm \ref{alg_psp_har} to get $d_{\overline{ER}}(s,t)$ should be a minor influence on the running time compared to exploring all shortest $s,t$ paths. As stated earlier, the exploration is the reason why both PSP algorithms can have exponential time and space complexity.

\scalebox{1}{\begin{minipage}{1\textwidth}
\begin{algorithm}[H]
	\LinesNumbered
	\DontPrintSemicolon
	\SetKw{KwParallel}{in parallel}
	\SetAlgorithmName{Algorithm}{algorithm}{List of Algorithms}
	\caption{$PSP$-$HarmonicAllNodes(\mathcal{G},\phi)$}\label{alg_psp_har}
	\KwData{ Uncertain graph $\mathcal{G}=(V,E,P)$, threshold $\phi$}
	\KwResult{ List $H$ with PSP-harmonic closeness for all nodes $v\in V$}
	$D$[ ][ ] $\leftarrow$ size $|V|-1\times |V|-2 \times \cdots \times 1$ \\
	$H$[ ] $\leftarrow$ size $|V|$, all values 0 \\
	\For{$s=0$ \KwTo $s=|V|-2$ \KwParallel}{
		\For{$t=s+1$ \KwTo $t=|V|-1$}{
			$\overline{p}_{s,t} = PSP$-$DistanceDistribution(\mathcal{G},s,t,\phi)$ \\
			\If{$\overline{p}_{s,t}(\infty)=1$}
				{$D[s][t-s-1]=\infty$}
			\Else{
				$D[s][t-s-1] = \frac{1}{1-\overline{p}_{s,t}(\infty)} \sum_{k=1}^{|V|-1}\overline{p}_{s,t}(k)k$
			}
		}
	}
	\For{$v=0$ \KwTo $v=|V|-1$ \KwParallel}{
		\For{$t=0$ \KwTo $t=v-1$}{
			\lIf{$D[t][v-t-1] = \infty$}{\textbf{continue}}
			$H[v] = H[v] + \frac{1}{D[t][v-t-1]}$
		}
		\For{$t=v+1$ \KwTo $t=|V|-1$}{
			\lIf{$D[v][t-v-1] = \infty$}{\textbf{continue}}
			$H[v] = H[v] + \frac{1}{D[v][t-v-1]}$
		}
		$H[v] = \frac{1}{|V|-1} H[v]$
	}
	\Return{H}
	\end{algorithm}
\end{minipage}} %scalebox,minipage

\subsection{All Shortest Paths Algorithm for Betweenness}
The algorithm to explore all shortest paths for PSP-betweenness is almost the same as $AllShortestPathsHarmonic(\mathcal{G},s,t,D)$ (algorithm \ref{alg_asph}). The only difference is that we now need the inner nodes on every shortest path. So, we additionally store a bit vector $B_S$ of size $|V|$ for each shortest path $S$, where we keep track of the traversed nodes on each path. I will not give the pseudocode for this, as it is almost the exact same as in algorithm \ref{alg_asph}. The returned data of the altered algorithm $AllShortestPathsBetweenness(\mathcal{G},s,t,D)$ is a pair $(paths[ ], E_{min})$, where $paths$ is a vector of pairs of the aforementioned kind $(Pr(S), B_S)$ for every retrieved shortest $s,t$ path $S$. $E_{min}$ is still the vector of edges that are to be deleted, which is calculated in the same way as before. This algorithm then enables the following heuristic algorithm \ref{alg_psp_bet} for PSP-betweenness. Again, note that both the time and space complexity could be exponential in worst case instances.

\subsection{PSP-Betweenness for All Nodes}
We now assume that we have $T$ threads wit ID's $0$ to $T-1$. Using successive calls of $AllShortestPathsBetweenness(\mathcal{G},s,t,D)$, each thread calculates a partial sum of $\sum_{s\neq v \neq t}\frac{\overline{\sigma}(s,t|v)}{\overline{\sigma}(s,t)}\varphi_{st}$ for all $v\in V$ while working on distinct node pairs in parallel.

\scalebox{0.96}{\begin{minipage}{1\textwidth}
\begin{algorithm}[H]
	\LinesNumbered
	\DontPrintSemicolon
	\SetKw{KwParallel}{in parallel}
	\SetAlgorithmName{Algorithm}{algorithm}{List of Algorithms}
	\caption{$PSP$-$BetweennessAllNodes(\mathcal{G},\phi)$}\label{alg_psp_bet}
	\KwData{ Uncertain graph $\mathcal{G}=(V,E,P)$, threshold $\phi$}
	\KwResult{ List $B$ with estimated PSP-betweenness for all nodes $v\in V$}
	$B_{Thread}$[ ][ ]$ \leftarrow$ size $T\times |V|$, init. $B_{Thread}[i][v]=0$ for all $v\in V$, $i=0,\cdots,T-1$. \\
	\For{$s=0$ \KwTo $s = |V|-2$ \KwParallel}{
		$i \leftarrow $ ID of current thread \\
		\For{$t=s+1$ \KwTo $t = |V|-1$}{
			$B_{s,t}$[ ] $ \leftarrow $ size $|V|$, all values 0 \\
			$D = \emptyset$ \\
			$\sigma(s,t) = 0$ \\
			$negatedProbabilityProduct = 1$ \\
			$\varphi_{st}=0$ \\
			\While{$\varphi_{st} < \phi$}{
				$(paths$[ ]$, E_{min}) = AllShortestPathsBetweenness(\mathcal{G},s,t,D)$ \\
				\lIf{$paths = $[ ]}{\textbf{break}}
				$temporaryNPP = negatedProbabilityProduct$ \\
				\ForEach{$(Pr(S), B_S)\in paths$}{
					$\overline{Pr}(S) = Pr(S)\cdot$negatedProbabilityProduct \\
					$\sigma(s,t) = \sigma(s,t) + \overline{Pr}(S)$ \\
					$temporaryNPP = temporaryNPP \cdot (1-Pr(S))$ \\
					\ForEach{$v\in V \setminus \{s,t\}$ }{
						\lIf{$B_S(v)=1$}{
							$B_{s,t}[v]= B_{s,t}[v]+ \overline{Pr}(S)$
						}
					}
				}
				$negatedProbabilityProduct = temporaryNPP$ \\
				$\varphi_{st} = 1 - negatedProbabilityProduct$ \\
				$ D = D \cup E_{min}$
			}
			\lIf{$\sigma(s,t) = 0$}{\textbf{continue}}
			\ForEach{$v\in V \setminus \{s,t\}$ }{
				$B_{Thread}[i][v] = B_{Thread}[i][v] + \frac{B_{s,t}[v]}{\sigma(s,t)}\varphi_{st}$
			}
		}	
	}
	$B$[ ] $ \leftarrow $ size $|V|$ \\
	\ForEach{$v\in V$}{
		$B[v] = \frac{2}{(|V|-1)(|V|-2)} \sum_{i=0}^{T-1} B_{Thread}[i][v]$
	}
	
	\Return{B}
	\end{algorithm}
\end{minipage}} %scalebox,minipage

\subsection{Monte Carlo}
The Monte Carlo method for harmonic closeness, seen in algorithm \ref{alg_mc}, was implemented by sampling $r$ graphs in parallel, i.e. one thread works on one sample at a time. Each sample is created by looping over all edges $\{s,t\}\in E$ and adding $(\min(s,t),\max(s,t))$ to a hash set if the given edge is included. We again assume $T$ threads with ID's 0 to $T-1$. To calculate the harmonic closeness of all nodes in a sample $G$, the given thread loops over all $v\in V$ and calculates a vector of distances $d[s]=d_G(v,s)$ for all $s\in V$. This is realized by traversing $G$ using breadth-first search, starting from $v$. The reciprocal distances are then added to a partial harmonic closeness sum $H_{Thread}[i][v]$ for each thread $i$ and node $v$. In the end, the partial results of all threads are summed up and divided by $r(|V|-1)$ to normalize the calculated values and get the average over all $r$ samples. The time complexity for each sample is dominated by running breadth-first search once for each node to calculate the distances, which yields an overall time complexity of $\mathcal{O}(r|V|(|V|+|E|))$ for $r$ Monte Carlo samples. %Every thread stores a hash set of edges (representing the current instance of $\mathcal{G}$), $|V|$ distance values (for the current node $v$ in line 5), $|V|$ partial harmonic closeness sums in the $H_{Thread}$ list and executing the BFS takes $\mathcal{O}(|V|)$ space. So, each thread uses $\mathcal{O}(|E|+|V|)$ space, giving a total space complexity of $\mathcal{O}(T(|E|+|V|))$.

\scalebox{0.96}{\begin{minipage}{1\textwidth}
\begin{algorithm}[H]
	\LinesNumbered
	\DontPrintSemicolon
	\SetKw{KwParallel}{in parallel}
	\SetAlgorithmName{Algorithm}{algorithm}{List of Algorithms}
	\caption{$MonteCarloHarmonic(\mathcal{G},r)$}\label{alg_mc}
	\KwData{ Uncertain graph $\mathcal{G}=(V,E,P)$, amount of samples $r$}
	\KwResult{ List $H$ of estimated harmonic closeness for all nodes $v\in V$}
	$H_{Thread}$[ ][ ] $\leftarrow$ size $T\times|V|$, init. $H_{Thread}[i][v]=0$ for all $v\in V$, $i=0,\cdots,T-1$. \\
	\For{$k=1$ \KwTo $k=r$ \KwParallel}{
		$i$ $\leftarrow$ ID of current thread \\
		$G$ $\leftarrow$ instance of $\mathcal{G}$, sampled uniformly at random \\
		\ForEach{$v\in V$}{
			$d$[ ] $\leftarrow$ size $|V|$, $d[s]=d_G(s,v)$ for all $s\in V$, calculated using BFS \\
			\ForEach{$s\in V \setminus\{ v\}$}{
				\lIf{$d[s]=\infty$}{\textbf{continue}}
				$H_{Thread}[i][v] = H_{Thread}[i][v] + \frac{1}{d[s]}$
			}
		}
	}
	$H$[ ] $\leftarrow$ size $|V|$ \\
	\ForEach{$v\in V$}{
		$H[v] = \frac{1}{r(|V|-1)} \sum_{i=0}^{T-1}H_{Thread}[i][v]$
	}
	\end{algorithm}
\end{minipage}} %scalebox,minipage

Monte Carlo for betweenness was implemented in a similar way. Now though, Brandes algorithm could be used to calculate the betweenness centrality of all nodes. This is a state of the art algorithm for this task with a time complexity of $\mathcal{O}(|V| |E|)$ \cite{Brandes}. This in turn yields an overall time complexity of $\mathcal{O}(r|V||E|)$ for $r$ Monte Carlo samples.

\section{Experiments}

\subsection{General Experimental Setting}

All experiments where run on computers with two Intel Xeon X6126 12 Core CPU's (hyperthreading was turned off, i.e. 24 threads where available) and 192GB of RAM. Each run of one of the PSP heuristics or the Monte Carlo algorithms on a single graph did have unshared access to one of those computers. The different runs where distributed over the available computers using the Slurm workload manager \cite{SLURM}.\\

In the first experimental stage, similarly to the experiments conducted in the PSP-betweenness publication by Chenxu Wang and Ziyuan Lin, I tested the efficacy of both PSP heuristics and the influence of the hyperparameter $\phi$. This was done by comparing the results to the Monte Carlo method as ground-truth, using randomized graphs with randomized edge probabilities. \\

Then, in the second experimental stage, I tested the efficacy of both algorithms on real world graphs. Again, Monte Carlo was used as ground-truth. Some of these graphs have predefined edge probabilities based on real world data, while I randomized the edge probabilities for the remaining graphs.\\

In the third stage, I tested both algorithms on much larger graphs without the use of the Monte Carlo method. This was done solely to test the scalability of their runtime. \\

I used the same two metrics as in \cite{PSP} to evaluate the efficacy of both algorithms. Namely, the Mean Absolute Error (MAE) and the Spearman Correlation Coefficient (SCC). For an uncertain graph $\mathcal{G}=(V,E,P)$ and two centrality measures on said graph, $C_1,C_2:V\rightarrow\mathbb{R}$, the MAE of these two measures is given by

\[ MAE = \frac{1}{|V|} \sum_{v\in V} | C_{1}(v) - C_{2}(v) | \]

In our case, one of these two centrality measures is the one given by the PSP heuristics, and the other one is given by the Monte Carlo method. 
To test the meaningfulness of said measures, we are not only interested in the MAE, but also in the ranking of the nodes being induced. Hence, the SCC is employed as a second metric. Here, we sort the nodes by their centrality scores, i.e. for $i=1,2$ we get bijective maps $R_i : V \rightarrow \{ 1, \cdots , |V| \}$ with $R_i(v)<R_i(u)$ if $C_i(v)<C_i(u)$ for all $u,v\in V$. Then, the SCC of these two rankings is given by

\[ SCC = 1 - \frac{6\sum_{v\in V} (R_{1}(v)-R_{2}(v))^2}{|V|(|V|^2-1)} \]
This yields a value in the range [-1,1], where 1 would mean that both measures produce the exact same ranking and $-1$ would mean that we get the opposite ranking. A value of 0 would mean no correlation of the rankings at all.

\subsection{Random Graph Generators}
In the publication by Chenxu Wang and Ziyuan Lin, two models of random graphs where used to test the PSP-betweenness heuristic.  \\

Firstly, they used the Erdős-Rényi (ER) model (or, to be more precise, the version presented by Gilbert \cite{Gilbert}, which is closely related to the original model by Erdős and Rényi \cite{RandomGraphs}). This model takes two parameters, $ER(n,p)$, to sample a graph $G=(V,E)$ with $|V|=n$, where every possible edge is sampled independently with probability $p$ to appear in $E$.\\

Secondly, they used the Barabási-Albert (BA) model \cite{RandomGraphs}, which requires two parameters, $BA(n,m)$, and a seed graph $G=(V,E)$. Then, $n$ nodes are added successively to $G$. Assuming that at least $m$ nodes existed in the initial seed graph, each new node $u$ is connected to $m$ already existent nodes, where the probability of the edge $\{u,v\}$ being added is given by

\[ p_{v} = \frac{deg(v)}{\sum_{ k \in V} deg(k)} \]

So, nodes with higher degree are more likely to gain new edges (preferential attachment). This creates scale-free graphs, i.e. the amount of nodes with degree $d$, $A(d)$, follows a power-law distribution: $A(d)\sim \gamma^{-d}$ for some $\gamma\in\mathbb{R}$. \\

Additionally, we will also include a third model, namely one kind of random hyperbolic graphs as described in \cite{RHG}. In this model, 3 parameters are used: $RH(n,k,\gamma)$. Based on these parameters, a radius $R$ is calculated. Then, $n$ points are randomly distributed over a 2-dimensional hyperbolic disk with said radius $R$ (each point associated with a unique node, i.e. we generate $|V|=n$ nodes). Two nodes are connected with an edge iff the hyperbolic distance of their associated points is not greater than $R$. The generated graphs have an average degree of $k$ and are scale free with $A(d)\sim \gamma^{-d}$.\\

For all generators, I used the implementation given in the NetworKit library \cite{NetworKit}. In the $BA(n,m)$ model I did not provide a seed graph. The given implementation then first creates a circle graph with $m$ nodes and afterwards adds $n-m$ new nodes by preferential attachment, producing a graph with $|V|=n$ and $|E|=(n-m)m+m$.

\subsection{Efficacy on Random Graphs}
\subsubsection{Setup and Results by Chenxu Wang and Ziyuan Lin}
\begin{figure}[H]
\centering
\begin{subfigure}{.5\textwidth}
  \centering
  \includegraphics[width=0.8\linewidth]{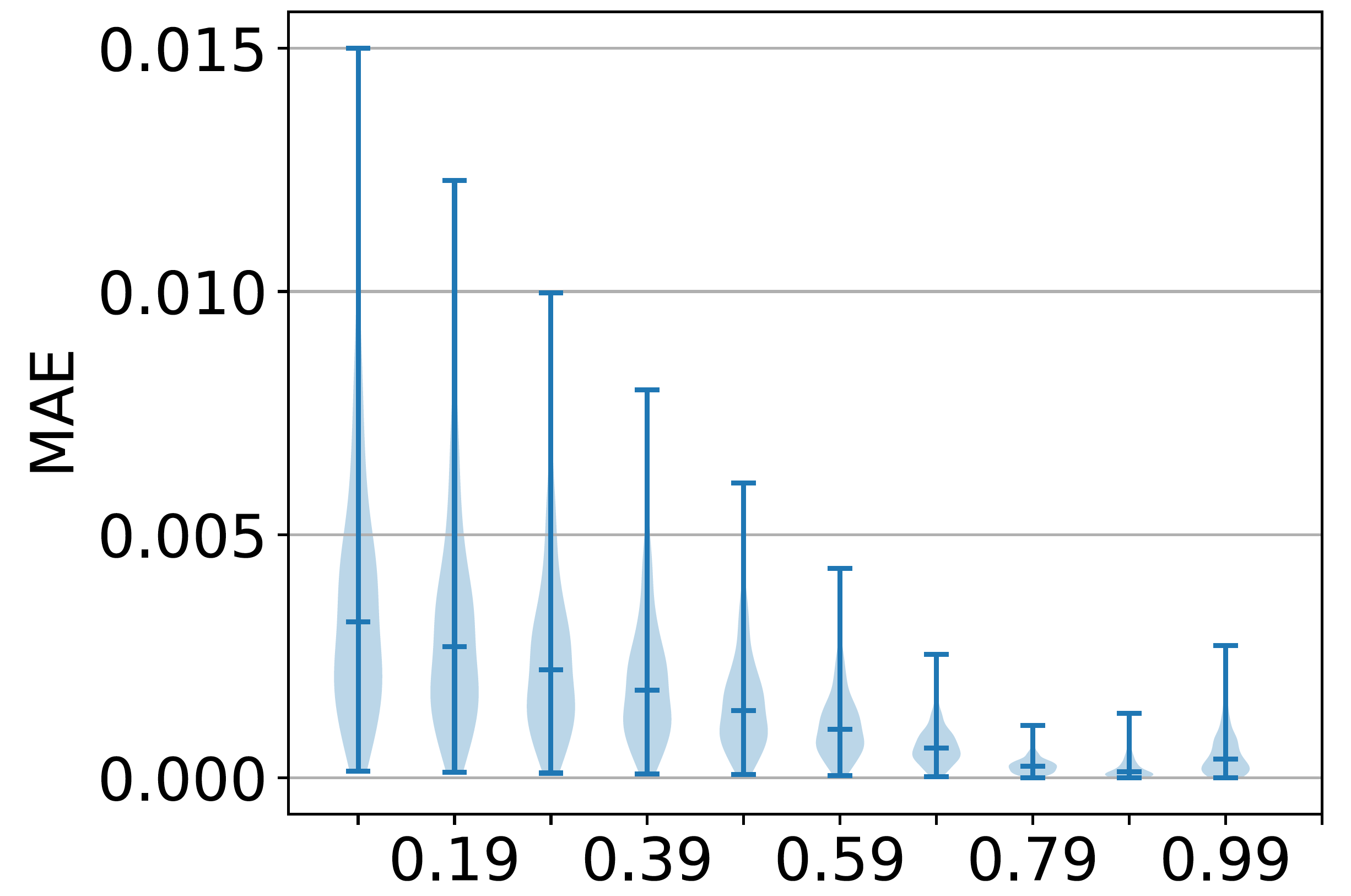}
  \caption{MAE versus $\phi$ in ER \cite{PSP}}
  \label{psp_mae_phi_er}
\end{subfigure}%
\begin{subfigure}{.5\textwidth}
  \centering
  \includegraphics[width=0.8\linewidth]{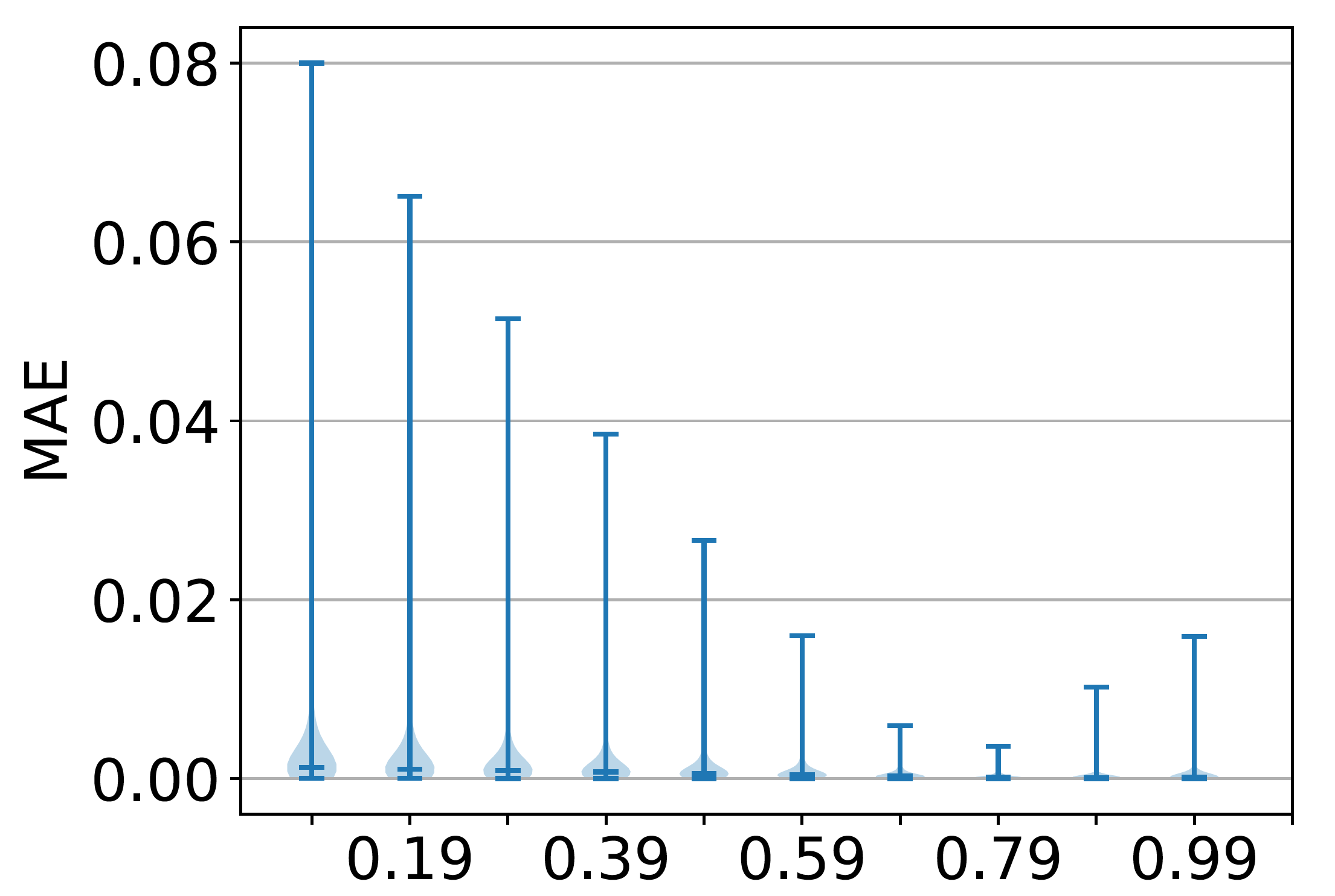}
  \caption{MAE versus $\phi$ in BA \cite{PSP}}
  \label{psp_mae_phi_ba}
\end{subfigure}
\caption{MAE versus $\phi$ for BA and ER, as Presented in \cite{PSP}}
\end{figure}

\begin{figure}[H]
	\begin{center}
		\includegraphics[width=0.4\linewidth]{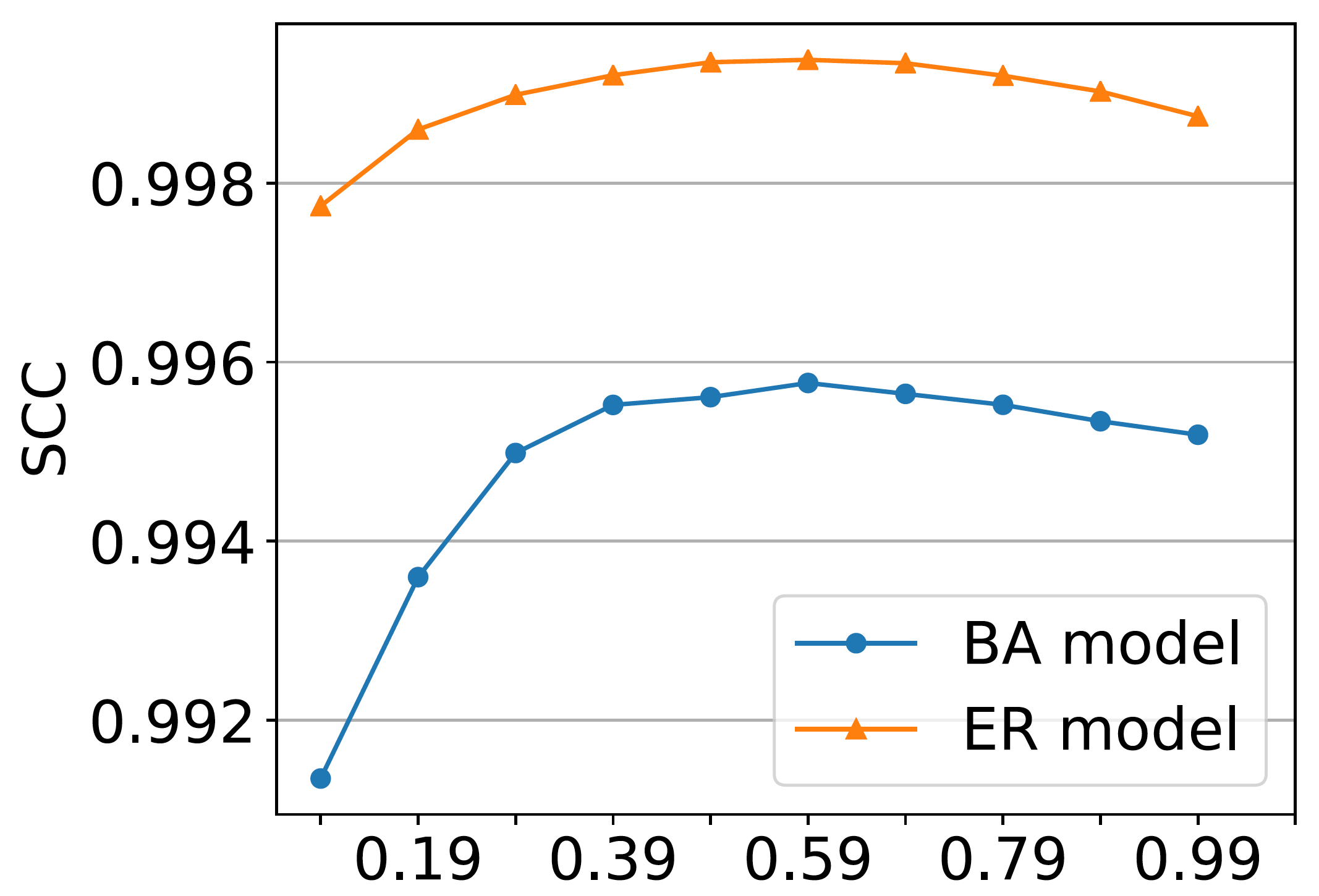}
	\caption{SCC versus $\phi$ for BA and ER, as Presented in \cite{PSP}}
	\label{psp_scc}
	\end{center}
\end{figure}

Chenxu Wang and Ziyuan Lin used 50 graphs of kind $ER(500,0.04)$ and 50 graphs of kind $BA(500,5)$ to test the PSP-betweenness heuristic and the influence of the hyperparameter $\phi$. The edge probabilities where drawn uniformly at random from [0,1], which we will now refer to as $\mathcal{U}[0,1]$. They then took the average results of the SCC and MAE over all 50 graphs of each kind and for each tested choice of $\phi$. Figure \ref{psp_mae_phi_er} shows their achieved average MAE for the ER model and figure \ref{psp_mae_phi_ba} their achieved average MAE for the BA model. The average SCC for both ER and BA is shown in figure \ref{psp_scc}.The value of $\phi$ is given on the $x$-axis. Monte Carlo with 73,777 samples was used as ground-truth.

They concluded that a value of $\phi = 0.8$ should be used, as the MAE increased for $\phi > 0.8$ and $\phi<0.8$. Most notably, the achieved average SCC was very close to one, even for small values of $\phi$, indicating a good efficacy in detecting central nodes (i.e. central in the notion of betweenness centrality). \\

\subsubsection{Setup in this Thesis}
I copied the experimental settings given in \cite{PSP} for the most part. I used the same amount of 73,777 samples for Monte Carlo to produce a fair accuracy comparison. For the exploration threshold, I tested the values of $\phi = 0.1, 0.2, \cdots, 1.0$. I used $ER(500,0.05)$ instead of $ER(500,0.04)$, though. This was due to a misread that I only noticed after evaluating the experiments. However, I decided to not rerun the experiments, as this should not influence the results by much. \\

\begin{figure}[H]
	\begin{center}
		\includegraphics[width=0.4\linewidth]{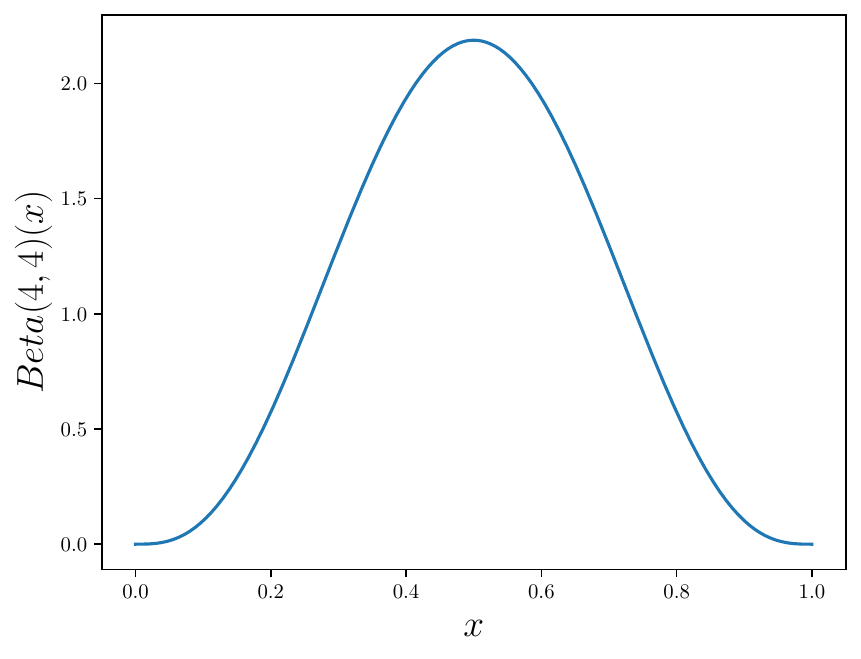}
	\end{center}
	\caption{$Beta(4,4)$ Probability Density Function}
	\label{beta_density}
\end{figure}

I restricted the amount of $BA(500,5)$ and $ER(500,0.05)$ graphs from 50 to 20 each. Though, I compensated for this limitation in two ways. Firstly by adding 20 graphs of kind $RH(500,6,3)$, $k=6$ and $\gamma=3$ being the default parameters in NetworKit. Secondly, I created two uncertain graphs from each of the generated graphs (instead of just one), yielding 120 experimental instances in total (compared to the 100 instances in \cite{PSP}). This was done by using two different kinds of randomized edge probabilities per graph: once from $\mathcal{U}[0,1]$, same as in \cite{PSP}, and once using the $Beta(4,4)$ distribution to see the impact of changing the edge probability distribution. For the parameters $p,q\in\mathbb{R}$, the $Beta(p,q)$ probability density function is defined on the interval (0,1) and given by
\[ Beta(p,q)(x) = \frac{\Gamma(p+q)}{\Gamma(p)\Gamma(q)}x^{p-1}(1-x)^{q-1} \]
Here, $\Gamma$ is the gamma function.
The reason for choosing this distribution is that it, loosely speaking, closely assembles a normal distribution for the chosen parameters $p=q=4$, but is limited to the domain (0,1). Though, numerous other choices would have been possible. Figure \ref{beta_density} shows the probability density function of $Beta(4,4)$, the expected value being $\frac{1}{2}$ with a standard deviation of $\frac{1}{6}$.

\subsubsection{Results for the PSP-Betweenness Heuristic}

\begin{figure}[H]
\centering
\begin{subfigure}{.5\textwidth}
  \centering
  \includegraphics[width=0.8\linewidth]{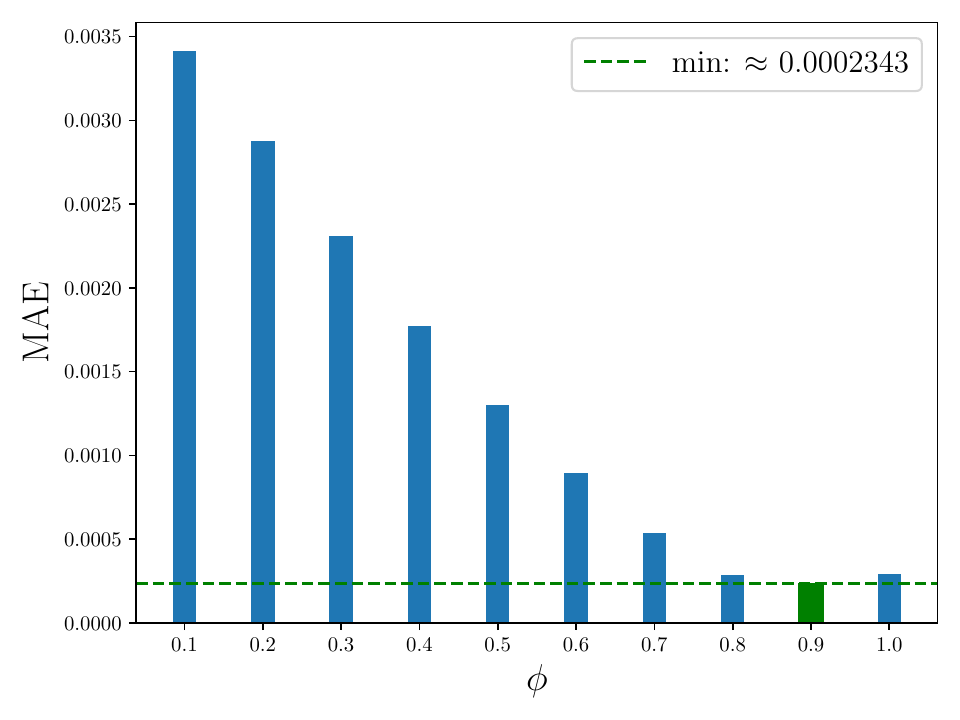}
  \caption{$Beta(4,4)$ distribution}
  \label{bet_phi_mae_ba_beta}
\end{subfigure}%
\begin{subfigure}{.5\textwidth}
  \centering
  \includegraphics[width=0.8\linewidth]{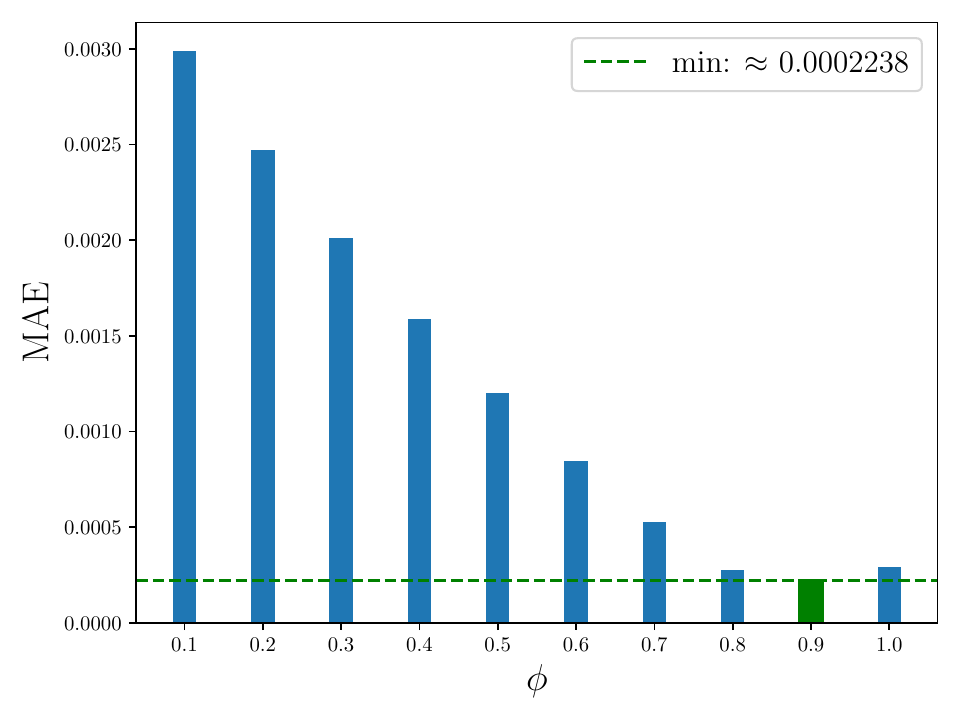}
  \caption{$\mathcal{U}[0,1]$ distribution}
  \label{bet_phi_mae_ba_uni}
\end{subfigure}
\caption{Average MAE, PSP-Betweenness Heuristic versus $\phi$, $BA(500,5)$}
\label{bet_mae_ba}
\end{figure}

Figure \ref{bet_mae_ba} shows the average MAE of the PSP-betweenness heuristic versus $\phi$ over all 40 tested BA graphs. For the beta edge distribution, shown in figure \ref{bet_phi_mae_ba_beta}, the best value was $\approx0.00023$. For the uniform edge distribution, shown in figure \ref{bet_phi_mae_ba_uni}, the best value was $\approx0.00022$. Both optimums where achieved at $\phi=0.9$, with only minor differences for $\phi=0.8,1.0$.

\begin{figure}[H]
\centering
\begin{subfigure}{.5\textwidth}
  \centering
  \includegraphics[width=0.8\linewidth]{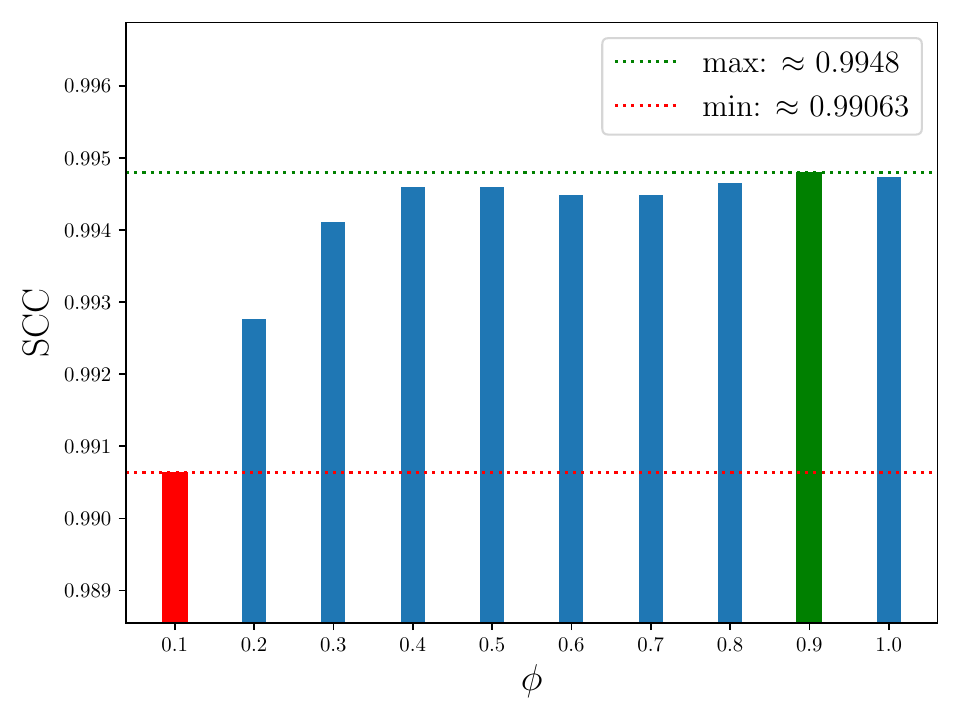}
  \caption{$Beta(4,4)$ distribution}
  \label{bet_scc_ba_beta}
\end{subfigure}%
\begin{subfigure}{.5\textwidth}
  \centering
  \includegraphics[width=0.8\linewidth]{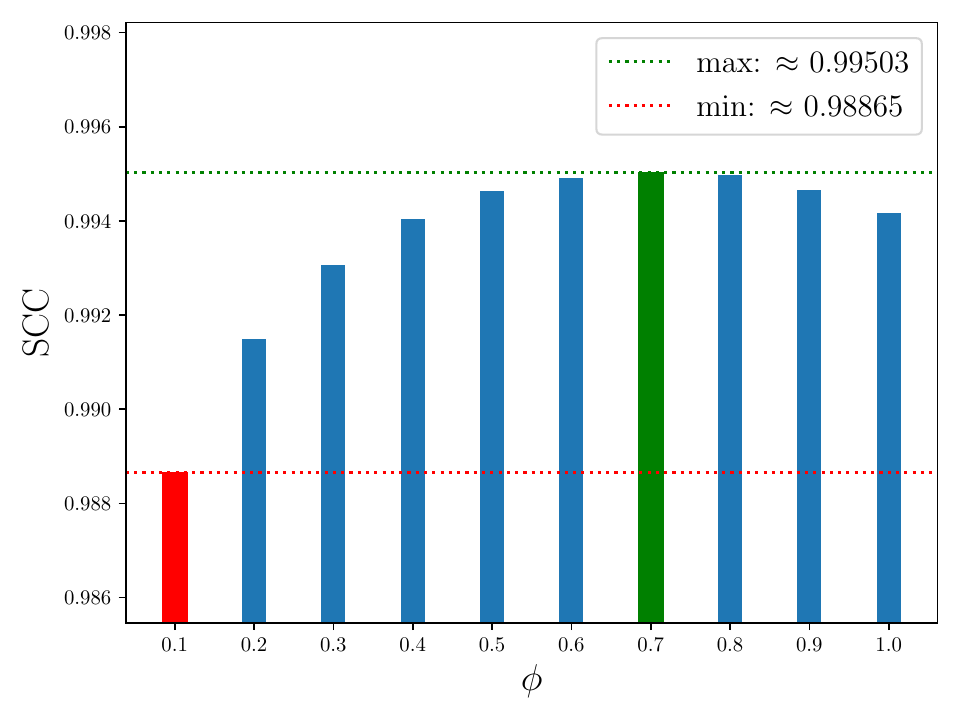}
  \caption{$\mathcal{U}[0,1]$ distribution}
  \label{bet_scc_ba_uni}
\end{subfigure}
\caption{Average SCC, PSP-Betweenness Heuristic versus $\phi$, $BA(500,5)$}
\label{bet_scc_ba}
\end{figure}

The achieved average SCC versus $\phi$ over all BA graphs is shown in figure \ref{bet_scc_ba}. The best value was $\approx 0.995$ at $\phi=0.9$ for the beta edge distribution, as seen in figure \ref{bet_scc_ba_beta}. For the uniform edge distribution, shown in figure \ref{bet_scc_ba_uni}, the best value was $\approx0.995$ as well, achieved at $\phi=0.7$. For every choice of $\phi$ though, the values where never worse than $\approx0.99$, with the minimum being achieved at $\phi=0.1$.

\begin{figure}[H]
\centering
\begin{subfigure}{.5\textwidth}
  \centering
  \includegraphics[width=0.8\linewidth]{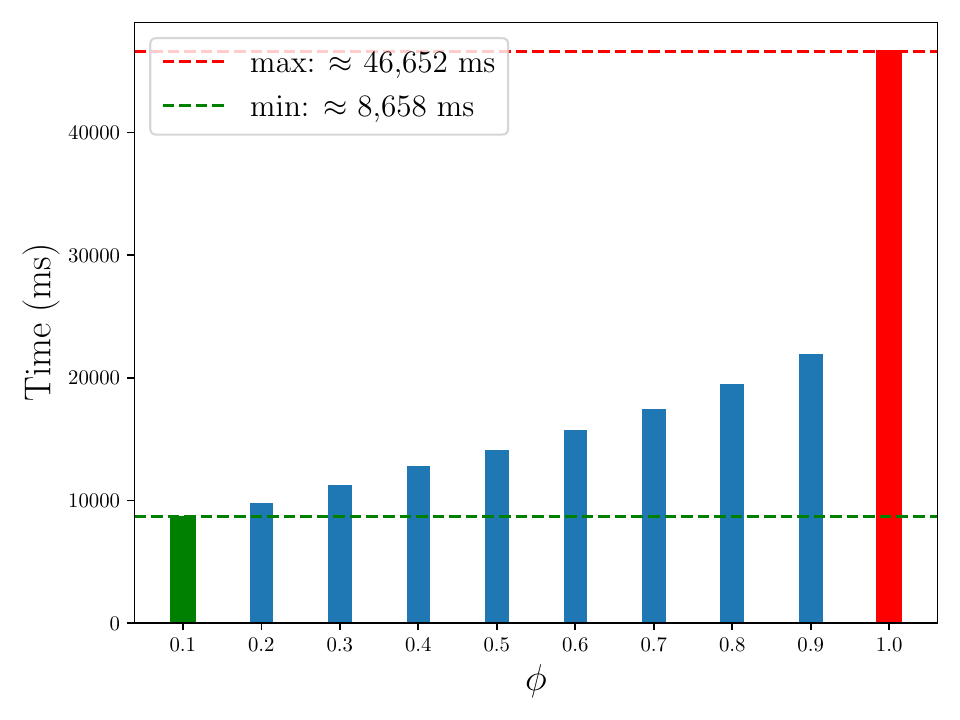}
  \caption{$Beta(4,4)$ distribution}
  \label{bet_runtime_ba_beta}
\end{subfigure}%
\begin{subfigure}{.5\textwidth}
  \centering
  \includegraphics[width=0.8\linewidth]{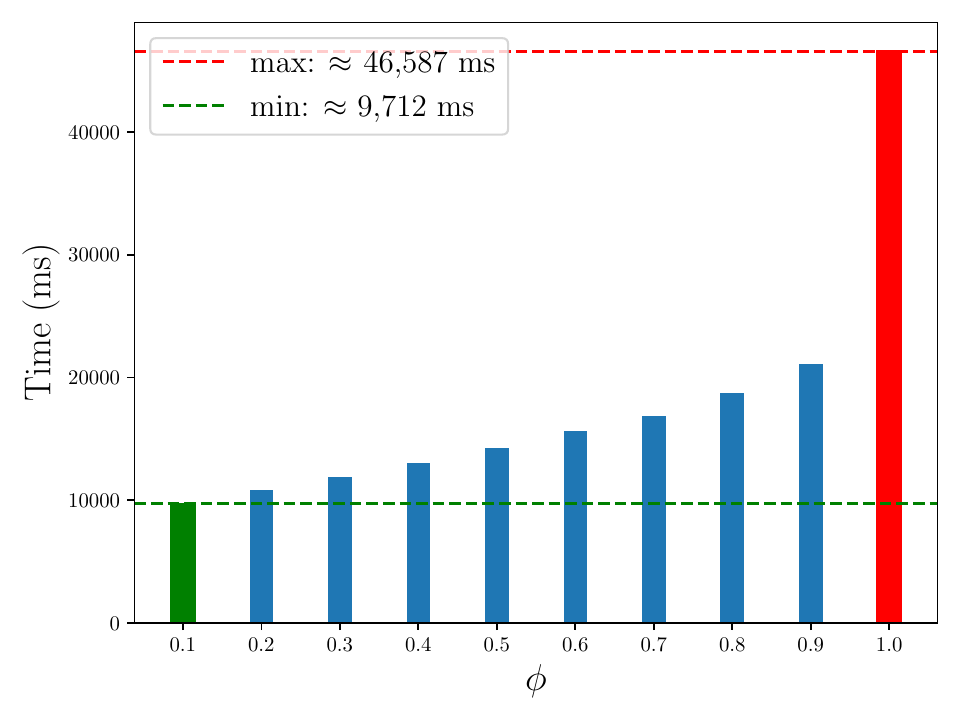}
  \caption{$\mathcal{U}[0,1]$ distribution}
  \label{bet_runtime_ba_uni}
\end{subfigure}
\caption{Running time, PSP-Betweenness Heuristic versus $\phi$, $BA(500,5)$}
\label{bet_time_ba}
\end{figure}

The running time versus $\phi$ of the PSP-betweenness heuristic on the BA graphs is shown in figure \ref{bet_time_ba}. For both edge probability distributions, we see a spike in the running time at $\phi=1.0$, while the running time did only slowly increase versus $\phi$ before that point. For the beta distribution, shown in figure \ref{bet_runtime_ba_beta}, the minimal running time was $\approx8,700$ ms, and for the uniform distribution, shown in figure \ref{bet_runtime_ba_uni}, the minimal running time was $\approx9,700$ ms. The maximal running time for both distributions was $\approx46,600$ ms at $\phi=1.0$. Independently of the chosen edge distribution, Monte Carlo with 73,777 samples had a running time of $\approx290,000$ ms.

\begin{figure}[H]
\centering
\begin{subfigure}{.5\textwidth}
  \centering
  \includegraphics[width=0.8\linewidth]{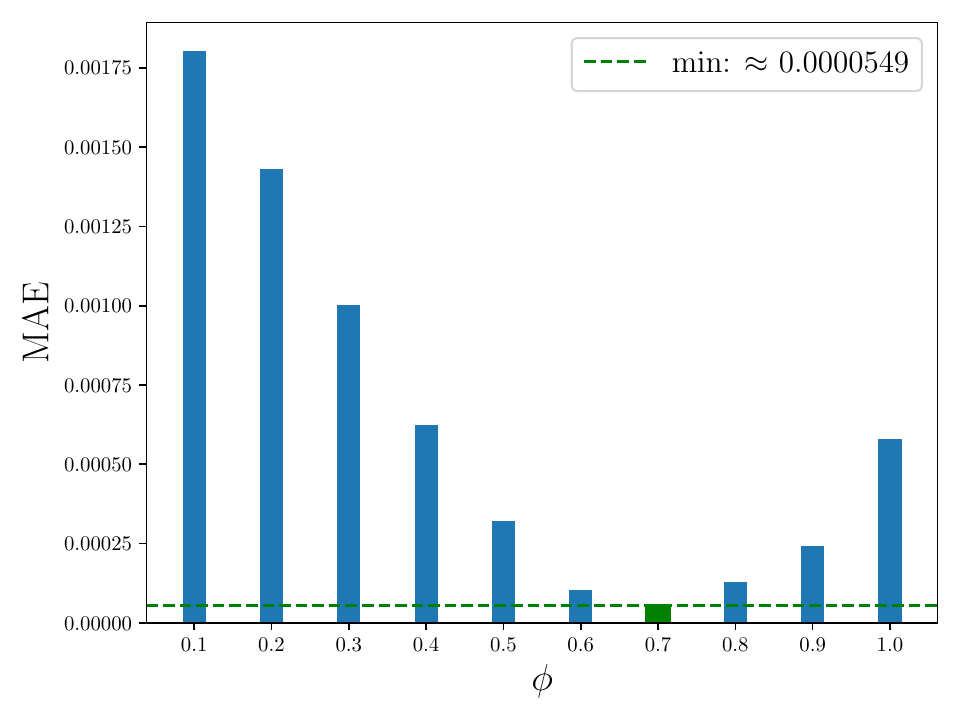}
  \caption{$Beta(4,4)$ distribution}
  \label{bet_phi_mae_er_beta}
\end{subfigure}%
\begin{subfigure}{.5\textwidth}
  \centering
  \includegraphics[width=0.8\linewidth]{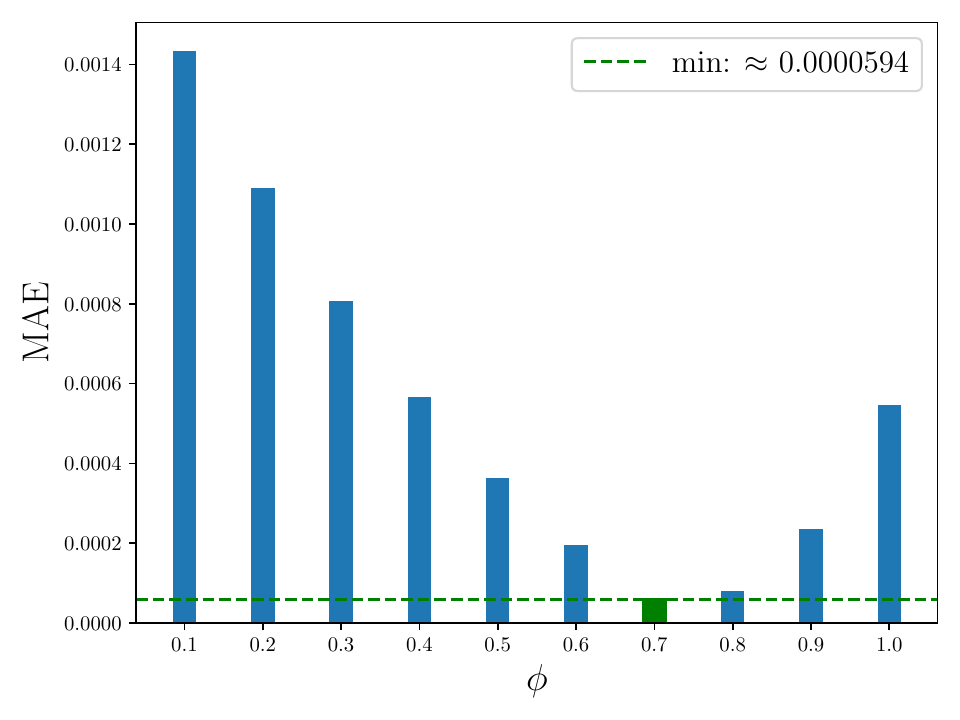}
  \caption{$\mathcal{U}[0,1]$ distribution}
  \label{bet_phi_mae_er_uni}
\end{subfigure}
\caption{Average MAE, PSP-Betweenness Heuristic versus $\phi$, $ER(500,0.05)$}
\label{bet_mae_er}
\end{figure}

Figure \ref{bet_mae_er} shows the achieved average MAE of the PSP-betweenness heuristic versus $\phi$ over all 40 tested ER graphs. For the beta edge distribution, shown in figure \ref{bet_phi_mae_er_beta}, the best value was $\approx0.00005$, and for the uniform edge distribution, shown in figure \ref{bet_phi_mae_er_uni}, the best value was $\approx0.00006$. Both minimal values where achieved at $\phi=0.7$.

\begin{figure}[H]
\centering
\begin{subfigure}{.5\textwidth}
  \centering
  \includegraphics[width=0.8\linewidth]{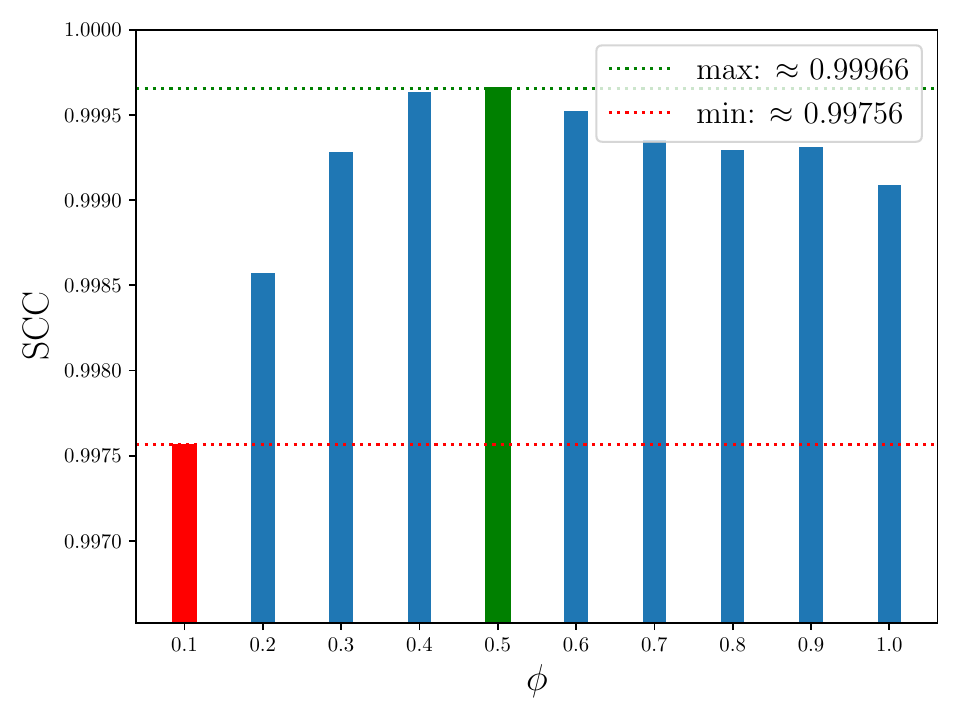}
  \caption{$Beta(4,4)$ distribution}
  \label{bet_scc_er_beta}
\end{subfigure}%
\begin{subfigure}{.5\textwidth}
  \centering
  \includegraphics[width=0.8\linewidth]{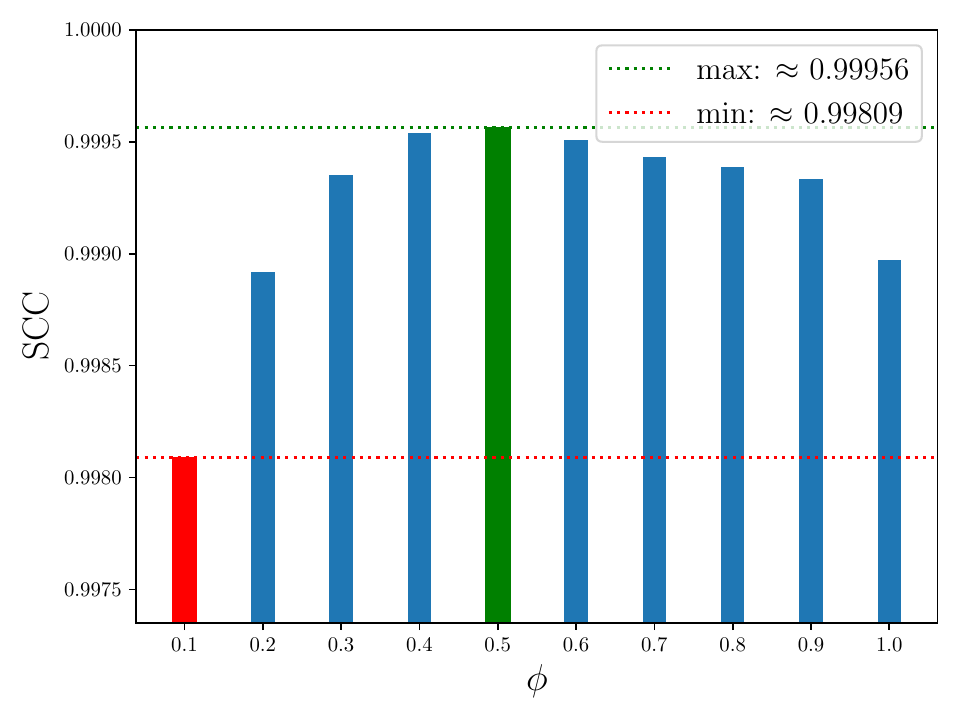}
  \caption{$\mathcal{U}[0,1]$ distribution}
  \label{bet_scc_er_uni}
\end{subfigure}
\caption{Average SCC, PSP-Betweenness Heuristic versus $\phi$, $ER(500,0.05)$}
\label{bet_scc_er}
\end{figure}

The achieved average SCC versus $\phi$ over all ER graphs is shown in figure \ref{bet_scc_er}. The best value was $\approx 0.999$ at $\phi=0.5$ for the beta edge distribution, as seen in figure \ref{bet_scc_er_beta}. For the uniform edge distribution, shown in figure \ref{bet_scc_er_uni}, the best value was $\approx0.999$ as well, also achieved at $\phi=0.5$. For every choice of $\phi$ though, the values where never worse than $\approx0.998$ for both distributions, with the minimum being achieved at $\phi=0.1$ in both cases.

\begin{figure}[H]
\centering
\begin{subfigure}{.5\textwidth}
  \centering
  \includegraphics[width=0.8\linewidth]{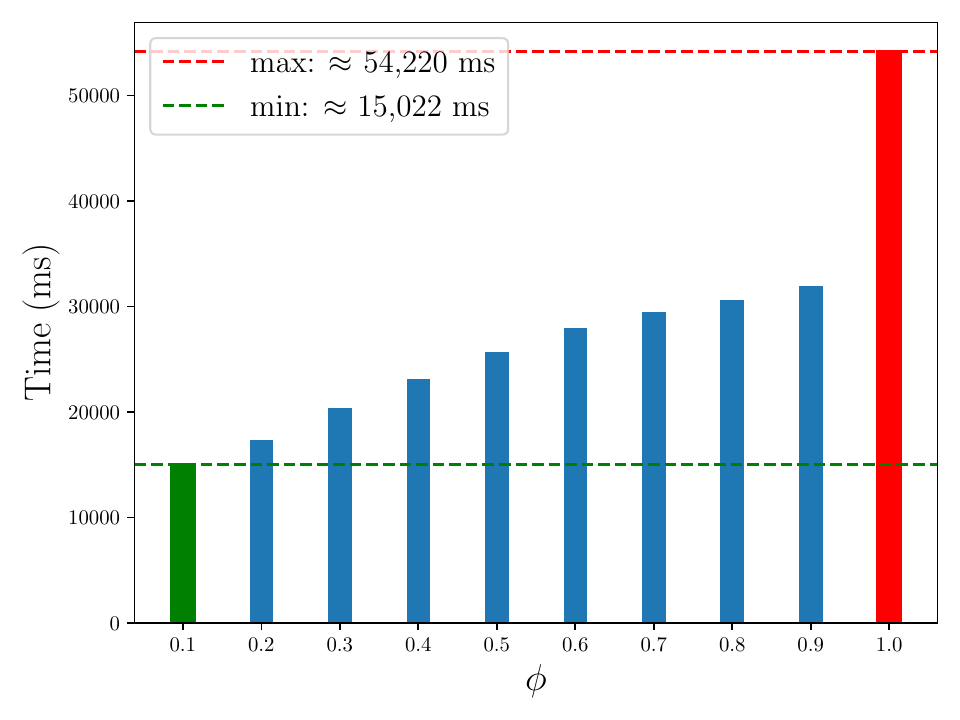}
  \caption{$Beta(4,4)$ distribution}
  \label{bet_runtime_er_beta}
\end{subfigure}%
\begin{subfigure}{.5\textwidth}
  \centering
  \includegraphics[width=0.8\linewidth]{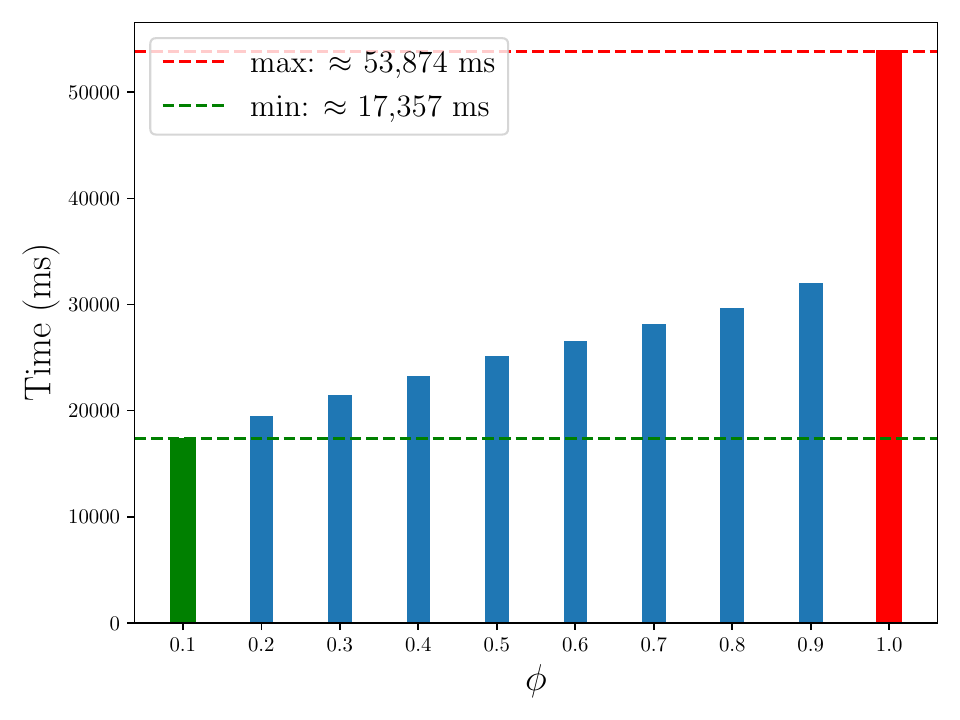}
  \caption{$\mathcal{U}[0,1]$ distribution}
  \label{bet_runtime_er_uni}
\end{subfigure}
\caption{Running time, PSP-Betweenness Heuristic versus $\phi$, $ER(500,0.05)$}
\label{bet_time_er}
\end{figure}

The running time versus $\phi$ of the PSP-betweenness heuristic on the ER graphs is shown in figure \ref{bet_time_er}. For both edge probability distributions, we see a spike in the running time at $\phi=1.0$ again, while the running time did only slowly increase versus $\phi$ before that point. For the beta distribution, shown in figure \ref{bet_runtime_er_beta}, the minimal running time was $\approx15,000$ ms, and for the uniform distribution, shown in figure \ref{bet_runtime_er_uni}, the minimal running time was $\approx17,400$ ms. The maximal running time for both distributions was $\approx54,000$ ms at $\phi=1.0$. Independently of the chosen edge distribution, Monte Carlo with 73,777 samples had a running time of $\approx685,000$ ms.

\begin{figure}[H]
\centering
\begin{subfigure}{.5\textwidth}
  \centering
  \includegraphics[width=0.8\linewidth]{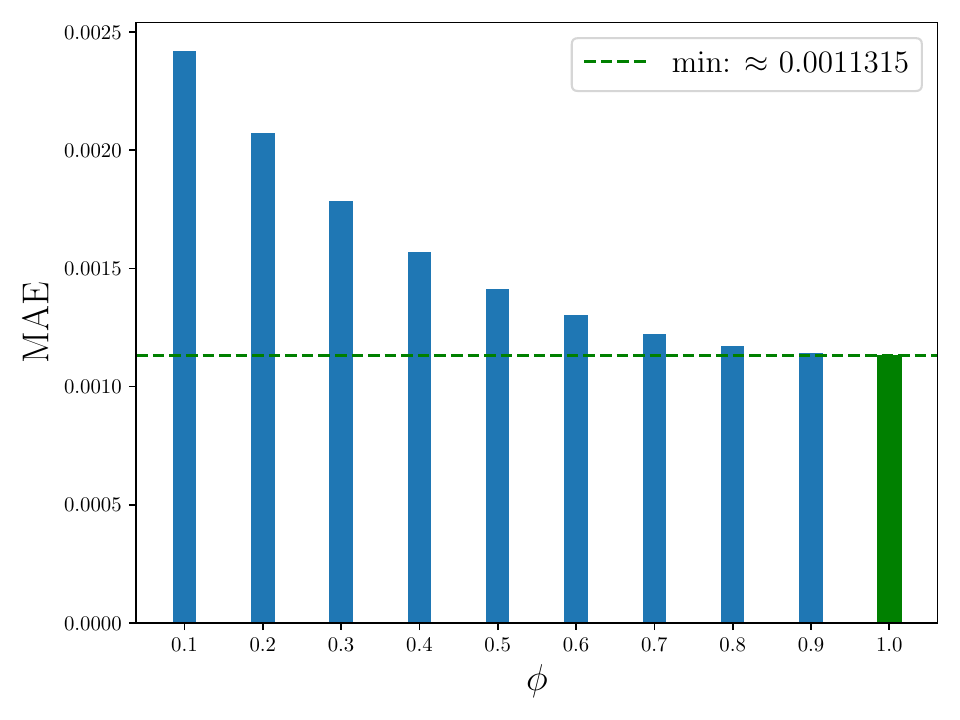}
  \caption{$Beta(4,4)$ distribution}
  \label{bet_phi_mae_rh_beta}
\end{subfigure}%
\begin{subfigure}{.5\textwidth}
  \centering
  \includegraphics[width=0.8\linewidth]{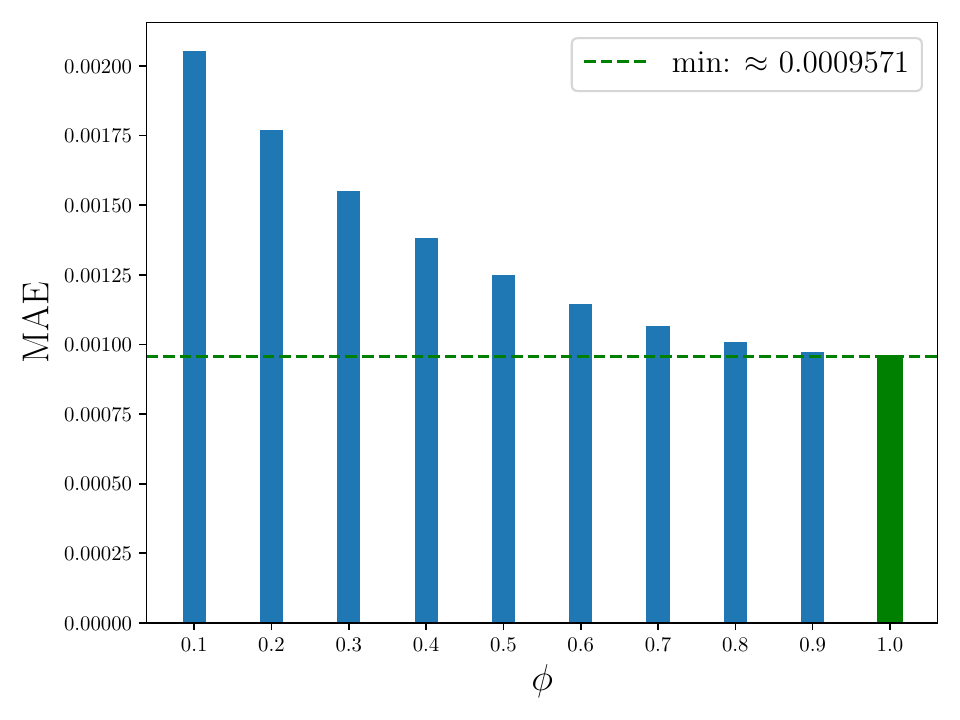}
  \caption{$\mathcal{U}[0,1]$ distribution}
  \label{bet_phi_mae_rh_uni}
\end{subfigure}
\caption{Average MAE, PSP-Betweenness Heuristic versus $\phi$, $RH(500,6,3)$}
\label{bet_mae_rh}
\end{figure}

Figure \ref{bet_mae_rh} shows the achieved average MAE of the PSP-betweenness heuristic versus $\phi$ over all 40 tested RH graphs. For the beta edge distribution, shown in figure \ref{bet_phi_mae_rh_beta}, the best value was $\approx0.0011$, and for the uniform edge distribution, shown in figure \ref{bet_phi_mae_rh_uni}, the best value was $\approx0.0009$. Both optimums where achieved at $\phi=1.0$, with almost equally good values for $\phi=0.8, 0.9$ though.

\begin{figure}[H]
\centering
\begin{subfigure}{.5\textwidth}
  \centering
  \includegraphics[width=0.8\linewidth]{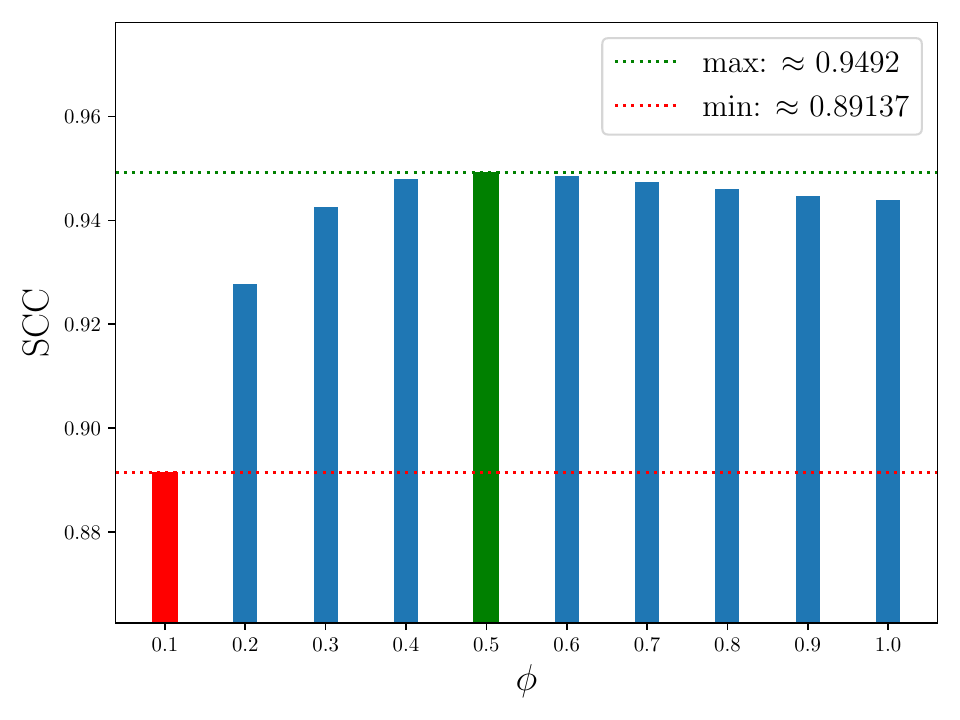}
  \caption{$Beta(4,4)$ distribution}
  \label{bet_scc_rh_beta}
\end{subfigure}%
\begin{subfigure}{.5\textwidth}
  \centering
  \includegraphics[width=0.8\linewidth]{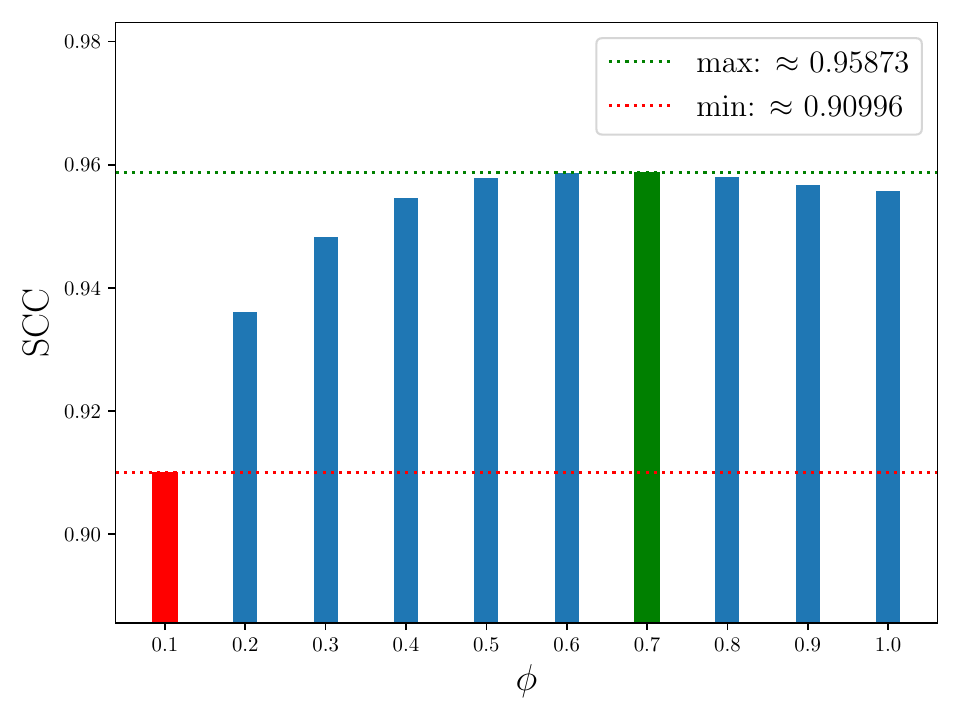}
  \caption{$\mathcal{U}[0,1]$ distribution}
  \label{bet_scc_rh_uni}
\end{subfigure}
\caption{Average SCC, PSP-Betweenness Heuristic versus $\phi$, $RH(500,6,3)$}
\label{bet_scc_rh}
\end{figure}

The achieved average SCC versus $\phi$ over all RH graphs is shown in figure \ref{bet_scc_rh}. The best value was $\approx 0.95$ at $\phi=0.5$ for the beta edge distribution, as seen in figure \ref{bet_scc_rh_beta}. For the uniform edge distribution, shown in figure \ref{bet_scc_rh_uni}, the best value was $\approx0.96$, achieved at $\phi=0.7$. For every choice of $\phi$ though, the values where never worse than $\approx0.9$ for both distributions, with the minimum being achieved at $\phi=0.1$ again.

\begin{figure}[H]
\centering
\begin{subfigure}{.5\textwidth}
  \centering
  \includegraphics[width=0.8\linewidth]{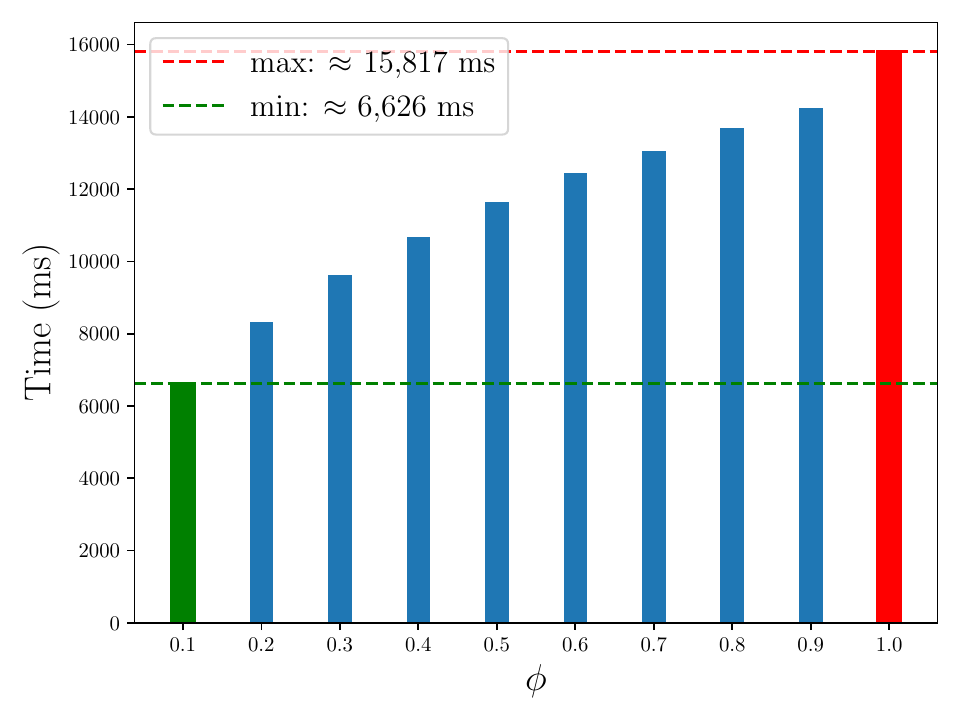}
  \caption{$Beta(4,4)$ distribution}
  \label{bet_runtime_rh_beta}
\end{subfigure}%
\begin{subfigure}{.5\textwidth}
  \centering
  \includegraphics[width=0.8\linewidth]{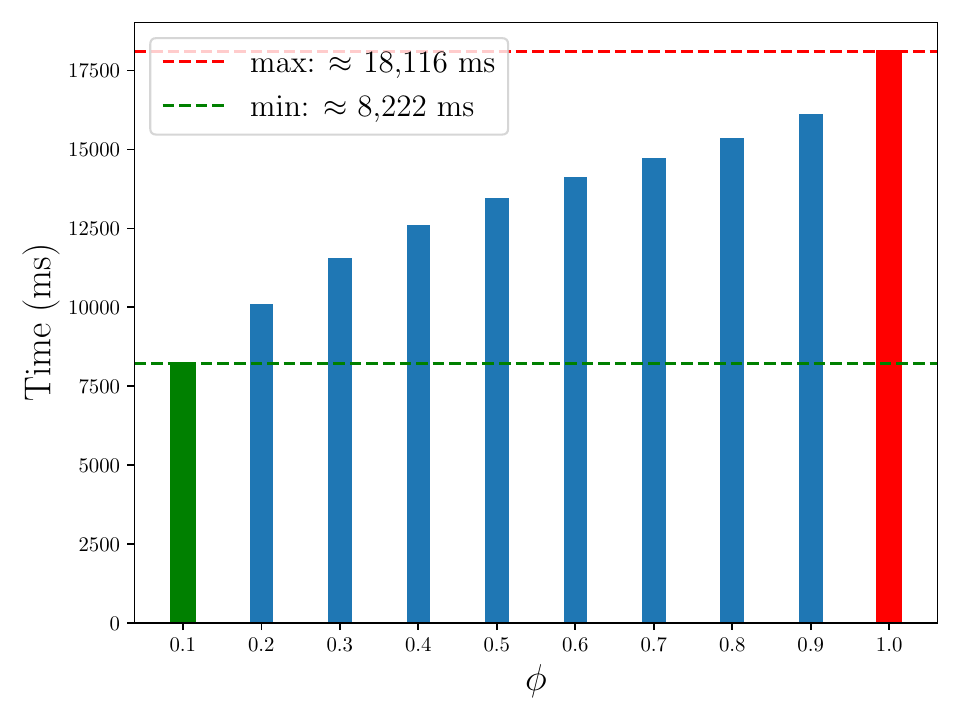}
  \caption{$\mathcal{U}[0,1]$ distribution}
  \label{bet_runtime_rh_uni}
\end{subfigure}
\caption{Running time, PSP-Betweenness Heuristic versus $\phi$, $RH(500,6,3)$}
\label{bet_time_rh}
\end{figure}

The running time versus $\phi$ of the PSP-betweenness heuristic on the RH graphs is shown in figure \ref{bet_time_rh}. For both edge probability distributions, we do not see a spike in the running time at $\phi=1.0$ this time, with the running time roughly increasing linearly versus $\phi$. For the beta distribution, shown in figure \ref{bet_runtime_rh_beta}, the minimal running time was $\approx6,600$ ms, and for the uniform distribution, shown in figure \ref{bet_runtime_rh_uni}, the minimal running time was $\approx8,200$ ms. The maximal running time for the beta distributions was $\approx15,800$ ms, and for the uniform edge distribution, it was $\approx18,100$ ms. Independently of the chosen edge distribution, Monte Carlo with 73,777 samples had a running time of $\approx74,000$ ms.

\subsubsection{Results for the PSP-Harmonic Heuristic}

\begin{figure}[H]
\centering
\begin{subfigure}{.5\textwidth}
  \centering
  \includegraphics[width=0.8\linewidth]{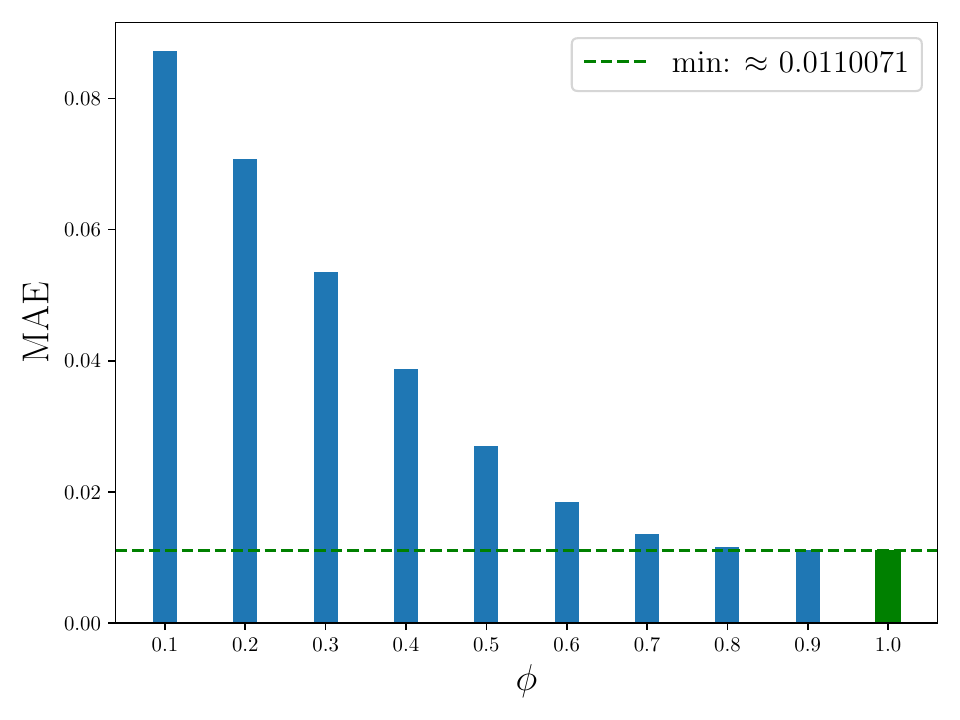}
  \caption{$Beta(4,4)$ distribution}
  \label{har_phi_mae_ba_beta}
\end{subfigure}%
\begin{subfigure}{.5\textwidth}
  \centering
  \includegraphics[width=0.8\linewidth]{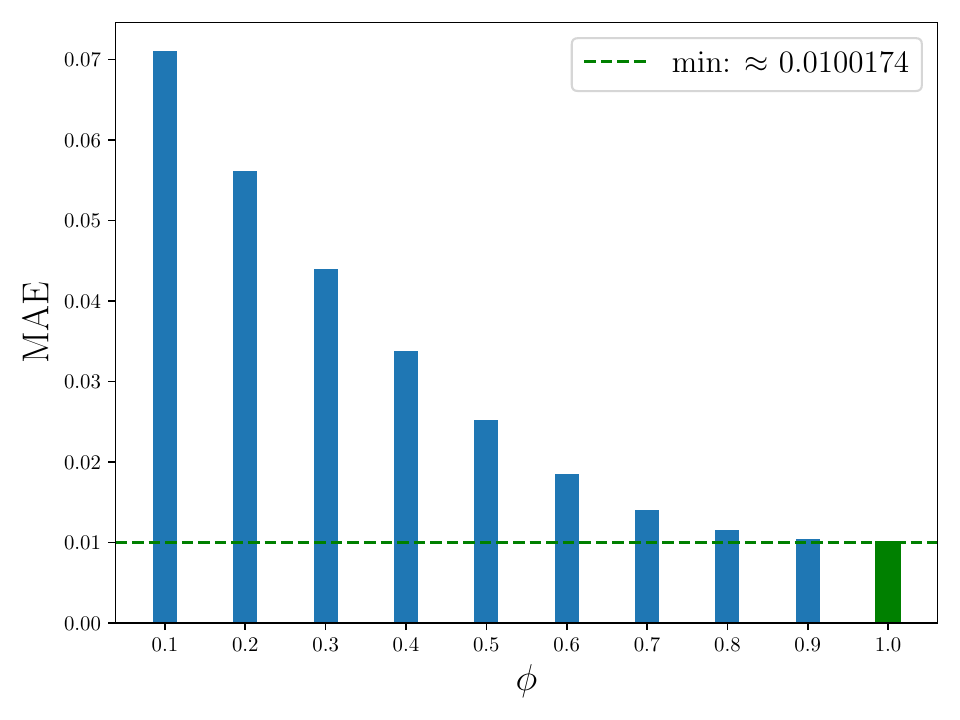}
  \caption{$\mathcal{U}[0,1]$ distribution}
  \label{har_phi_mae_ba_uni}
\end{subfigure}
\caption{Average MAE, PSP-Harmonic Heuristic versus $\phi$, $BA(500,5)$}
\label{har_mae_ba}
\end{figure}

Figure \ref{har_mae_ba} shows the achieved average MAE of the PSP-harmonic heuristic versus $\phi$ over all 40 tested BA graphs. For the beta edge distribution, shown in figure \ref{har_phi_mae_ba_beta}, the best value was $\approx0.01$, and for the uniform edge distribution, shown in figure \ref{har_phi_mae_ba_uni}, the best value also $\approx0.01$. Both minimal values where achieved at $\phi=1.0$. Same as for betweenness though, the differences where minor for $\phi=0.8, 0.9$

\begin{figure}[H]
\centering
\begin{subfigure}{.5\textwidth}
  \centering
  \includegraphics[width=0.8\linewidth]{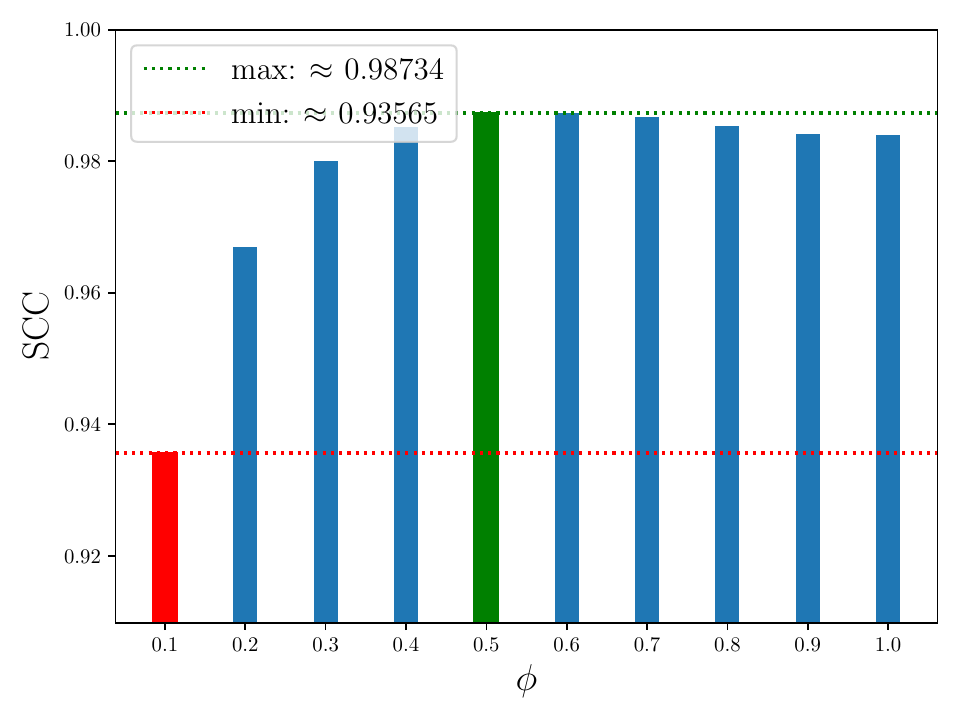}
  \caption{$Beta(4,4)$ distribution}
  \label{har_scc_ba_beta}
\end{subfigure}%
\begin{subfigure}{.5\textwidth}
  \centering
  \includegraphics[width=0.8\linewidth]{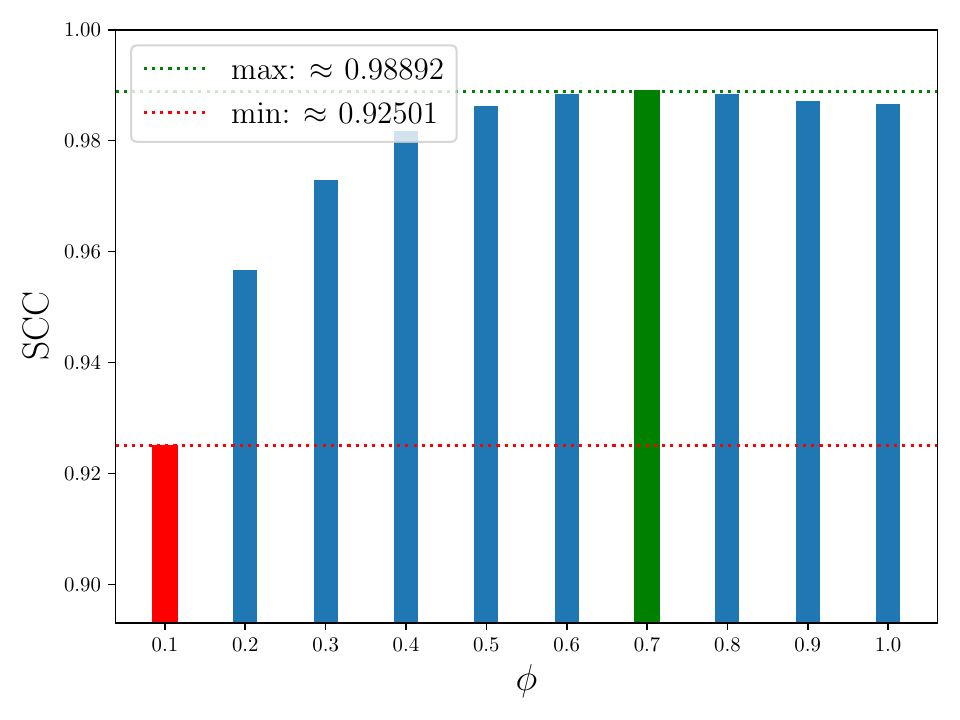}
  \caption{$\mathcal{U}[0,1]$ distribution}
  \label{har_scc_ba_uni}
\end{subfigure}
\caption{Average SCC, PSP-Harmonic Heuristic versus $\phi$, $BA(500,5)$}
\label{har_scc_ba}
\end{figure}

The achieved average SCC versus $\phi$ on the BA graphs is shown in figure \ref{har_scc_ba}. The best value was $\approx 0.988$ at $\phi=0.5$ for the beta edge distribution, as seen in figure \ref{har_scc_ba_beta}. For the uniform edge distribution, shown in figure \ref{har_scc_ba_uni}, the best value was $\approx0.989$, achieved at $\phi=0.7$. For every choice of $\phi$ though, the values where never worse than $\approx0.93$, with the minimum being achieved at $\phi=0.1$.

\begin{figure}[H]
\centering
\begin{subfigure}{.5\textwidth}
  \centering
  \includegraphics[width=0.8\linewidth]{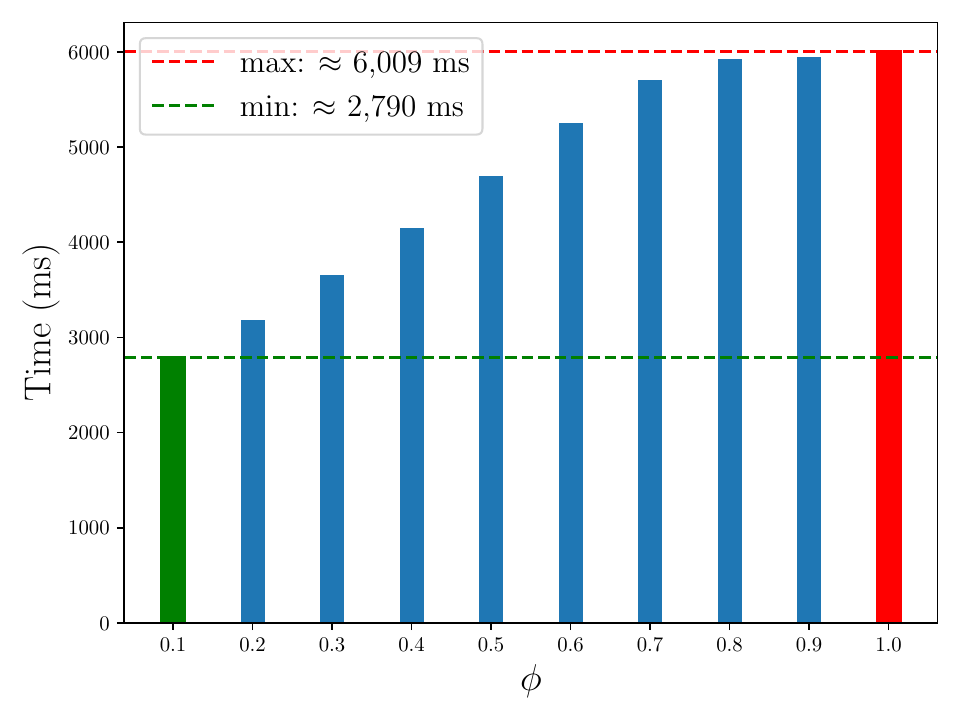}
  \caption{$Beta(4,4)$ distribution}
  \label{har_runtime_ba_beta}
\end{subfigure}%
\begin{subfigure}{.5\textwidth}
  \centering
  \includegraphics[width=0.8\linewidth]{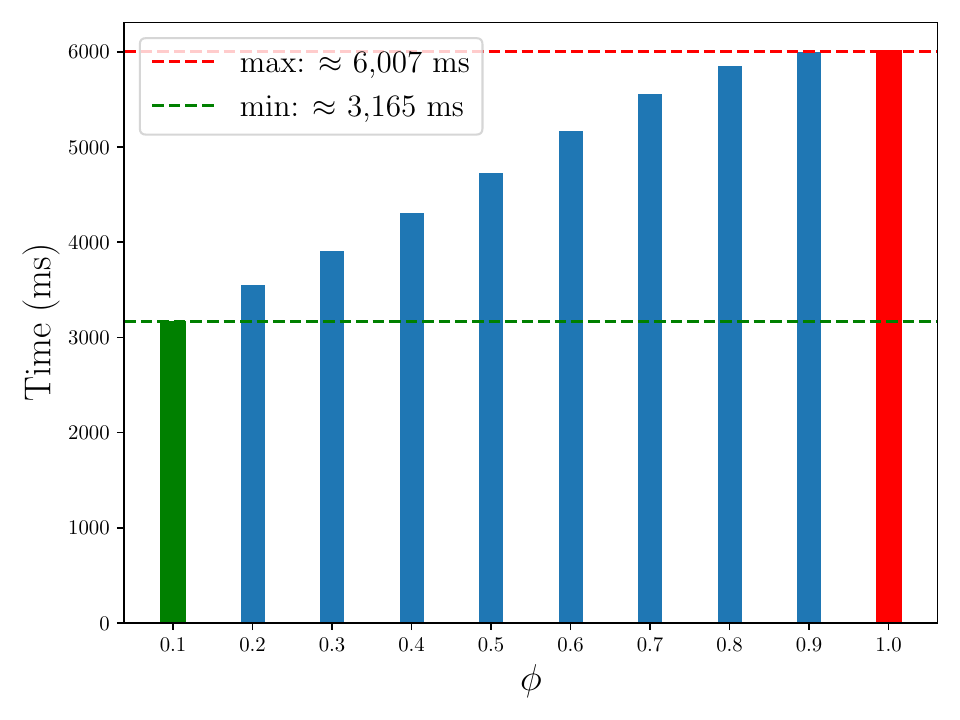}
  \caption{$\mathcal{U}[0,1]$ distribution}
  \label{har_runtime_ba_uni}
\end{subfigure}
\caption{Running time, PSP-Harmonic Heuristic versus $\phi$, $BA(500,5)$}
\label{har_time_ba}
\end{figure}

The running time versus $\phi$ of the PSP-harmonic heuristic on the BA graphs is shown in figure \ref{har_time_ba}. For both edge probability distributions, we now do not see a spike at $\phi=1.0$ and roughly equal running time for $\phi=0.8,0.9,1.0$. Before that point, the running time did roughly increase linearly versus $\phi$. For the beta distribution, shown in figure \ref{har_runtime_ba_beta}, the minimal running time was $\approx2,800$ ms, and for the uniform distribution, shown in figure \ref{har_runtime_ba_uni}, the minimal running time was $\approx3,200$ms. The maximal running time for both distributions was $\approx6,000$ ms at $\phi=1.0$. Independently of the chosen edge distribution, 73,777 Monte Carlo samples had a running time of $\approx250,000$ ms.

\begin{figure}[H]
\centering
\begin{subfigure}{.5\textwidth}
  \centering
  \includegraphics[width=0.8\linewidth]{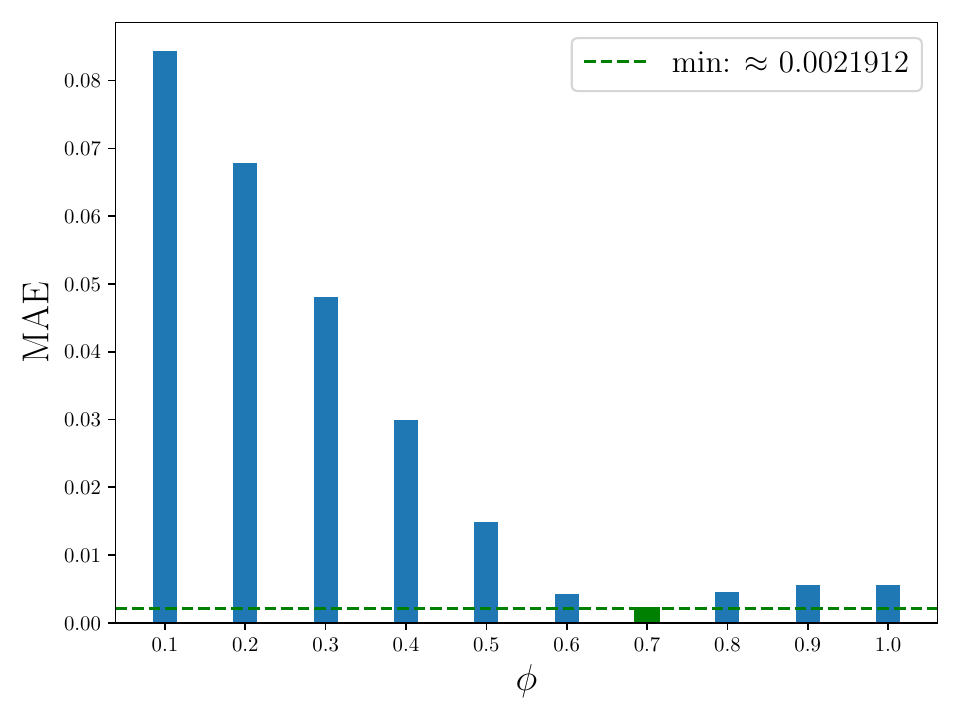}
  \caption{$Beta(4,4)$ distribution}
  \label{har_phi_mae_er_beta}
\end{subfigure}%
\begin{subfigure}{.5\textwidth}
  \centering
  \includegraphics[width=0.8\linewidth]{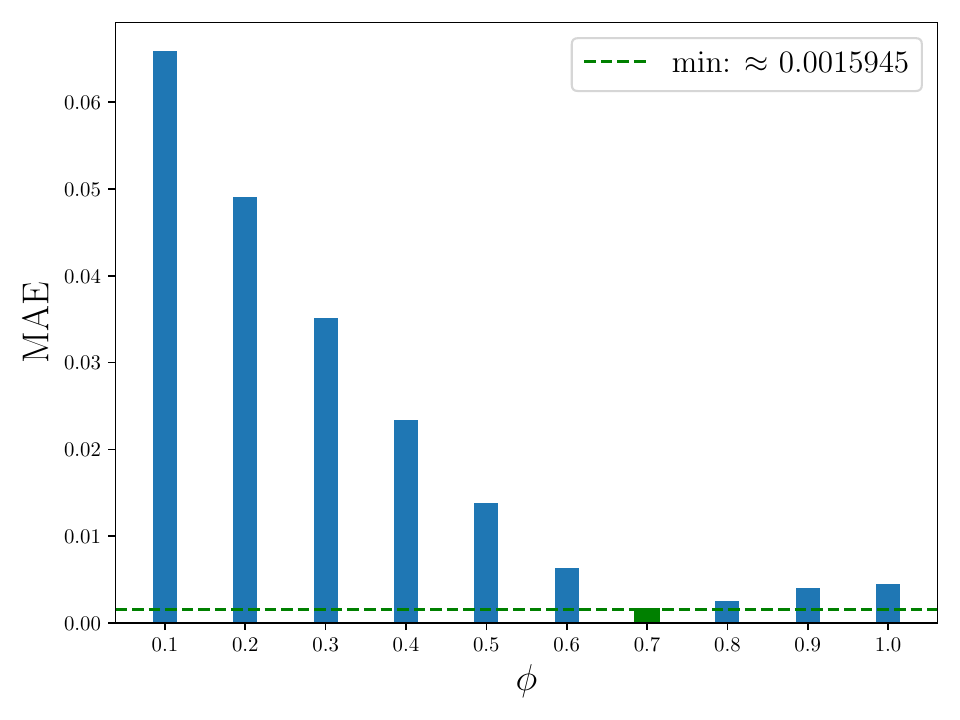}
  \caption{$\mathcal{U}[0,1]$ distribution}
  \label{har_phi_mae_er_uni}
\end{subfigure}
\caption{Average MAE, PSP-Harmonic Heuristic versus $\phi$, $ER(500,0.05)$}
\label{har_mae_er}
\end{figure}

Figure \ref{har_mae_er} shows the achieved average MAE versus $\phi$ over all 40 tested ER graphs. For the beta edge distribution, shown in figure \ref{har_phi_mae_er_beta}, the best value was $\approx0.002$. For the uniform edge distribution, shown in figure \ref{har_phi_mae_er_uni}, the best value was $\approx0.002$ as well, both achieved at $\phi=0.7$.

\begin{figure}[H]
\centering
\begin{subfigure}{.5\textwidth}
  \centering
  \includegraphics[width=0.8\linewidth]{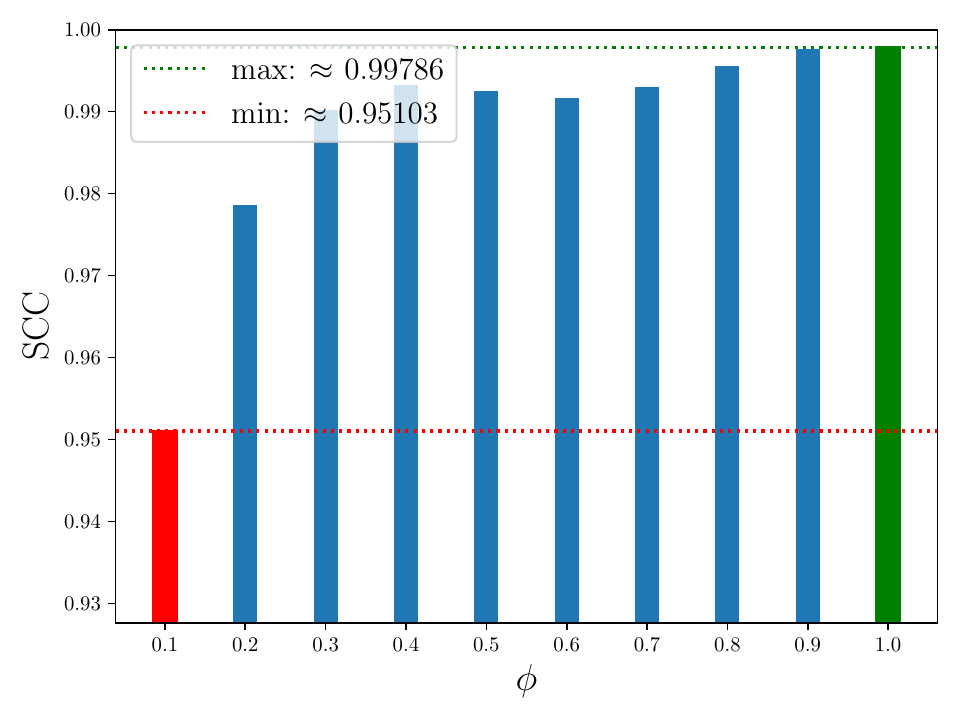}
  \caption{$Beta(4,4)$ distribution}
  \label{har_scc_er_beta}
\end{subfigure}%
\begin{subfigure}{.5\textwidth}
  \centering
  \includegraphics[width=0.8\linewidth]{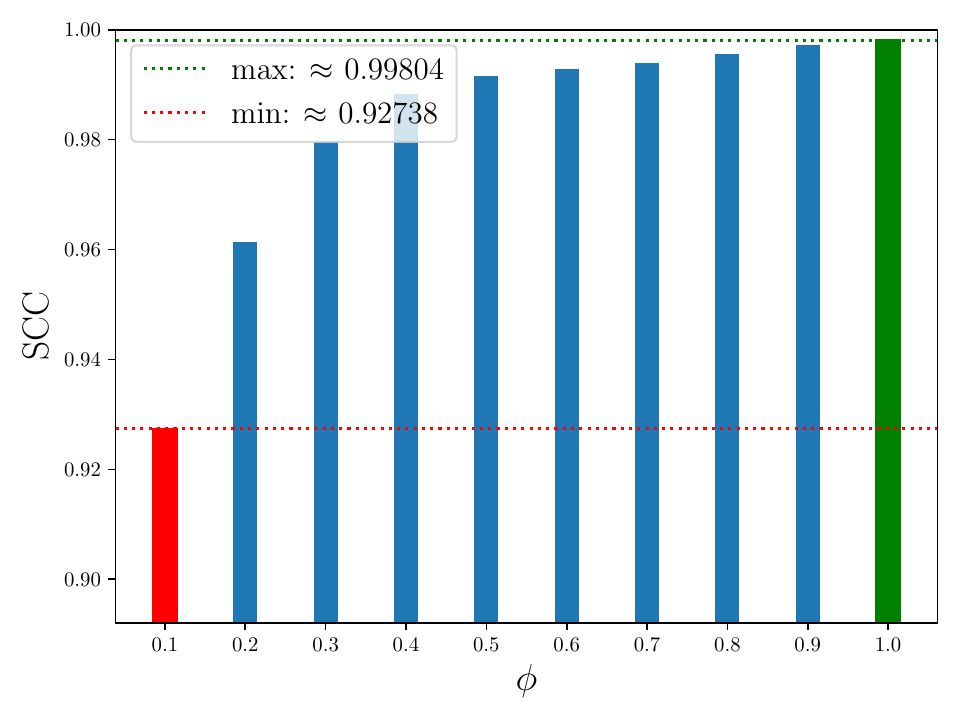}
  \caption{$\mathcal{U}[0,1]$ distribution}
  \label{har_scc_er_uni}
\end{subfigure}
\caption{Average SCC, PSP-Harmonic Heuristic versus $\phi$, $ER(500,0.05)$}
\label{har_scc_er}
\end{figure}

The achieved average SCC versus $\phi$ on the ER graphs is shown in figure \ref{har_scc_er}. The best value was $\approx 0.998$ at $\phi=1.0$ for the beta edge distribution, as seen in figure \ref{har_scc_er_beta}. For the uniform edge distribution, shown in figure \ref{har_scc_er_uni}, the best value was $\approx0.998$ as well, also achieved at $\phi=1.0$. For every choice of $\phi$ though, the values where never worse than $\approx0.93$ for both distributions, with the minimum being achieved at $\phi=0.1$ in both cases.

\begin{figure}[H]
\centering
\begin{subfigure}{.5\textwidth}
  \centering
  \includegraphics[width=0.8\linewidth]{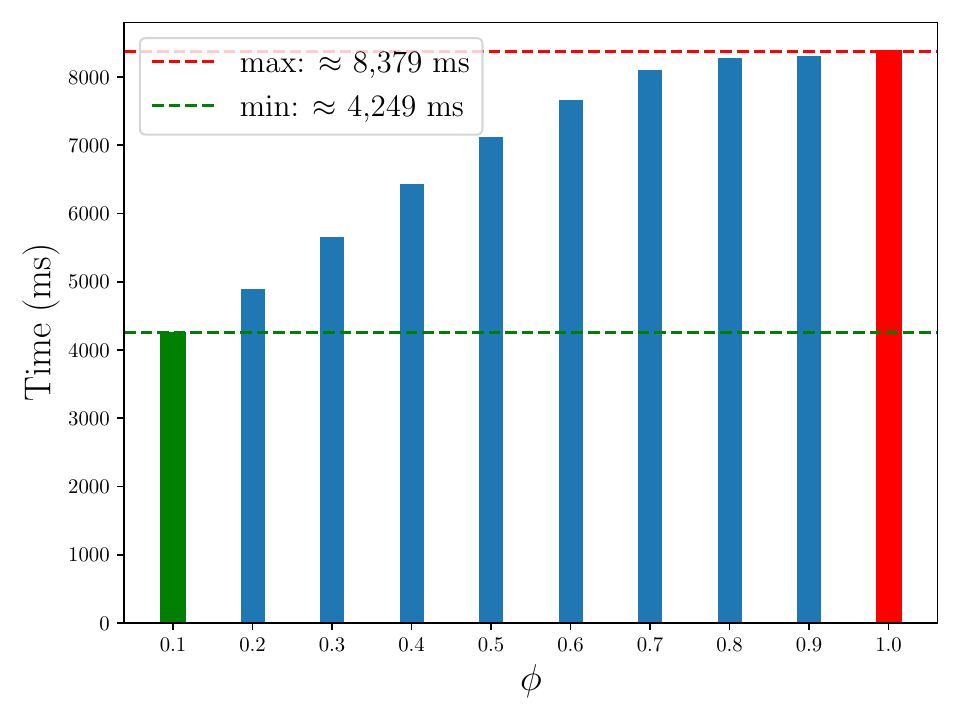}
  \caption{$Beta(4,4)$ distribution}
  \label{har_runtime_er_beta}
\end{subfigure}%
\begin{subfigure}{.5\textwidth}
  \centering
  \includegraphics[width=0.8\linewidth]{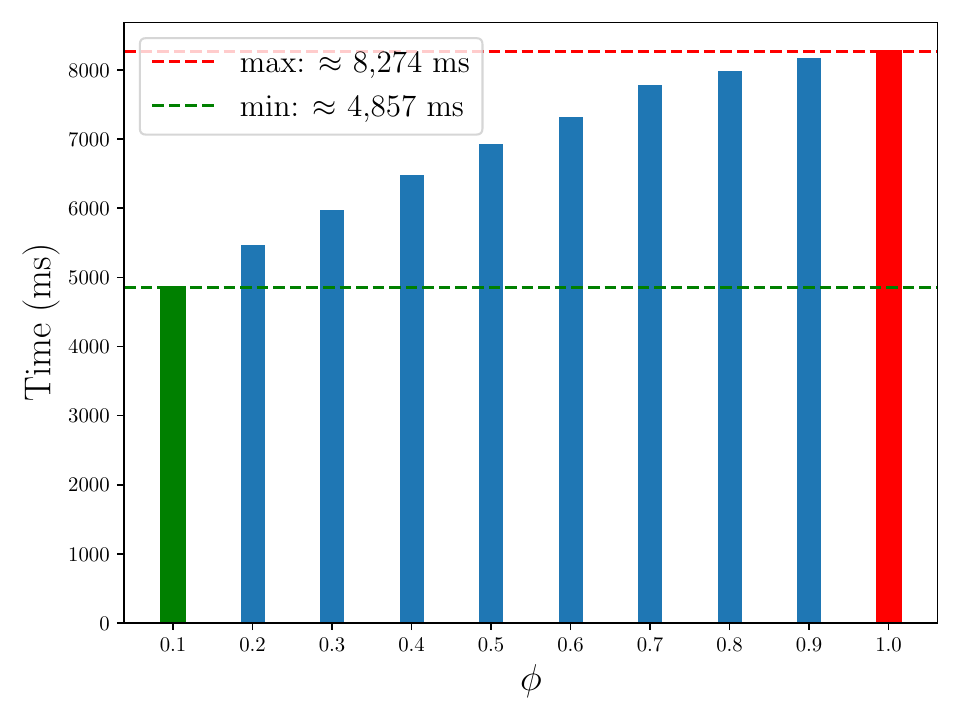}
  \caption{$\mathcal{U}[0,1]$ distribution}
  \label{har_runtime_er_uni}
\end{subfigure}
\caption{Running time, PSP-Harmonic Heuristic versus $\phi$, $ER(500,0.05)$}
\label{har_time_er}
\end{figure}

The running time versus $\phi$ of the PSP-harmonic heuristic on the ER graphs is shown in figure \ref{har_time_er}. Again, we do not see the spike at $\phi=1.0$ and roughly equal running time for $\phi=0.8,0.9,1.0$. For the beta distribution, shown in figure \ref{har_runtime_er_beta}, the minimal running time was $\approx4,200$ ms, and for the uniform distribution, shown in figure \ref{har_runtime_er_uni}, the minimal running time was $\approx4,900$ ms. The maximal running time for both distributions was $\approx8,300$ ms. Independently of the chosen edge distribution, Monte Carlo with 73,777 samples had a running time of $\approx630,000$ ms.

\begin{figure}[H]
\centering
\begin{subfigure}{.5\textwidth}
  \centering
  \includegraphics[width=0.8\linewidth]{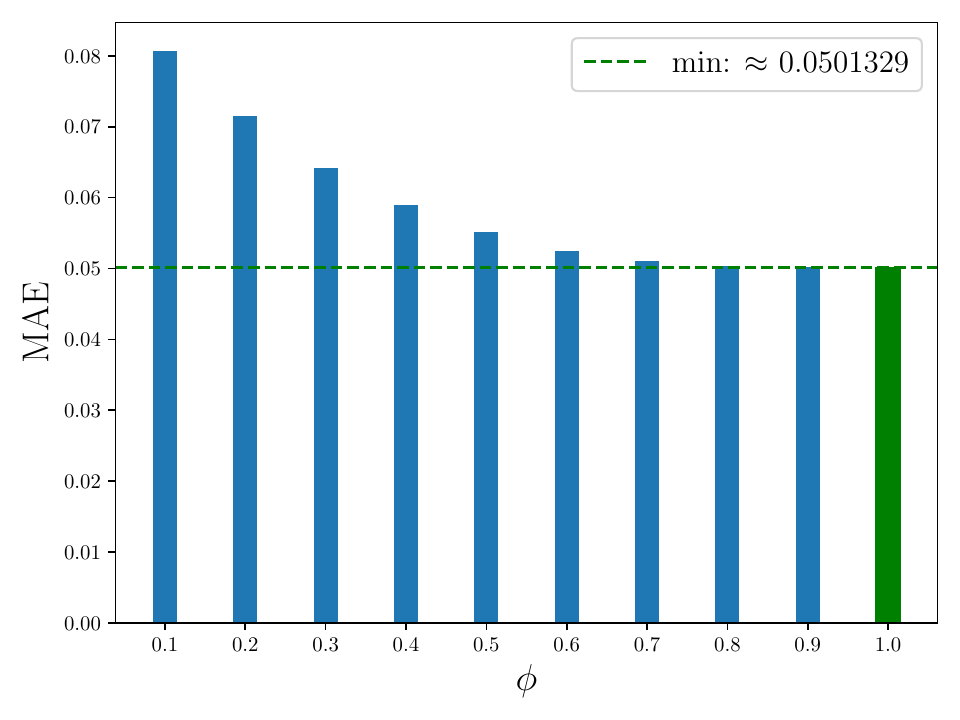}
  \caption{$Beta(4,4)$ distribution}
  \label{har_phi_mae_rh_beta}
\end{subfigure}%
\begin{subfigure}{.5\textwidth}
  \centering
  \includegraphics[width=0.8\linewidth]{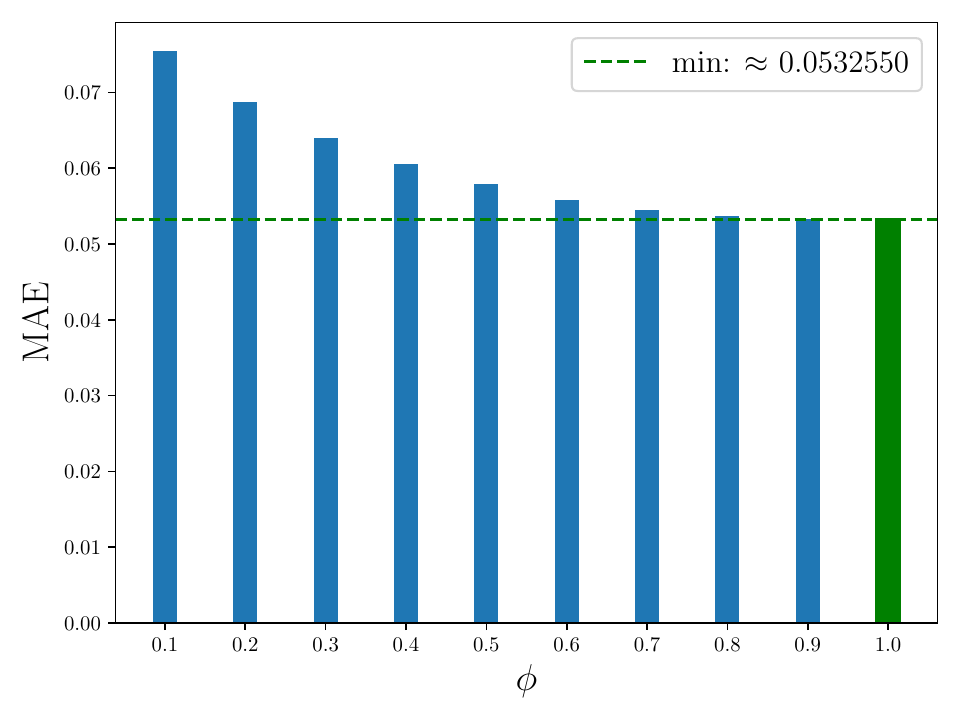}
  \caption{$\mathcal{U}[0,1]$ distribution}
  \label{har_phi_mae_rh_uni}
\end{subfigure}
\caption{Average MAE, PSP-Harmonic Heuristic versus $\phi$, $RH(500,6,3)$}
\label{har_mae_rh}
\end{figure}

Figure \ref{har_mae_rh} shows the achieved average MAE versus $\phi$ over all 40 tested RH graphs. For the beta edge distribution, shown in figure \ref{har_phi_mae_rh_beta}, the best value was $\approx0.05$, and for the uniform edge distribution, shown in figure \ref{har_phi_mae_rh_uni}, the best value was $\approx0.05$ as well. Both optimums where achieved at $\phi=1.0$, though $\phi=0.8, 0.9$ gave almost equally good results.

\begin{figure}[H]
\centering
\begin{subfigure}{.5\textwidth}
  \centering
  \includegraphics[width=0.8\linewidth]{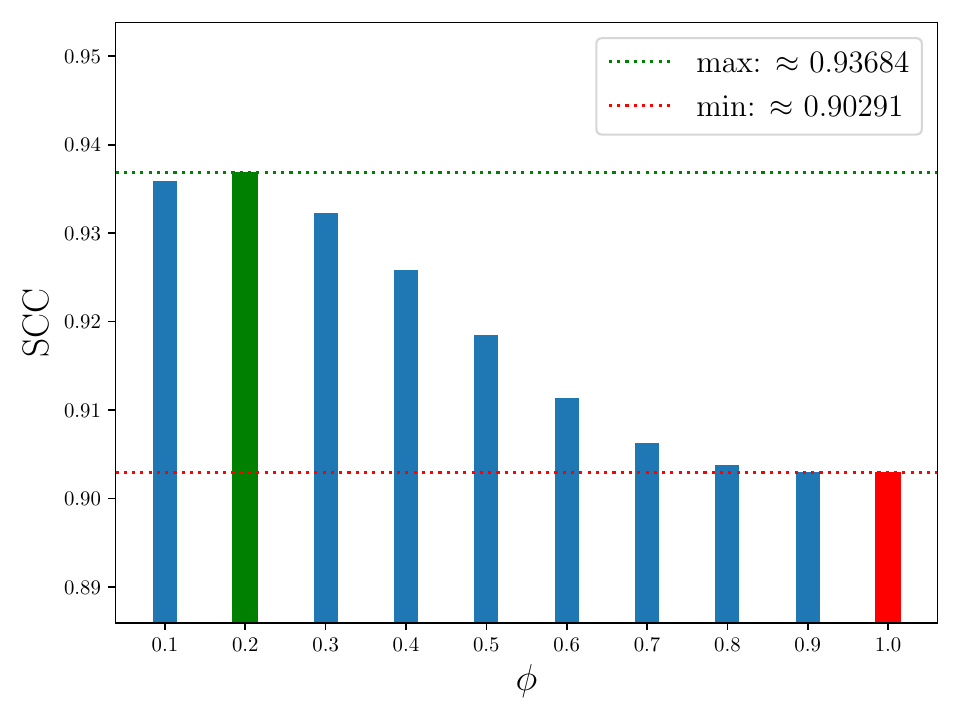}
  \caption{$Beta(4,4)$ distribution}
  \label{har_scc_rh_beta}
\end{subfigure}%
\begin{subfigure}{.5\textwidth}
  \centering
  \includegraphics[width=0.8\linewidth]{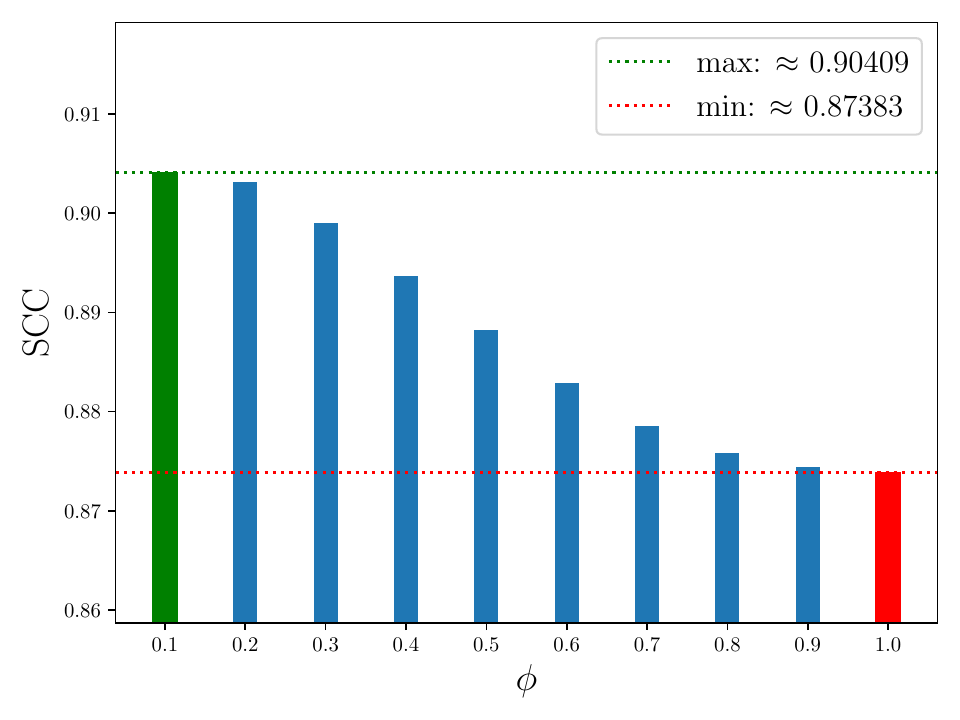}
  \caption{$\mathcal{U}[0,1]$ distribution}
  \label{har_scc_rh_uni}
\end{subfigure}
\caption{Average SCC, PSP-Harmonic Heuristic versus $\phi$, $RH(500,6,3)$}
\label{har_scc_rh}
\end{figure}

The achieved average SCC versus $\phi$ on the RH graphs is shown in figure \ref{har_scc_rh}. Interestingly, we now see better scores for lower values of $\phi$. For the beta edge distribution, shown in figure \ref{har_scc_rh_beta}, the best value was $\approx0.94$ at $\phi=0.2$ and the worst value was $\approx0.9$ at $\phi=1.0$. For the uniform edge distribution, shown in figure \ref{har_scc_rh_uni}, the best value was $\approx0.9$ at $\phi=0.1$ and the worst value was $\approx0.87$ at $\phi=1.0$.

\begin{figure}[H]
\centering
\begin{subfigure}{.5\textwidth}
  \centering
  \includegraphics[width=0.8\linewidth]{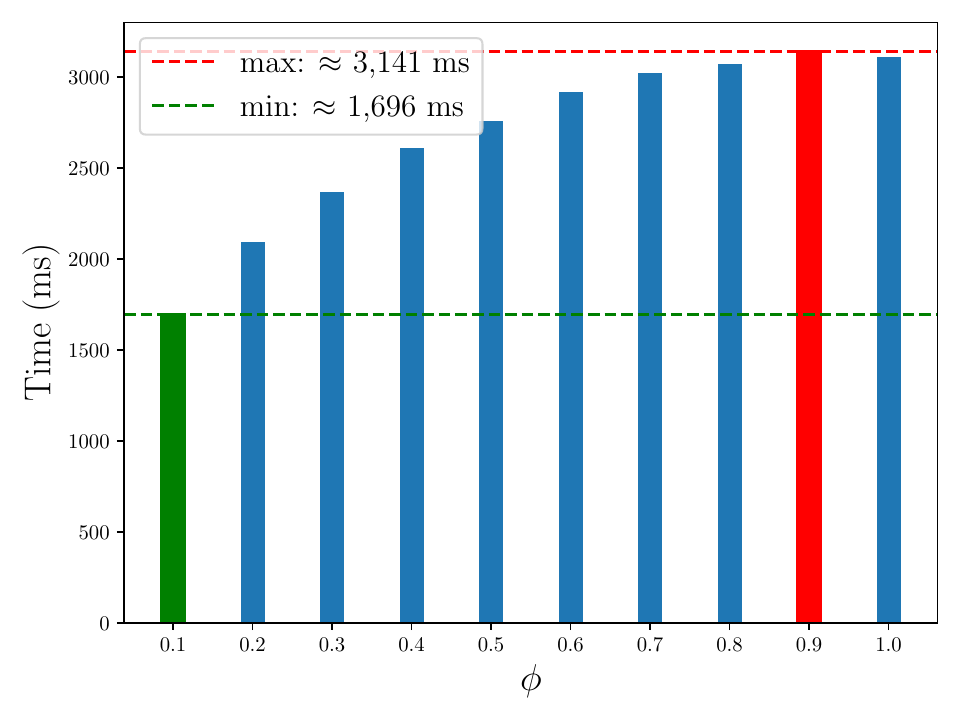}
  \caption{$Beta(4,4)$ distribution}
  \label{har_runtime_rh_beta}
\end{subfigure}%
\begin{subfigure}{.5\textwidth}
  \centering
  \includegraphics[width=0.8\linewidth]{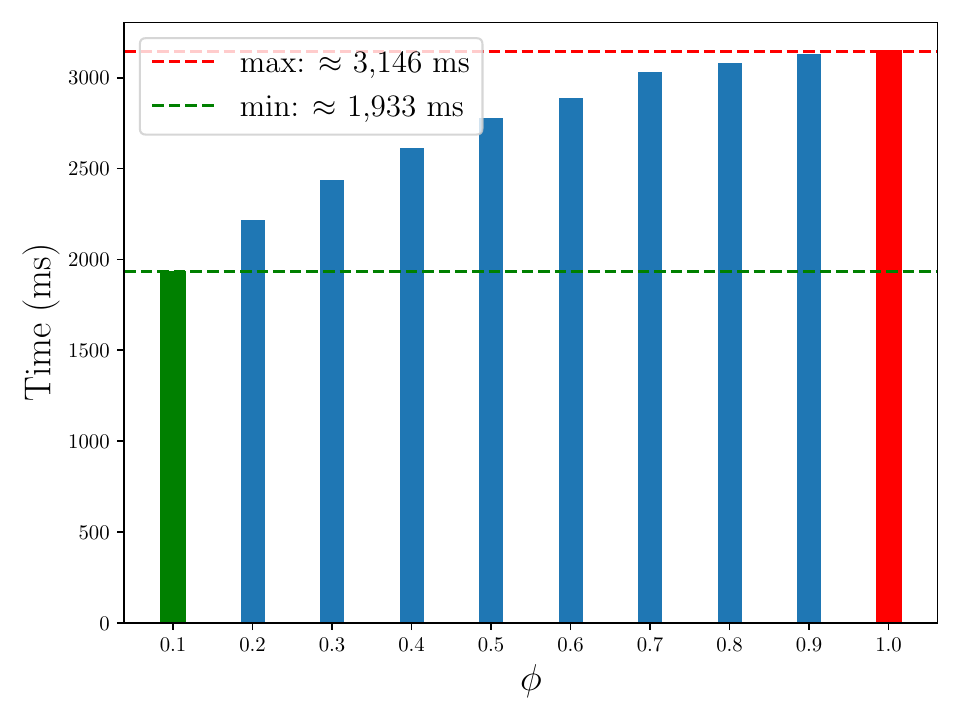}
  \caption{$\mathcal{U}[0,1]$ distribution}
  \label{har_runtime_rh_uni}
\end{subfigure}
\caption{Running time, PSP-Harmonic Heuristic versus $\phi$, $RH(500,6,3)$}
\label{har_time_rh}
\end{figure}

The running time versus $\phi$ of the PSP-harmonic heuristic on the RH graphs is shown in figure \ref{har_time_rh}. Again, we do not see a spike in the running time and roughly equal running time around $\phi=0.8,0.9,1.0$. For the beta edge distribution, shown in figure \ref{har_runtime_rh_beta}, the minimal running time was $\approx1,700$ ms, and the maximal running time was $\approx3,100$ ms. The fact that $\phi=0.9$ had a higher running time than $\phi=1.0$ cannot occur in theory and should be an anomaly caused by e.g. some network interrupt. For the uniform edge distribution, shown in figure \ref{har_runtime_rh_uni}, the minimal running time was $\approx1,900$ ms and the maximal running time was $\approx3,100$ ms. Independently of the chosen edge distribution, Monte Carlo with 73,777 samples had a running time of $\approx64,000$ ms.

\subsubsection{Conclusion on Randomized Graphs versus Monte Carlo}
Overall, I was able to replicate the promising efficacy results on randomized graphs that where presented by Chenxu Wang and Ziyuan Lin for the PSP-betweenness heuristic. Also, the addition of random hyperbolic graphs and a second edge probability distribution did not worsen the results. Especially, on every tested kind of graph and for every tested choice of $\phi$ (even for $\phi=0.1$), the achieved SCC scores show a great efficacy in detecting central nodes when comparing against the centrality ranking induced by the Monte Carlo method (i.e. central in the notion of the expected betweenness centrality), as already indicated in \cite{PSP}. \\

The PSP-harmonic heuristic generally showed slightly worse MAE and SCC scores on the same set of graphs. Still, the achieved values where in a similar range (compared to the achieved MAE and SCC scores of the PSP-betweenness heuristic). Hence, on the given kinds of randomized graphs, the PSP approach seems to also be highly effective when using it to estimate the expected harmonic closeness.\\

Even in the best case for PSP-betweenness, its lowest running time was still more than three times slower compared to PSP-harmonic on all tested graphs. This was of course expected, as, for every explored possible shortest path, just one floating point number is stored in the exploration algorithm for PSP-harmonic (the existence probability of said path), while for PSP-betweenness the inner nodes of every path are additionally stored in a bit vector, which creates more memory accesses and worse cache locality. Additionally, the running time of the PSP-betweenness heuristic always increased versus $\phi$ (as expected), and this increase was even very sharp for the BA and ER graphs at $\phi=1.0$, indicating a highly extensive exploration of paths as $\phi$ approaches 1.0 in some graph structures. For the PSP-harmonic heuristic on the other hand, the increase in running time versus $\phi$ roughly stopped around $\phi=0.8$ or $\phi=0.9$. As, for both PSP heuristics, the same paths are explored in every call of the respective \textit{AllShortestPaths} algorithms, this should be due to the second stopping condition when calculating the estimated distance probability distribution $\overline{p}_{s,t}$ in the PSP-harmonic heuristic. Namely, the special case where we would have $\sum_{D=1}^{|V|-1} \overline{p}_{s,t} (D) > 1$ and  hence set $\overline{p}_{s,t}(k)=1-\sum_{D=1}^{k-1}\overline{p}_{s,t}(k)$ and terminate the exploration at depth $k$, evading the extensive exploration of long paths until the two current nodes are disconnected (which is the only stopping condition of the PSP-betweenness exploration algorithm at $\phi=1.0$).\\

Regarding the choice of the parameter $\phi$ for the PSP-betweenness heuristic, I could not replicate the fact that $\phi\approx0.8$ is an optimal choice to minimize the MAE. Instead, every graph generator produced a different optimal $\phi$. The difference in the results compared to \cite{PSP} might be due to the randomness of the tested graphs, though. A bigger sample size could show similar results for the MAE versus $\phi$. \\

For the PSP-harmonic heuristic, there was also no clear best choice of $\phi$ versus the MAE. Furthermore, both for the PSP-betweenness heuristic and the PSP-harmonic heuristic, no value of $\phi$ consistently produced the best SCC scores. Still, in the following section, I decided to stick to the proposed choice of $\phi=0.8$ when evaluating the efficacy of both algorithms on real world graphs. Besides the fact that this was optimal in \cite{PSP} for the MAE, it was almost optimal for the MAE in all of the just presented experiments for both algorithms. For the SCC, $\phi=0.8$ was also very close to the optimal value on all tested graphs, only excluding the RH graphs for PSP-harmonic. On these graphs, the SCC was best for small values of $\phi=0.1$ (uniform distribution) or $\phi=0.2$ (Beta distribution) and worse at $\phi=1.0$, while, for both algorithms, $\phi=0.1$ did yield the minimal SCC in all other cases. However, the SCC was generally relatively stable when changing $\phi$. In the special case of the aforementioned experiments on RH graphs for PSP-harmonic, $\phi=0.8$ was still within a proximity of $\approx0.03$ to the optimal achieved SCC.

\subsection{Efficacy on Real World Graphs}
Now, I will present experimental results for both PSP heuristics on a set of 13 real world graphs. Again, they are compared to the Monte Carlo method. As stated earlier, some graphs (the first five) have given edge existence probabilities based on real world data. For all the other ones, i.e. all graphs where it is not explicitly pointed out differently, I randomized the edge probabilities using the $\mathcal{U}[0,1]$ distribution. As just established, I did use $\phi=0.8$ in all following experiments for both algorithms. 100,000 Monte Carlo samples where used as ground-truth. \\

Only if the SCC of PSP-harmonic was less than 0.9, I reran Monte Carlo with 1,000,000 samples to test if this would change the MAE or SCC by any significant amount. In every such case though, the SCC did only change by $10^{-4}$ or less, and the MAE by $10^{-5}$ or less. So, it seems fair to assume that 100,000 samples are sufficiently accurate. Hence, the MAE and SCC are always calculated based on the result given by 100,000 samples, and the Monte Carlo running time also always refers to 100,000 samples. I did not perform these reruns for the PSP-betweenness heuristic, since the SCC scores for betweenness where never below 0.95 in the worst case. \\

The following instances where used in the experiments:
\begin{itemize}
	\item \textbf{Graph 1 (\textit{Collins})}: a protein-protein interaction network (PPI), as described in \cite{Collins} by Collins et al., where the probabilities of edges between proteins (nodes) is given by empirical data. The graph has $|V|=1,622$ nodes and $|E|=9,074$ edges. The existence probability of the edges are relatively high, with a mean probability of $\approx0.78$ and a standard deviation of $\approx 0.18$.
	\item \textbf{Graph 2 (\textit{Collins-Rescaled})}: the same as the previous graph, but all edge probabilities are rescaled to values in the range [0,1], i.e. we have a new edge probability function $P_{new}$ based on the given probability function $P$ of the previous experiment, where for every $e\in E$ we set
\[ P_{new}(e)=\frac{P(e)-\min(P)}{\max(P)-\min(P)} \]
	\item \textbf{Graph 3 (\textit{Gavin})}: another PPI, again with empirical edge probabilities, as described in \cite{Gavin} by Gavin et al. It consists of $|V|=1,727$ nodes and $|E|=7,534$ edges. Here though, the edges have relatively low probabilities. Their mean is $\approx0.35$ with a standard deviation of $\approx0.14$.
	\item \textbf{Graph 4 (\textit{Gavin-Rescaled})}: again, the same as the previous graph but with the edge probabilities rescaled to values in the range [0,1] (in the same way as described for \textit{Collins-Rescaled}).
	\item \textbf{Graph 5 (\textit{Krogan})}: one more PPI with empirical edge probabilities, as described in \cite{Krogan} by Krogan et al. Here, we have $|V|=2,708$ and $|E|=7,123$. About 25 \% of the edges have existence probabilities greater than 0.9, while the remaining 75 \% are roughly uniformly distributed between 0.27 and 0.9.
	\item \textbf{Graph 6 (\textit{Facebook-Polit})}: a graph consisting of $|V|=5,908$ nodes, each representing a \textit{facebook} page of a politician. Two nodes are connected if the two respective pages did mutually 'like' one another, producing $|E|=41,729$ edges. The graph is available at the Stanford Network Analysis Project (SNAP) \cite{SNAP-Facebook}. 
	\item \textbf{Graph 7 (\textit{Facebook-TV})}: a graph from the same dataset as \textit{Facebook-Polit}. Nodes also represent \textit{facebook} pages (now pages of TV shows) and edges once again represent mutual 'likes'. We have $|V|=3,892$ and $|E|=17,262$ \cite{SNAP-Facebook}.
	\item \textbf{Graph 8 (\textit{LastFM})}: another social network from the SNAP repository. It consists of $|V|=7,624$ nodes, representing users of \textit{last.fm} from Asian countries, where the $|E|=27,806$ edges show a mutual follower relationship \cite{SNAP-lastfm}.
	\item \textbf{Graph 9 (\textit{Euroroads})}: a road network from the KONECT Network Data Repository \cite{KONECT-euroroad}. $|V|=1,174$ nodes represent European cities, mutually connected by $|E|=1,417$ roads.
	\item \textbf{Graph 10 (\textit{Power-Grid})}: one more infrastructure graph from the KONECT database. The $|V|=4,941$ nodes represent transformers, substations and generators in Western States of the USA. The $|E|=6,594$ edges represent high-voltage power connections between them \cite{KONECT-opsahl}.
	\item \textbf{Graph 11 (\textit{Co-Author})}: a collaboration network where the $|V|=1,461$ nodes represent scientist in the area of network science. They are connected by one of the $|E|=2,742$ edges if they had at least one mutual publication. This graph was again available at the KONECT database \cite{KONECT-dimacs}.
	\item \textbf{Graph 12 (\textit{Copperfield})}: another graph from the KONECT database. The $|V|=112$ nodes represent common nouns and adjectives from the 19th century novel \textit{David Copperfield} by Charlies Dickens. They are connected by one of the $|E|=425$ edges if they appeared at least once in adjacent positions \cite{KONECT-copperfield}.
	\item \textbf{Graph 13 (\textit{Jazz})}: lastly, an additional collaboration network from the KONECT database. Each of the $|V|=198$ nodes represent a jazz musician. They are connected by one of the $|E|=2,742$ edges if they played together in a band at some point in their career \cite{KONECT-jazz}.
\end{itemize}

\subsubsection{Results}
Table \ref{bet_real_1} shows the achieved MAE, SCC and the running time of the PSP-betweenness heuristic and Monte Carlo (MC) on the five tested PPI graphs. Both the MAE and SCC show a similarly good efficacy compared to the results on randomized graphs. However, except for the \textit{Krogan} graph, Monte Carlo did outperform the PSP-betweenness heuristic in running time. This was especially evident in the rescaled PPI graphs, Monte Carlo being more than 11 times faster on \textit{Collins-Rescaled} and more than 22 times faster on \textit{Gavin-Rescaled}. The faster running time is marked in bold text. 
\begin{table}[H]
\begin{tabular}{l|l|l|l|l|l|}
\cline{2-6}
                                     & \textit{Collins} & \textit{Collins-Rescaled} & \textit{Gavin}          & \textit{Gavin-Rescaled} & \textit{Krogan}         \\ \hline
\multicolumn{1}{|c|}{$|V|$}            & 1,622          & 9,074            & 1,727          & 1,727          & 2,708          \\ \hline
\multicolumn{1}{|c|}{$|E|$}            & 9,074          & 9,074            & 7,534          & 7,534          & 7,123          \\ \hline
\multicolumn{1}{|c|}{$\approx$ MAE}  & 0.0002         & 0.0003           & 0.0017         & 0.0001         & 0.0002         \\ \hline
\multicolumn{1}{|c|}{$\approx$ SCC}  & 0.96           & 0.95             & 0.96           & 0.96           & 0.99           \\ \hline
\multicolumn{1}{|c|}{Time PSP (sec)} & 4,807          & 16,273           & 3,830          & 3,871          & \textbf{2,347} \\ \hline
\multicolumn{1}{|c|}{Time MC (sec)}  & \textbf{2,246} & \textbf{1,448}   & \textbf{2,201} & \textbf{172}   & 4,577           \\ \hline
\end{tabular}
\caption{PSP-Betweenness Efficacy on Real World Graphs 1-5}
\label{bet_real_1}
\end{table}

Table \ref{bet_real_2} shows the results of the PSP-betweenness heuristic on the graphs 6 to 9. Again, we see a consistently good efficacy in detecting central nodes. Still, in every of these graphs, Monte Carlo did have a lower running time than PSP-betweenness. For the \textit{Euroroads} graph, we again see a surprisingly big difference, Monte Carlo being more than 7 times faster on said graph. The other graphs give speedup factors ranging from $\approx1.68$ (\textit{Facebook-Polit}) to $\approx3.3$ (\textit{LastFM}) when using Monte Carlo instead of the PSP heuristic.

\begin{table}[H]
\begin{tabular}{c|l|l|l|l|}
\cline{2-5}
\multicolumn{1}{l|}{}           		& \textit{Facebook-Polit} 	& \textit{Facebook-TV} & \textit{LastFM} & \textit{Euroroads} \\ \hline
\multicolumn{1}{|c|}{$|V|$}          	& 5,908                  	 	& 3,892                	& 7,624           		& 1,174              \\ \hline
\multicolumn{1}{|c|}{$|E|$}          	& 41,729                   		& 17,262                	& 27,806           	& 1,417              \\ \hline
\multicolumn{1}{|c|}{$\approx$ MAE}            	& 0.00007                   	& 0.0003                	& 0.00007             	& 0.00005            \\ \hline
\multicolumn{1}{|c|}{$\approx$ SCC}            	& 0.986                    		& 0.97                 	& 0.99            		& 0.99               \\ \hline
\multicolumn{1}{|c|}{Time PSP (sec)} & 155,147            		& 45,464         		& 224,396   		& 179      \\ \hline
\multicolumn{1}{|c|}{Time MC (sec)}  & \textbf{92,079}               	& \textbf{17,715}       & \textbf{67,378}        & \textbf{23} \\ \hline
\end{tabular}
\caption{PSP-Betweenness Efficacy on Real World Graphs 6-9}
\label{bet_real_2}
\end{table}

Table \ref{bet_real_3} shows the results of PSP-betweenness on the remaining graphs 10 to 13. The achieved MAE and SCC are similar to the prior results. Though, for the \textit{Power-Grid} graph, we now get almost a 268 times faster running time when using Monte Carlo compared to the PSP-betweenness heuristic. On the \textit{Co-Author} graph, PSP runs slightly faster than Monte Carlo, but only gives a speedup factor of $\approx1.15$. The only graphs where PSP runs significantly faster than Monte Carlo are the last two graphs, \textit{Copperfield} and \textit{Jazz}. These are also by far the smallest graphs though, both having less than 200 nodes. This strongly indicates that 100,000 Monte Carlo samples are unreasonably many for these two graphs, as the same sample size was sufficient for all the other (larger) tested graphs. Using less samples should close the gap and could potentially even make Monte Carlo more efficient than the PSP-betweenness heuristic for these two graphs as well.

\begin{table}[H]
\begin{tabular}{c|l|l|l|l|}
\cline{2-5}
\multicolumn{1}{l|}{}           		& \textit{Power-Grid} 	& \textit{Co-Author} & \textit{Copperfield} & \textit{Jazz} \\ \hline
\multicolumn{1}{|c|}{$|V|$}          	& 4,941                  	 	& 1,461                	& 112           		& 198              \\ \hline
\multicolumn{1}{|c|}{$|E|$}          	& 6,594                   		& 2,742                	& 425           		& 2,742              \\ \hline
\multicolumn{1}{|c|}{$\approx$ MAE}            	& 0.00002                   	& 0.00007                	& 0.001             	& 0.0009            \\ \hline
\multicolumn{1}{|c|}{$\approx$ SCC}            	& 0.98                    		& 0.98                 	& 0.99            		& 0.98               \\ \hline
\multicolumn{1}{|c|}{Time PSP (sec)} & 76,591            		& \textbf{48}         	& \textbf{0.2}   		& \textbf{1.4}      \\ \hline
\multicolumn{1}{|c|}{Time MC (sec)}  & \textbf{286}               	& 55      			& 9.5        			& 151	 \\ \hline
\end{tabular}
\caption{PSP-Betweenness Efficacy on Real World Graphs 10-13}
\label{bet_real_3}
\end{table}

Table \ref{har_real_1} shows the results of the PSP-harmonic heuristic on the five tested PPI graphs. Contrary to PSP-betweenness, we now get consistently better running time than Monte Carlo (at least with 100,000 samples). However, the MAE and SCC scores are of varying quality. Both versions of the \textit{Collins} graph give SCC scores above 0.9. \textit{Gavin} and \textit{Krogan} still yield SCC scores of $\approx0.78$ and $\approx0.84$ respectively. Though, for the rescaled PPI \textit{Gavin-Rescaled} we now get a vastly lower SCC score of only $\approx$ 0.47. Additionally, this graph achieved the maximal MAE of $\approx$ 0.16 when comparing all 13 tested graphs.

\begin{table}[H]
\begin{tabular}{l|l|l|l|l|l|}
\cline{2-6}
                                     & \textit{Collins} & \textit{Collins-Rescaled} & \textit{Gavin} & \textit{Gavin-Rescaled} & \textit{Krogan} \\ \hline
\multicolumn{1}{|c|}{$|V|$}            & 1,622            & 9,074                     & 1,727          & 1,727                   & 2,708           \\ \hline
\multicolumn{1}{|c|}{$|E|$}            & 9,074            & 9,074                     & 7,534          & 7,534                   & 7,123           \\ \hline
\multicolumn{1}{|c|}{$\approx$ MAE}  & 0.016            & 0.037                     & 0.1            & 0.16                    & 0.07            \\ \hline
\multicolumn{1}{|c|}{$\approx$ SCC}  & 0.98             & 0.91                      & 0.78           & 0.47                    & 0.84            \\ \hline
\multicolumn{1}{|c|}{Time PSP (sec)} & \textbf{161}     & \textbf{551}              & \textbf{437}   & \textbf{103}            & \textbf{475}    \\ \hline
\multicolumn{1}{|l|}{Time MC (sec)}  & 2,048            & 1,330                     & 1,987          & 439                     & 4,020           \\ \hline
\end{tabular}
\caption{PSP-Harmonic Efficacy on Real World Graphs 1-5}
\label{har_real_1}
\end{table}

In table \ref{har_real_2} we see the PSP-harmonic results for graphs 6 to 9. We do not have an extreme outlier like the \textit{Gavin-Rescaled} graph here, with the MAE being relatively stable at $\approx$ 0.06 and the SCC being in the range of $\approx$ 0.7 to $\approx$ 0.9. PSP-harmonic did beat the Monte Carlo running time in the graphs 6 to 8, the ninth graph \textit{Euroroads} did give a more than 5 times faster running time though when using Monte Carlo instead of the PSP-harmonic heuristic. However, similar speedup factors are achieved the other way around, both for \textit{Facebook-Polit} and \textit{Facebook-TV}, while \textit{LastFM} was still almost 3 times faster when using the PSP-harmonic heuristic versus 100,000 Monte Carlo samples.

\begin{table}[H]
\begin{tabular}{c|l|l|l|l|}
\cline{2-5}
\multicolumn{1}{l|}{}                & \textit{Facebook-Polit} 	& \textit{Facebook-TV} & \textit{LastFM} & \textit{Euroroads} \\ \hline
\multicolumn{1}{|c|}{$|V|$}          & 5,908                   		& 3,892                	& 7,624           		& 1,174              \\ \hline
\multicolumn{1}{|c|}{$|E|$}          & 41,729                   	& 17,262                	& 27,806           	& 1,417              \\ \hline
\multicolumn{1}{|c|}{$\approx$ MAE}            & 0.05                   		& 0.06                	& 0.07             		& 0.05               \\ \hline
\multicolumn{1}{|c|}{$\approx$ SCC}            & 0.896                    	& 0.8                 	& 0.74            		& 0.7               \\ \hline
\multicolumn{1}{|c|}{Time PSP (sec)} & \textbf{13,635}      & \textbf{3,230}         & \textbf{19,416}    	& 72       \\ \hline
\multicolumn{1}{|c|}{Time MC (sec)}  & 72,022                   & 15,585                	& 57,375           	& \textbf{14}                \\ \hline
\end{tabular}
\caption{PSP-Harmonic Efficacy on Real World Graphs 6-9}
\label{har_real_2}
\end{table}

The results for PSP-harmonic on the last four graphs are shown in table \ref{har_real_3}. Here, we get a second outlier for the SCC: the \textit{Power-Grid} graph did also yield the overall minimal achieved SCC of $\approx$ 0.47. Though, the MAE was not an outlier here, giving a value of $\approx$ 0.05. However, \textit{Power-Grid} is a second graph where Monte Carlo did beat PSP-harmonic in their respective running time. This time, the difference was much bigger: Monte Carlo was almost 50 times faster. Hence, the graph \textit{Power-Grid} seems to generally be a bad case in structure for the PSP approach, as it was also the worst case for PSP-betweenness when comparing the running time to Monte Carlo (PSP-betweenness being almost 268 times slower than Monte Carlo).

\begin{table}[H]
\begin{tabular}{c|l|l|l|l|}
\cline{2-5}
\multicolumn{1}{l|}{}           		& \textit{Power-Grid} 	& \textit{Co-Author} & \textit{Copperfield} & \textit{Jazz} \\ \hline
\multicolumn{1}{|c|}{$|V|$}          	& 4,941                  	 	& 1,461                	& 112           		& 198              \\ \hline
\multicolumn{1}{|c|}{$|E|$}          	& 6,594                   		& 2,742                	& 425           		& 2,742              \\ \hline
\multicolumn{1}{|c|}{$\approx$ MAE}            	& 0.05                   		& 0.007                	& 0.05	             	& 0.016            \\ \hline
\multicolumn{1}{|c|}{$\approx$ SCC}            	& 0.47                    		& 0.93                 	& 0.93            		& 0.99               \\ \hline
\multicolumn{1}{|c|}{Time PSP (sec)} & 8,690            			& \textbf{11}         	& \textbf{0.09}   	& \textbf{0.5}      \\ \hline
\multicolumn{1}{|c|}{Time MC (sec)}  & \textbf{174}               	&  40      			& 12,943		       	&  134\\ \hline
\end{tabular}
\caption{PSP-Harmonic Efficacy on Real World Graphs 10-13}
\label{har_real_3}
\end{table}

% fast 286

\subsubsection{Conclusion on Real World Graphs versus Monte Carlo}
The efficacy of the PSP-betweenness heuristic was similar on the tested real world graphs compared to the achieved results on randomized graphs. Especially, we once again see a very good correlation of the produced centrality rankings given by PSP-betweenness and Monte Carlo. The average MAE over all 13 tested graphs was $\approx0.0004$ and the average SCC was $\approx0.98$. Even the worst case MAE and SCC scores where still good, giving $\approx$ 0.001 for the maximal MAE and $\approx$ 0.95 for the minimal SCC. However, its running time did not hold up well compared to Monte Carlo. It was faster on the last two graphs, which was of course expected. As already stated, they are small (less than 200 nodes), hence 100,000 samples should be way more than actually needed for a sufficient Monte Carlo result. Besides that, it was only faster on the \textit{Krogan} PPI (2,347 sec versus 4,577 sec) and on the \textit{Co-Author} collaboration network (48 sec versus 55 sec). On all other tested graphs, Monte Carlo was faster than the PSP-betweenness heuristic, once even yielding a tremendously shorter running time on the \textit{Power-Grid} graph, beating PSP by almost a factor of 268. On average, when excluding the last two small graphs and the extreme case of the \textit{Power-Grid} graph, Monte Carlo was $\approx6.24$ times faster than the PSP-betweenness heuristic. Especially on the smaller graphs out of the tested instances, but possibly also on the larger ones, the running time of Monte Carlo could become even better compared to the PSP-betweenness heuristic, as (possibly many) less samples could suffice. \\

The PSP-harmonic heuristic did produce variable results depending on the graph. The two worst cases for the SCC, the \textit{Gavin-Rescaled} and \textit{Power-Grid} graphs, did only give SCC scores of $\approx$ 0.47. The \textit{Gavin-Rescaled} PPI also gave the worst case MAE of $\approx$ 0.16. The conducted Monte Carlo rerun, whenever the SCC was below 0.9, using 1,000,000 samples instead of 100,000, did never change the results by a significant value. So, we can assume with high confidence that 100,000 samples converged close to the expected harmonic closeness, substantiating the fact that the PSP-harmonic heuristic is sometimes much worse than on the tested randomized graphs (when aiming to get close to the expected harmonic closeness). The MAE of $\approx$ 0.16 in the aforementioned worst case also indicates that the estimated distance function can produce quite inaccurate values for some graph structures. Still, the average SCC over all 13 graphs was $\approx0.8$ and, when excluding the two aforementioned worst case graphs, it did never drop below 0.7. Furthermore, 8 out of 13 graphs achieved SCC scores of at least 0.8 and 5 graphs achieved a SCC of at least 0.9. When excluding the two worst cases for the MAE (\textit{Gavin} with $\approx$ 0.1 and \textit{Gavin-Rescaled} with $\approx$ 0.16), the MAE was never above 0.07, and its average over all 13 graphs was $\approx0.058$. Also, except for \textit{Euroroads} and \textit{Power-Grid}, i.e. in 11 out of 13 cases, the PSP-harmonic heuristic was faster than Monte Carlo with 100,000 samples. If we once again exclude the two small graphs \textit{Copperfield} and \textit{Jazz} for fairness, PSP-harmonic was $\approx4.14$ times faster than Monte Carlo when taking the average over the remaining 11 graphs. Though, as 100,000 samples seemed to suffice for Monte Carlo on all 13 graphs, (maybe many) less samples could also produce adequate results. This could close the gap or make Monte Carlo faster on some or all of the 13 tested graphs. More research would though be needed to test how many Monte Carlo samples are actually needed in general, which should be highly dependent on the given graph structure and edge probability distribution.

\subsection{Scalability Results on Large Graphs}
Initially, while hoping for better results in the second experimental stage, this third set of experiments was planned just to see the scalability of the running time for both PSP-heuristics on much larger graphs. For this, I collected a variety of large scale graphs. Fully randomized graphs, real world graphs with randomized edge probabilities and also large scale real world uncertain graphs. To name a few, the scalability experiments included random graphs of kind $BA(1'000'000,5)$, $ER(1'000'000,0.00005)$ and $RH(1'000'000,6,3)$ with randomized edge probabilities. Furthermore, a set of multiple dense graphs based on climate data, where only 7,320 nodes exist per graph (grid points on earth) but up to around 1,000,000 edges, with the edge existence probability being given by experimental correlation for different measurements (e.g. temperature) \cite{Climate}. But also real world graphs with more nodes and less density, e.g. a graph of the Californian road network with 1,965,206 nodes (intersections and endpoints) and 2,766,607 edges (connecting roads) \cite{SNAP-road}, where I randomized the edge probabilities using $\mathcal{U}[0,1]$. \\

However, for any of those large scale graphs, both the PSP-betweenness and the PSP-harmonic heuristic where terminated by the Slurm workload manager within just a few minutes due to running out of memory. As stated earlier, each run of one of the algorithms had unshared access to a  computer with 192GB of RAM. 4.5GB where reserved for the system, but additionally 3.2GB of Swap are available before Slurm terminates the run. So, both PSP algorithms used 190.7GB of memory when being terminated. The fact that this occurred within just one up to only a few minutes shows that for large scale graphs, the calculation of all shortest paths between two given nodes is generally not practicable. Especially when considering that the path exploration algorithm for PSP-harmonic only stores a single 64 bit floating point number per shortest path (the existence probability of said path). I even implemented alternative versions of the exploration algorithms. Here, during the breadth-first traversal, a vector kept track of all nodes that where found at the current depth $k$. Once the first node with depth $k+1$ was dequeued, all stored existence probabilities of shortest paths to nodes with depth $k$ (and also the stored inner nodes in the case of PSP-betweenness) where released from memory. Additionally, I discarded of the bit vector of size $|V|$ in the PSP-betweenness exploration. Instead, to keep track of the inner nodes on a path, it was replaced by a (dynamically sized) vector of node indices, as a given shortest path should usually traverse much less than $|V|$ nodes. Still, this did not change the fact that early on, during the first traversal of all shortest paths, both algorithms ran out of memory. So, unfortunately both of these algorithms are not generally suitable for application on large scale graphs, except maybe on systems with access to vastly greater amounts of memory (which I was not able to test during the experiments for this thesis).

\section{Overall Conclusion and Outlook}
Overall, the PSP-betweenness heuristic seems to be highly effective in detecting central nodes, indicated by the SCC consistently being close to one and the MAE close to zero respectively on all tested graphs. Though, the experiments on real world graphs showed no benefit in running time over Monte Carlo. It did run faster than Monte Carlo on the tested generated graphs. Still, it remains questionable if this advantage could be retained when repeating said experiments while limiting the amount of Monte Carlo samples (e.g. by testing the actually needed sample size to reach a sufficient convergence of the SCC). Furthermore, for large scale graphs, it seems to either not even be applicable in its current state or require a significantly larger amount of memory. \\

The PSP-harmonic heuristic was not as consistent in its efficacy. It took less running time than the PSP-betweenness heuristic though and is possibly running faster than Monte Carlo, as indicated in the experiments on real world graphs. To test the actual benefit in running time compared to Monte Carlo, more experiments would though be needed as well, limiting the sample size of Monte Carlo to a fair amount based on some empirical measure of its convergence rate. However, same as for PSP-betweenness, the PSP-harmonic heuristic might not even be applicable to large scale graphs in its current state, as the amount of potential shortest paths between nodes can be too large to handle efficiently. Still, a system with more memory could be able to handle large scale graphs, even with the current implementation. Especially as the PSP-harmonic heuristic does not need to store any paths (just path probabilities). Hence, it should be able to handle much more potential shortest paths than the PSP-betweenness heuristic. This could be investigated in future research. \\

Further research could also show whether the PSP approach might be improved on, e.g. by introducing a second hyperparameter $k$, only calculating at most $k$ shortest paths between two given nodes in the possible shortest path exploration phase. This would evidently eliminate the danger of running out of memory for sufficiently small values of $k$. Also, instead of giving $k$ as a second hyperparameter, other approaches would be possible. E.g. some polynomial to calculate $k$ based on the given input graph. \\

Additionally, to produce more consistent efficacy for harmonic closeness centrality, different ways to calculate distances based on the concept of possible shortest paths could be investigated. For instance, experiments could show the impact of using a different distance measure based on the same approach of estimating the distance probability distribution $p_{s,t}$. Also, calculating the distance probability distribution differently could be researched (e.g., as we seemed to reach to case of $\sum_{D=1}^{k}p_{s,t}(k)>1$ in the experiments, different mechanism could be tried to avoid this situation or to handle it differently, once it occurs). Of course, totally different distance measures based on possible shortest paths are also conceivable.

\newpage
\bibliographystyle{unsrt}
\bibliography{references}

\begin{thebibliography}{10}

\bibitem{Potamias}
Michalis Potamias, Francesco Bonchi, Aristides Gionis, and George Kollios.
\newblock k-nearest neighbors in uncertain graphs.
\newblock {\em PVLDB}, 3:997--1008, 09 2010.

\bibitem{PSP}
Chenxu Wang and Ziyuan Lin.
\newblock An efficient approximation of betweenness centrality for uncertain
  graphs.
\newblock {\em IEEE Access}, 7:61259--61272, 2019.

\bibitem{Betweenness}
Linton~C. Freeman.
\newblock A set of measures of centrality based on betweenness.
\newblock {\em Sociometry}, 40(1):35--41, 1977.

\bibitem{Harmonic}
Massimo Marchiori and Vito Latora.
\newblock Harmony in the small-world.
\newblock {\em Physica A: Statistical Mechanics and its Applications},
  285(3-4):539--546, oct 2000.

\bibitem{MPSP}
Arkaprava Saha, Ruben Brokkelkamp, Yllka Velaj, Arijit Khan, and Francesco
  Bonchi.
\newblock Shortest paths and centrality in uncertain networks.
\newblock {\em Proc. VLDB Endow.}, 14(7):1188--1201, mar 2021.

\bibitem{OMP}
L.~Dagum and R.~Menon.
\newblock Openmp: an industry standard api for shared-memory programming.
\newblock {\em IEEE Computational Science and Engineering}, 5(1):46--55, 1998.

\bibitem{boost}
Boost {C}++ {L}ibraries.
\newblock \url{https://www.boost.org/}.
\newblock Accessed: 2023-28-03.

\bibitem{Brandes}
Ulrik Brandes.
\newblock A faster algorithm for betweenness centrality.
\newblock {\em The Journal of Mathematical Sociology}, 25, 03 2004.

\bibitem{SLURM}
M~Jette, C~Dunlap, J~Garlick, and M~Grondona.
\newblock Slurm: Simple {L}inux {U}tility for {R}esource {M}anagement.

\bibitem{Gilbert}
E.~N. Gilbert.
\newblock {Random Graphs}.
\newblock {\em The Annals of Mathematical Statistics}, 30(4):1141 -- 1144,
  1959.

\bibitem{RandomGraphs}
Manuel Penschuck, Ulrik Brandes, Michael Hamann, Sebastian Lamm, Ulrich Meyer,
  Ilya Safro, Peter Sanders, and Christian Schulz.
\newblock Recent advances in scalable network generation.
\newblock {\em CoRR}, abs/2003.00736, 2020.

\bibitem{RHG}
Moritz von Looz, Mustafa Özdayi, Sören Laue, and Henning Meyerhenke.
\newblock Generating massive complex networks with hyperbolic geometry faster
  in practice, 2016.

\bibitem{NetworKit}
Networ{K}it.
\newblock \url{https://networkit.github.io/}.
\newblock Accessed: 2023-15-04.

\bibitem{Collins}
Sean~R. Collins, Patrick Kemmeren, Xue-Chu Zhao, Jack~F. Greenblatt, Forrest
  Spencer, Frank~C.P. Holstege, Jonathan~S. Weissman, and Nevan~J. Krogan.
\newblock Toward a comprehensive atlas of the physical interactome of
  saccharomyces cerevisiae*.
\newblock {\em Molecular \& Cellular Proteomics}, 6(3):439--450, 2007.

\bibitem{Gavin}
Anne-Claude Gavin, Patrick Aloy, Paola Grandi, Roland Krause, Markus Boesche,
  Martina Marzioch, Christina Rau, Lars Jensen, Sonja Bastuck, Birgit
  Dümpelfeld, Angela Edelmann, Marie-Anne Heurtier, Verena Hoffman, Christian
  Hoefert, Karin Klein, Manuela Hudak, Anne-Marie Michon, Malgorzata Schelder,
  Markus Schirle, and Giulio Superti-Furga.
\newblock Proteome survey reveals modularity of the yeast cell machinery.
\newblock {\em Nature}, 440:631--6, 04 2006.

\bibitem{Krogan}
Nevan Krogan, Gerard Cagney, Haiyuan Yu, Gouqing Zhong, Xinghua Guo, Alexandr
  Ignatchenko, Joyce Li, Shuye Pu, Nira Datta, Aaron Tikuisis, Thanuja Punna,
  José Peregrín-Alvarez, Michael Shales, Xin Zhang, Mike Davey, Mark
  Robinson, Alberto Paccanaro, James Bray, Anthony Sheung, and Jack Greenblatt.
\newblock Krogan, n. j., cagney, g., haiyuan, y., zhong, g. \& guo, x. global
  landscape of protein complexes in the yeast saccharomyces cerevisiae. nature
  440, 637-643.
\newblock {\em Nature}, 440:637--43, 04 2006.

\bibitem{SNAP-Facebook}
Benedek Rozemberczki, Ryan Davies, Rik Sarkar, and Charles Sutton.
\newblock Gemsec: Graph embedding with self clustering.
\newblock In {\em Proceedings of the 2019 IEEE/ACM International Conference on
  Advances in Social Networks Analysis and Mining 2019}, pages 65--72. ACM,
  2019.

\bibitem{SNAP-lastfm}
Benedek Rozemberczki and Rik Sarkar.
\newblock Characteristic functions on graphs: Birds of a feather, from
  statistical descriptors to parametric models, 2020.

\bibitem{KONECT-euroroad}
Euroroads network dataset -- {KONECT}.
\newblock \url{http://konect.cc/networks/subelj_euroroad}, October 2017.
\newblock Accessed: 2023-15-04.

\bibitem{KONECT-opsahl}
Us power grid network dataset -- {KONECT}.
\newblock \url{http://konect.cc/networks/opsahl-powergrid}, October 2017.
\newblock Accessed: 2023-15-04.

\bibitem{KONECT-dimacs}
Network science network dataset -- {KONECT}.
\newblock \url{http://konect.cc/networks/dimacs10-netscience}, January 2018.
\newblock Accessed: 2023-15-04.

\bibitem{KONECT-copperfield}
David copperfield network dataset -- {KONECT}.
\newblock \url{http://konect.cc/networks/adjnoun_adjacency}, October 2017.
\newblock Accessed: 2023-15-04.

\bibitem{KONECT-jazz}
Jazz musicians network dataset -- {KONECT}.
\newblock \url{http://konect.cc/networks/arenas-jazz}, October 2017.
\newblock Accessed: 2023-15-04.

\bibitem{Climate}
Zhen Su, Jürgen Kurths, and Henning Meyerhenke.
\newblock Network sparsification via degree- and subgraph-based edge sampling,
  2023.

\bibitem{SNAP-road}
{SNAP} {C}alifornia road network.
\newblock \url{https://snap.stanford.edu/data/roadNet-CA.html}.
\newblock Accessed: 2023-30-04.

\end{thebibliography}

\begin{spacing}{0.93}
\listofalgorithms
\addcontentsline{toc}{section}{List of Algorithms, Figures and Tables}
\listoffigures
\listoftables
\end{spacing}

\end{document}